\documentclass[%
 reprint,
nofootinbib,
 amsmath,amssymb,
 aps,
]{revtex4-2}

\pdfoutput=1
\usepackage{graphicx}
\usepackage{dcolumn}
\usepackage{bm}
\usepackage{tabularx}
\usepackage{appendix}
\usepackage{graphicx}
\usepackage{colortbl}
\usepackage{color}
\usepackage{cancel}
\usepackage{comment}
\usepackage{subfig}
\usepackage[countmax]{subfloat}
\usepackage{framed}
\usepackage{mdframed}
\usepackage{multirow}
\usepackage{slashed}
\usepackage{booktabs} 
\usepackage{hyperref}
\usepackage{caption}
\usepackage{ragged2e}
\DeclareCaptionJustification{justified}{\justifying}
\usepackage{comment}
\usepackage{breakurl}

\newcolumntype{L}[1]{>{\hsize=#1\hsize\raggedright\arraybackslash}X}%
\newcolumntype{R}[1]{>{\hsize=#1\hsize\raggedleft\arraybackslash}X}%
\newcolumntype{C}[2]{>{\hsize=#1\hsize\columncolor{#2}\centering\arraybackslash}X}%

\begin{document}
\title{Gravitational Wave Gastronomy}

\author{David I. Dunsky}
\email{ddunsky@berkeley.edu}
\affiliation{ Berkeley Center for Theoretical Physics, University of California, Berkeley, CA 94720, USA}
\affiliation{Theoretical Physics Group, Lawrence Berkeley National Laboratory, Berkeley, California 94720, USA}
\author{Anish Ghoshal}
\affiliation{Institute of Theoretical Physics, Faculty of Physics, University of Warsaw,ul.  Pasteura 5, 02-093 Warsaw, Poland}
\affiliation{INFN - Sezione Roma “Tor Vergata”,
Via della Ricerca Scientifica 1, 00133, Roma, Italy}
\author{Hitoshi Murayama}
\email{hitoshi@berkeley.edu, hitoshi.murayama@ipmu.jp, Hamamatsu Professor}
\affiliation{ Berkeley Center for Theoretical Physics, University of California, Berkeley, CA 94720, USA}
\affiliation{ Theory Group, Lawrence Berkeley National Laboratory, Berkeley, CA 94720, USA}
\affiliation{ Kavli IPMU (WPI), UTIAS, The University of Tokyo, Kashiwa, Chiba 277-8583, Japan}
\author{Yuki Sakakihara}
\email{yukis@mail.sysu.edu.cn}
\affiliation{School of Physics and Astronomy, Sun Yat-sen University, Zhuhai 519082, China}
\author{Graham White}
\email{graham.white@ipmu.jp}
\affiliation{ Kavli IPMU (WPI), UTIAS, The University of Tokyo, Kashiwa, Chiba 277-8583, Japan}
\preprint{}
\begin{abstract}
The symmetry breaking of grand unified gauge groups in the early Universe often leaves behind relic topological defects such as cosmic strings, domain walls, or monopoles. 
For some symmetry breaking chains, hybrid defects can form where cosmic strings attach to domain walls or monopoles attach to strings. In general, such hybrid defects are unstable, with one defect ‘eating’ the other via the conversion of its rest mass into the other’s kinetic energy and subsequently decaying via gravitational waves. 
In this work, we determine the gravitational wave spectrum from 1) the destruction of a cosmic string network by the nucleation of monopoles which cut up and `eat' the strings, 2) the collapse and decay of a monopole-string network by strings that ‘eat’ the monopoles, 3) the destruction of a domain wall network by the nucleation of string-bounded holes on the wall that expand and `eat' the wall,  and 4) the collapse and decay of a string-bounded wall network by walls that `eat' the strings. We call the gravitational wave signals produced from the `eating' of one topological defect by another \textit{gravitational wave gastronomy}. We find that the four gravitational wave gastronomy signals considered yield unique spectra that can be used to narrow down the SO$(10)$ symmetry breaking chain to the Standard Model and the scales of symmetry breaking associated with the consumed topological defects. Moreover, the systems we consider are unlikely to have a residual monopole or domain wall problem.

\end{abstract}
\date{\today}

\maketitle
\flushbottom

\newpage

\section{Introduction}
The Universe is transparent to gravitational waves, even at very early times. Therefore, the search for a cosmological gravitational wave background provides a new way of observing our early cosmic history. Furthermore, the Hubble scale $H$ for cosmic inflation in the primordial universe could be as large as $5\times 10^{13}$ GeV \cite{Planck:2018jri} (for a review see \cite{Baumann:2009ds}), implying the early Universe could have reached energies 
far beyond that of Earth based colliders. Therefore, gravitational wave physics is a unique probe of extremely high scale physics.  \par 
A particularly promising class of sources for primordial gravitational waves arises from topological defects produced during certain types of transitions that spontaneously break a symmetry. Cosmic strings, domain walls and 
textures all produce a gravitational wave power spectrum with an amplitude that monotonically increases with the scale of the symmetry breaking \cite{Vilenkin:1981bx}. This implies that gravitational waves from topological defects are a unique probe of very high scale physics. We are coming into a golden age of gravitational wave cosmology, with new experiments using pulsar timing arrays \cite{Janssen:2014dka,Lentati:2015qwp,NANOGrav:2020spf}, astrometry \cite{boehm2017theia,Moore:2017ity,Garcia-Bellido:2021zgu}, space and ground based interferometry \cite{TheLIGOScientific:2014jea,Moore:2014lga,Inomata:2018epa,Maggiore:2019uih,Caprini:2019egz,Reitze:2019iox,Kawamura:2020pcg} all due to come online in the next few decades and probing frequencies from the nanohertz to kilohertz range. Indeed, NANOGrav and PPTA might have already seen evidence of a primordial gravitational wave background \cite{NANOGrav:2020spf, Goncharov:2021oub} which can be corroborated by future pulsar timing arrays and astrometry \cite{Garcia-Bellido:2021zgu}. Information about the Universe at very early times and very high energy could be just over the horizon. \par 
Of particular interest at the high scale is the possibility that the gauge groups in the Standard Model could unify to a single gauge group, perhaps through a series of intermediate steps (for a review see \cite{Croon:2019kpe}). There are two remarkable hints that this might be the case: First, the gauge anomalies of the Standard Model miraculously cancel 
- a miracle that is necessary for the consistency of the theory and can be explained by an anomaly-free unified gauge group that has been spontaneously broken.
Second, the gauge coupling constants in the standard model approximately unify at a scale of around $10^{15}$ GeV. On top of these hints, if $B-L$ local symmetry is embedded in a unified group, the Baryon asymmetry can be generated through leptogenesis when this U(1)${}_{B-L}$ is spontaneously broken in the early Universe \cite{Buchmuller:2005eh}. \par 
While elegant, these Grand Unified Theories (GUTs) are notoriously difficult to test due to the high scales involved. Many symmetry breaking paths predict topological defects that are in conflict with present day cosmology unless their relic abundances are heavily diluted. For example, even a small flux of monopoles can destroy the magnetic fields of galaxies or potentially catalyze proton decays \cite{Parker:1970xv,Turner:1982ag,Callan:1982au,Rubakov:1982fp}. Moreover, domain walls, which dilute slowly with the expansion of the Universe, can come to dominate the energy density of the Universe which conflicts with the standard $\Lambda$CDM cosmology \cite{Zeldovich:1974uw}. \par 
A solution to these problematic defects is for inflation to dilute their abundance \cite{Guth:1980zm,Linde:1981mu}, which puts a qualitative constraint on the cosmological history of the Universe.  Another possibility is for the problematic defects to be `eaten' by another defect which is determined solely by the symmetry breaking. For example, for some symmetry breaking chains, strings can be cut by the Schwinger nucleation of monopole-antimonopole pairs \cite{Langacker:1980kd,Lazarides:1981fv,vilenkin1982cosmological} which `eat' the string before annihilating themselves. Similarly, in other symmetry breaking chains, domain walls can be consumed by the Schwinger nucleation of strings on their surface or can be cut into pieces of string-bounded walls by a pre-existing string network \cite{,Kibble:1982ae,Kibble:1982dd,Everett:1982nm} and later decay via gravitational waves. We call the  gravitational wave signatures from the `eating' of one defect by another \textit{gravitational wave gastronomy}.

We for the first time derive the gravitational wave spectrum that arises from symmetry breaking paths that form walls bounded by strings. The case where the domain walls destroy a pre-existing string network and where the walls are consumed by string nucleation generate distinct gravitational wave spectra. The former scenario is particularly interesting since it always occurs in chains that allow hybrid wall-bounded strings when inflation occurs prior to string formation. The latter scenario arises when inflation occurs between string and wall formation scale. Moreover, string nucleation on the wall is a tunneling process exponentially sensitive to the degeneracy between the cube of the string tension, $\mu^{3}$, and square of the wall tension, $\sigma^{2}$, and hence requires a coincidence of scales that is unnecessary in the first case. We also revisit the gravitational wave spectrum predicted from monopoles consuming strings \cite{Buchmuller:2019gfy,Buchmuller:2020lbh} and derive, for the first time, the gravitational wave spectrum that arises from strings eating a pre-existing monopole network. This case is again particularly interesting since it always occurs in chains that allow hybrid monopole-bounded strings when inflation occurs before monopole formation. 
\footnote{During the writing of this manuscript, the power spectrum we predict was independently derived in Ref. \cite{Buchmuller:2021mbb} which confirms the results in this paper for the monopole eating strings gastronomy scenario of Sec. \ref{sec:schwingerStringsMonopole}.}. 

Overall, we find that all types of hybrid defects generate distinguishable gravitational wave signals, implying that gravitational wave gastronomy is a remarkably promising method for learning information about the symmetry breaking chain that Nature chose to follow.  Such a program is at the very least complimentary to other probes of high scale physics including searches for lepton number violation in neutrinoless double beta decay \cite{Schechter:1981bd,Dolinski:2019nrj}, searches for non-Gaussianities in the CMB \cite{Chen:2009zp,Baumann:2011nk,Assassi:2012zq,Chen:2012ge,Noumi:2012vr,Arkani-Hamed:2015bza,Dimastrogiovanni:2015pla,Lee:2016vti,Kehagias:2017cym,An:2017hlx,Kumar:2017ecc,Baumann:2017jvh,Franciolini:2017ktv,Arkani-Hamed:2018kmz} and searches for proton decay \cite{Weinberg:1979sa,Wilczek:1979hc,Weinberg:1980bf,Sakai:1981pk,Super-Kamiokande:2014otb,Super-Kamiokande:2016exg,King:2020hyd,King:2021gmj}. Finally, in the case where monopoles are produced alongside strings, the possibility was raised that the strings dilute slowly enough that they can be replenished after there is enough $e$-foldings of inflation to dilute the monopoles. This results in a unique signal \cite{Cui:2019kkd}. We show explicit symmetry breaking chains that can accommodate this signal in Sec. \ref{sec:pre-inflation} and discuss how both strings and domain walls can sometimes replenish after monopoles are diluted away to reform a scaling network.\par 
The structure of this paper is as follows. In Section \ref{sec:topologicalDefectsFromGUTs} we review the menu of topological defects that can be generated from symmetry breaking and give an overview of all possible symmetry breaking paths from the SO$(10)$ GUT group that generate that can generate an observable gravitational wave signature. 
Finally, we make more general statements about all gauge groups by deriving a set of homotopy selection rules in order to argue that our menu of possible signals is complete and general. In Section \ref{sec:GWDetectors}, we review upcoming prospects for gravitational wave detection, including possible ways of constraining or detecting high frequency signals. In section \ref{sec:schwingerStringsMonopole} we consider the gravitational wave spectrum of monopoles consuming strings via Schwinger nucleation, and in Section \ref{sec:stringDestruction}, strings consuming a pre-existing monopoles network. In Section \ref{sec:dwconsumedbystrings} we consider strings consuming domain walls via Schwinger nucleation, and in Section \ref{sec:stringboundedWalls} domain walls consuming a pre-existing string network. In Section
\ref{sec:pre-inflation} we briefly discuss 
topological defects that are washed out by inflation before summarizing our results and discussing how each gravitational wave signal from hybrid defects can be distinguished in Section \ref{sec:summary}.

\section{Topological defects generated from Grand Unified Theories}
\label{sec:topologicalDefectsFromGUTs}
In this section, we review the menu of topological defects that can be produced by symmetry breaking chains. We then derive a set of topological selection rules and discuss four types of hybrid defects that commonly appear in $SO(10)$ GUTs.

\subsection{Menu of Topological Defects in Symmetry Breaking Paths
}
Let us begin by discussing the full set of defects that can occur in a symmetry breaking chain. As well as overviewing the defects conceptually, we will discuss the connection between the scale of symmetry breaking and the physical quantities - the domain wall surface tension, the string tension, and the monopole mass. We will find there is substantial flexibility in surface tension of the domain wall and the monopole mass, up to naturalness concerns, and only a moderate amount of flexibility in the relationship of the string tension and the associated symmetry breaking scales. \par 
Consider a gauge group $G$ spontaneously breaking to $H$.
In four dimensional spacetime, we have four possible topological defects that can arise during such a transition. Depending on the characteristics of the vacuum manifold $\mathcal{M}=G/H$, one can produce domain walls, cosmic strings, monopoles, and textures. The vacuum manifold is characterized by its homotopy class, that is the equivalence class of the maps from an $n$-dimensional sphere $S^n$ into $\mathcal{M}$, denoted as $\pi_n(\mathcal{M})$. We use the notation $I$ for trivial homotopy groups. If $\mathcal{M}$ is disconnected, $\pi_0(\mathcal{M})\neq I$, and two dimensional topological defects (domain walls) are formed through the symmetry breaking. Similarly, $\pi_1(\mathcal{M})\neq I$ predicts one dimensional defects (cosmic strings), $\pi_2(\mathcal{M})\neq I$ predicts point-like defects (monopoles), and $\pi_3(\mathcal{M})\neq I$ predicts three-dimensional defects (textures). \par 
Let us begin with a qualitative discussion of domain walls. A standard Mexican hat potential with a ${\mathbb Z}_2$ discrete symmetry
\begin{equation}
    V(\phi) = \frac{\lambda _\sigma }{4} (\phi ^2 - v^2 _\sigma)^2 \ ,
\end{equation}
will have a vacuum manifold that satisfies $\pi_0(\mathcal{M})\neq I$ and therefore admits domain walls. Consider a kink solution to the equation of motion between two degenerate vacuua
\begin{equation}
    \phi(x) = v _\sigma \tanh \left( \sqrt{\frac{\lambda _\sigma }{2}} v _\sigma x \right) \ .
\end{equation}
The surface tension of the wall is
\begin{equation}
    \sigma = \int _{-\infty } ^\infty dx \left( \frac{1}{2} \left[ \frac{\partial \phi (x)}{dx} \right]^2 + V(\phi(x)) \right)=\sqrt{\frac{8 \lambda _\sigma }{9}} v^3 _\sigma ,
  \label{eq:derivation_sigma}
\end{equation}
which, depending on the value of $\lambda_\sigma$, can in principle vary from an order of magnitude above $v_\sigma^3$ to arbitrarily small values (for a review see \cite{Nakayama:2016gxi}). To avoid committing to a particular form of a potential, throughout this paper we will parametrize the flexibility of the relationship between the surface tension and the symmetry breaking scale as
\begin{equation}
    \sigma = \epsilon v_\sigma ^3 \ .
\end{equation}
Note that although $\epsilon $ in principle can be arbitrarily small, naturalness will require $\epsilon \gtrsim g^2/4\pi \gtrsim 10^{-3}$. with the lower limit arising from Coleman-Weinberg one-loop quantum corrections, where $g \sim 0.1$ is the grand unified gauge coupling associated with the $Z_2$ symmetry above the scale $v_{\sigma}$. \par 
Next let us consider the case where the first homotopy group of the vacuum manifold is non-trivial, that is when strings can form. Consider a scalar theory with a $U(1)$ gauge symmetry,
\begin{equation}
    L= |D \phi |^2+V(\phi) +\frac{1}{4} F^2,  
\end{equation}
where $D_\mu=\partial _\mu - i e A_\mu$ is the covariant derivative. Again, we use the same form of the potential
\begin{equation}
    V=\frac{\lambda _\mu }{4} (|\phi |^2 - v^2_\mu )^2 \ .
\end{equation}
The classical equations of motion have the form
\begin{eqnarray}
  D^2 \phi + \frac{\lambda _\mu }{2}(|\phi |^2 - v^2_\mu  )\phi &=& 0\\
  \partial ^\mu F_{\mu \nu}-ie (\phi ^* D_\nu \phi - D_\nu \phi^* \phi ) &=& 0 
\end{eqnarray}
and admit a non-trivial solution of the form 
\begin{equation}
    \phi(r) = f(r) v _\mu e^{i \theta}, \ A_i = \frac{1}{er} A_\theta (r) \hat{\theta } _i,
\end{equation}
where $A(\infty)=f(\infty)=1$ and $A(0)=f(0)=0$.  The string tension can be found by substituting the string solution into the classical equations of motion into the Hamiltonian and integrating over the loop
\begin{eqnarray}
  \mu &=& \int r dr d \phi  \left[ \left| \frac{ \partial \phi}{\partial r} \right| ^2  + \left| \frac{1}{4} \frac{d \phi}{d \theta } - i q A_\theta ^\prime \phi \right| ^2  \right. \nonumber \\
  && \quad \quad \quad \quad  \left. + V(\phi) + \frac{B^\prime {} ^2}{2} \right] \\
  &=& 2 \pi  v^2 B\left(\frac{2 \lambda}{e^2}\right) \ .
  \label{eq:derivation_mu}
\end{eqnarray}
where $B ^\prime$ is the magnetic field related to the cosmic string. $B(x)$ is a slowly varying function that is equal to $1$ when $x=1$ and \cite{Hill:1987qx}
\begin{equation}
    B(x) \sim  \left\{ \begin{array}{cc}
        2.4/\ln (2/x ) & x < 10^{-2}  \\
        1.04 x^{0.195} & 10^{-2} < x \ll 1.
    \end{array} \right.
\end{equation}
for $x < 1$.
Since $2\lambda _\mu / e^2$ can in principle take a large range of values, there are many orders of magnitude that the argument of $B$ can take. However, as the function is so slowly varying, $\mu \sim v^2_\mu $ within an order of magnitude.\par 

Finally, let us consider monopoles which exist in the case where the second homotopy group of the vacuum manifold is non-trivial. That is, the vacuum is topologically equivalent to a sphere. For a simple example, consider a model with an $SU(2)$ gauge symmetry
\begin{equation}
    {\cal L}= \frac{1}{2}D^\mu \phi D_\mu \phi  - \frac{1}{4} B_{\mu \nu} B^{\mu \nu} - \frac{\lambda}{4} \left(\phi ^2 - v^2_m \right)^2 \ ,
\end{equation}
where $\phi$ is a real $SU(2)$ triplet.
The `t Hooft-Polyakov monopole \cite{tHooft:1974kcl,Polyakov:1974ek} has the behavior 
\begin{eqnarray}
  \phi &=& \hat{r} \frac{h(v_m e r)}{er} \nonumber \\ 
  W^i _a &=& \epsilon _{aij} \hat{x} ^j \frac{1-f(v_m er)}{er} \nonumber \\
  W_a^0 &=& 0 \, ,
\end{eqnarray}
where, using the shorthand $\xi = v_m e r$ for the product of $v_m$ with the gauge coupling constant $e$ and radial coordinate $r$, the functions $f$ and $h$ are solutions to the equations \cite{tHooft:1974kcl,Polyakov:1974ek}
\begin{eqnarray}
  \xi ^2 \frac{d^2f}{d \xi^2} &=& f(\xi) h(\xi )^2 + f(\xi) (f(\xi)^2 -1) \\
  \xi^2  \frac{d^2 h}{d \xi ^2} &=& 2 f(\xi )^2 h(\xi) + \frac{\lambda}{e^2} h(\xi) (h(\xi)^2 - \xi ^2) \ .
\end{eqnarray}
The boundary conditions satisfy $\lim _{\xi \to 0} f(\xi)-1= \lim _{\xi \to 0}h(\xi ) \sim {\cal O}(\xi )$ and $\lim _{\xi \to \infty}f(\xi) = 0 $, $\lim _{\xi \to \infty} h(\xi )  \sim \xi$. The monopole mass again comes from solving the equations of motion and then calculating the static Hamiltonian,
\begin{eqnarray}
\label{eq:monopoleMass}
  E=m&=& \frac{4 \pi v_m}{e} \int _0 ^\infty \frac{d \xi}{\xi ^2 }\left[ \xi^2 \left( \frac{d f}{d\xi} \right)^2  +\frac{1}{2} \left( \xi \frac{dh}{d \xi} - h \right)^2 \right.  \nonumber \\ 
  && \left. + \frac{1}{2}(f^2-1)^2+ f^2h^2 + \frac{\lambda}{4 e^2}(h^2- \xi ^2)^2 \right] \ . 
\end{eqnarray}
 It has the form
\begin{equation}
    m=\frac{4 \pi v_m}{e} f(\lambda / e^2) \ .
\end{equation}
The solution \eqref{eq:monopoleMass} has been calculated numerically for multiple values, and one finds that for $0.1<\lambda /e^2<10^1$, $f(\lambda /e^2)$ is slowly varying ${\cal O}(1)$ function \cite{tHooft:1974kcl}.

In conclusion there is a reasonably tight relationship between the symmetry breaking scale and the string tension $\mu \sim v_\mu ^2$. However, domain walls can have a significantly smaller surface tension than the cube of the symmetry breaking scale and the monopole mass can be well above $v_m$. Even still, one should expect from naturalness considerations for all relevant quantities to be within a few orders of magnitude of the relevant powers of the symmetry breaking scale.
\begin{figure}
    \centering
    \includegraphics[width=0.49\textwidth]{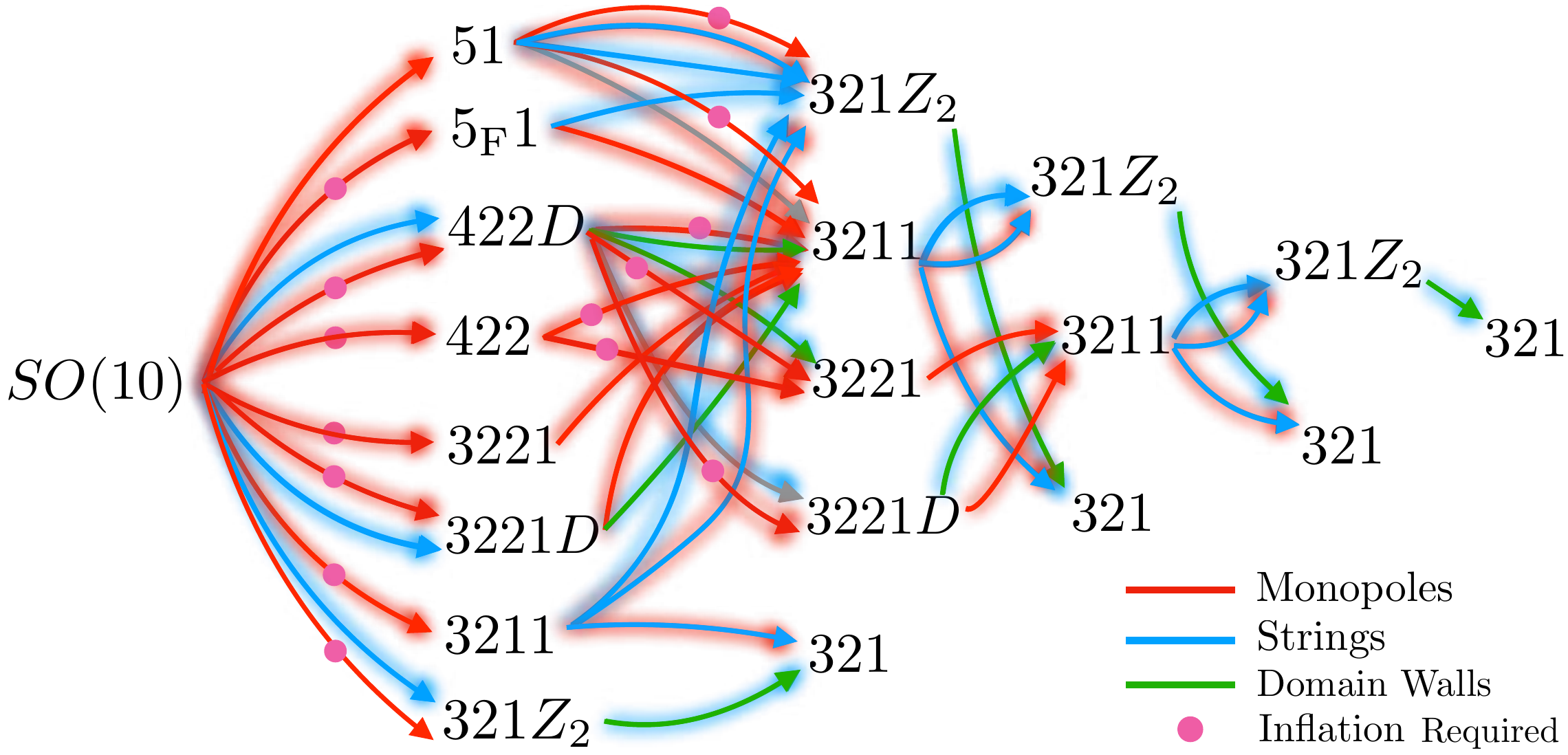}
    \caption{A sample of $SO(10)$ symmetry breaking paths down to the Standard Model that produce hybrid defects. 
    The color of the arrows denote the type of topological defect produced; red corresponds to magnetic monopoles, blue to cosmic strings, and green to domain walls. A glowing red path indicates magnetic monopoles attaching to cosmic strings while a glowing blue path indicates cosmic strings attaching to domain walls.   Note that the lower dimensional (boundary) defect of a hybrid defect always arises from an earlier stage of symmetry breaking than the higher dimensional (bulk) defect. As shown in Appendix \ref{ap:homotopySelectionRules}, this is generic feature of hybrid defects. 
    Red arrows with a dot indicate the monopoles need to be inflated away after the formation.
    These chains can still generate a gastronomy signal via nucleation of monopoles on strings or strings on walls after inflation. Gray arrows with red and blue glow indicate that the symmetry breaking leads to no defects associated with strings bounded by monopoles and domain walls bounded by strings, respectively.
    } 
    \label{fig:chains}
\end{figure}

\subsection{Hybrid Defects in Grand Unified Theories}
In the previous subsection, we considered the various types of topological defects that can be generated during a single symmetry breaking $G \rightarrow H$. Now, with the table set, we consider how a sequence of multiple transitions,
\begin{align}
   G \rightarrow H \rightarrow K ,
\end{align}
can give rise to hybrid topological defects composed of two different dimensional defects. For these hybrid defects, the bulk topological defect converts its rest mass to the kinetic energy of the boundary defect, leading to the appearance of one defect consuming the other. The relativistic motion of these defects leads to gravitational wave emission and eventual decay of the composite defect.

Consider first the case when the vacuum manifold $H/K$ is not simply connected but the full vacuum manifold, $G/K$ is. Then $\pi_1(H/K) \neq I$ and strings form at the transition $H \rightarrow K$. However, these strings are topologically unstable since in the full theory, $\pi_1(G/K) = I$, which does not permit stable strings below $K$. The topological instability of strings manifests itself by the nucleation of  magnetic monopole pairs that cut and `eat' the string \cite{Lazarides:1981fv,vilenkin1982cosmological} (see Fig. \ref{fig:WKBString}).  A set of homotopy selection rules proven in Appendix \ref{ap:homotopySelectionRules} show that the monopoles that nucleate on the string boundaries must always arise from the earlier phase transition $G \rightarrow H$ so that $\pi_2(G/H) \neq I$ and $v_m \geq v_\mu$. The gastronomy scenario of monopoles nucleating and eating a string network is discussed in Sec. \ref{sec:schwingerStringsMonopole}. 

\begin{figure}
    \centering
    \includegraphics[width=0.45\textwidth]{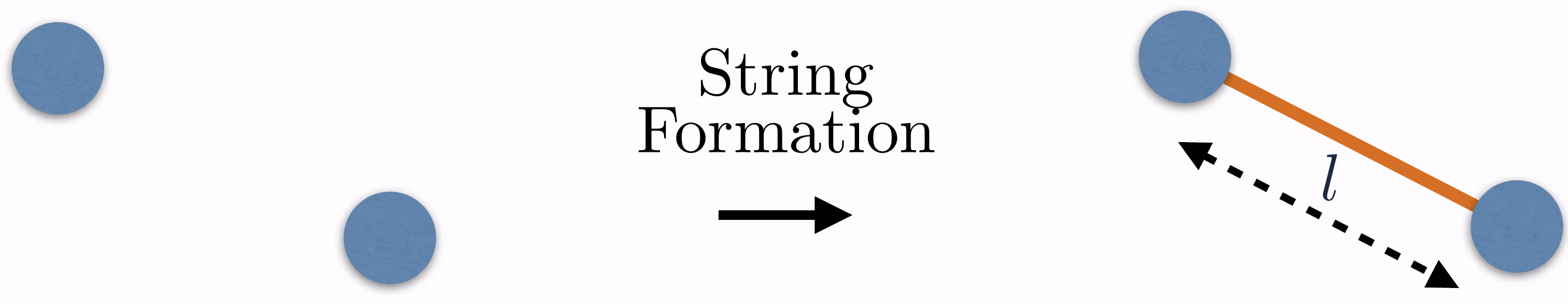}
    \caption{Illustration of monopoles connecting to strings below the string formation scale, $v_\mu$. At $v_\mu$, the magnetic field of the monopoles squeezes into flux tubes (strings) with the typical string length $l$ set by the monopole density at $v_\mu.$}
    \label{fig:StringDestructionDiagrom}
\end{figure}

The requirement that $G \rightarrow H$ generates monopoles that can attach to strings implies that, if inflation occurs before monopole formation, a significant number of monopoles can already be in the horizon at the time of string formation.  In this scenario, the magnetic field lines between monopole and antimonopole pairs squeeze into flux tubes (strings) after $H \rightarrow K$ (see Fig. \ref{fig:StringDestructionDiagrom}) and hence strings bounded by monopoles form right at the string formation scale $v_\mu$ \cite{Langacker:1980kd,Lazarides:1981fv,vilenkin1982cosmological}. The gastronomy scenario of strings attaching to and eating a pre-existing monopole network is discussed in Sec. \ref{sec:stringDestruction}.

Similarly, consider now the case when the vacuum manifold $H/K$ is disconnected but the full vacuum manifold, $G/K$ is connected. Then $\pi_0(H/K) \neq I$ and domain walls form at the transition $H \rightarrow K$. However, these domain walls are topologically unstable since in the full theory, $\pi_0(G/K) = I$, which does not permit stable domain walls below $K$. The topological instability of walls manifests itself by the nucleation of string-bounded holes on the wall (see Fig. \ref{fig:WKBWalls}) which expand and `eat' the wall \cite{Kibble:1982dd}. The same set of homotopy selection rules derived in Appendix \ref{ap:homotopySelectionRules} shows that the strings that nucleate on the wall must always arise from the earlier phase transition $G \rightarrow H$ so that $\pi_1(G/H) \neq I$ and $v_\mu \geq v_\sigma$. The gastronomy scenario of strings nucleating and eating a domain wall network  is discussed in Sec. \ref{sec:dwconsumedbystrings}.

\begin{figure}
    \centering
    \includegraphics[width=0.45\textwidth]{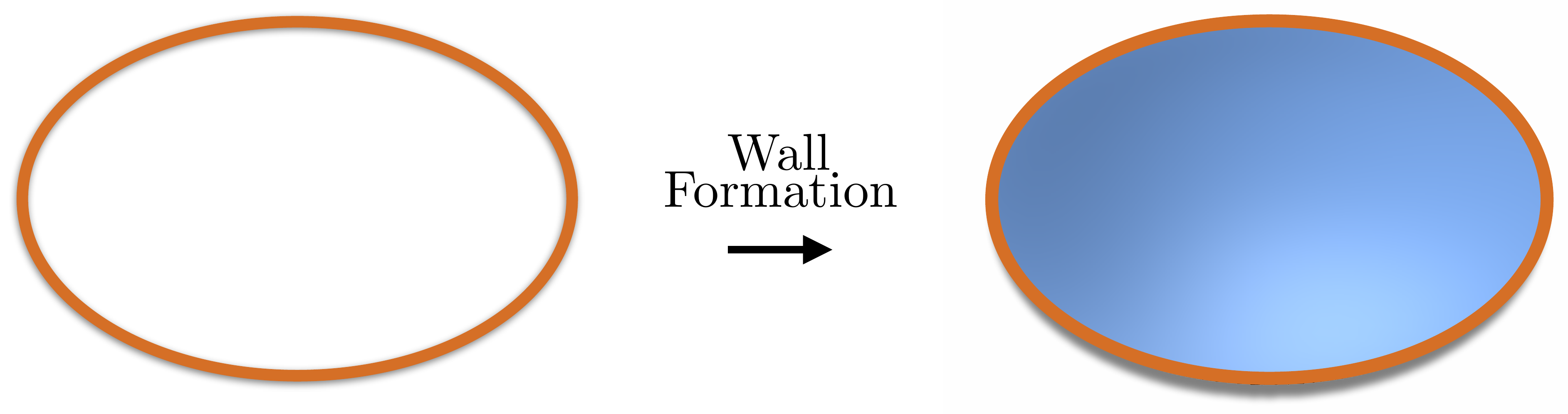}
    \caption{Illustration of strings connecting to walls below the wall formation scale, $v_\sigma$. The walls fill in the area between strings because winding the Higgs field, $\phi$, responsible for the symmetry breaking at $v_\sigma$ around a string necessarily generates a discontinuity in $\phi$ \cite{Preskill:1986kp,Hindmarsh:1994re}. As a result, a structure must abruptly change $\phi$ back to ensure the continuity of $\phi$. This structure is the domain wall.}
    \label{fig:WallFormationDiagrom}
\end{figure}

The requirement that $G \rightarrow H$ generates strings that can attach to walls implies that, if inflation occurs before string formation, a significant number of strings can already be in the horizon at the time of wall formation.  In this scenario, the space between strings is filled with a wall after $H \rightarrow K$  (see Fig. \ref{fig:WallFormationDiagrom}) and hence walls bounded by strings form right at the wall formation scale $v_\sigma$ \cite{Kibble:1982dd,Everett:1982nm}. The gastronomy scenario of walls attaching to and eating a pre-existing string network is discussed in Sec. \ref{sec:stringboundedWalls}.


In many GUT symmetry breaking chains to the Standard Model gauge group $G_{\rm SM}$, these type of homotopy sequences occur and hybrid defects form. Indeed, both $\pi_1(SO(10)/G_{\rm SM}) = I$ and $\pi_0(SO(10)/G_{\rm SM}) = I$ so that \textit{at least} one string or domain wall that forms during the intermediate breaking of $SO(10)$ down to $G_{\rm SM}$ must become part of a composite defect which can lead to the gastronomy signals of Sec. \ref{sec:schwingerStringsMonopole}-\ref{sec:stringboundedWalls}.

To see how ubiquitous hybrid topological defects are, we depict in Fig. \ref{fig:chains} a sample of possible cosmic histories of $SO(10)$ breaking and the topological defects produced at each stage.
The color of the arrows in Fig. \ref{fig:chains} denotes which type of defect is produced at each stage of breaking, with strings in blue, walls in green, and monopoles in red. The chains which produce monopoles that become attached to strings are shown by the glowing red paths while the chains which produce strings that become attached to walls are shown by the glowing blue paths. The meaning of each gauge group abbreviation is as follows:
\begin{align}
   \label{eq:symmetryNotation}
    {\rm 51}&=SU(5)\times U(1)_X / {\mathbb Z}_5 \ , \nonumber \\
    {\rm 5_{F}1}&= SU(5)_{\rm flipped} \times U(1)_{\rm flipped} / {\mathbb Z}_5\ , \nonumber \\
    {\rm 422}&= SU(4)_c \times SU(2)_L \times SU(2)_R / {\mathbb Z}_2 \ , \nonumber \\
    {\rm 3221}&= SU(3)_c\times SU(2)_L \times SU(2)_R\times U(1)_{B-L} / {\mathbb Z}_6  \ ,\nonumber \\
    {\rm 3211}&= 
        SU(3)_c\times SU(2)_L \times  U(1)_Y \times U(1)_{X} / {\mathbb Z}_6  \ ,\nonumber \\
    321 &= SU(3)_c\times SU(2)_L \times U(1)_{Y} / {\mathbb Z}_6 .
\end{align}
$D$ refers to D-parity, a discrete charge conjugation symmetry ~\cite{Kibble:1982ae,Kibble:1982dd}, 
$Z_2$ refers to matter parity, 
and $G_{\rm SM}  = 321$.


Note that the sequence for forming strings bounded by monopoles in Fig. \ref{fig:chains} is typically realized by the two-stage sequence \cite{vilenkin2000cosmic}
\begin{align}
    G \xrightarrow{\rm monopoles} H \times U(1) \xrightarrow{\rm strings} H,
\end{align}
with $\pi_1(G/H) = I$.
Monopoles form in the first transition when $G$ breaks to a subgroup containing a $U(1)$ and strings form and connect to the monopoles when this \textit{same} $U(1)$ is later broken.  
Likewise, walls bounded by strings are typically realized in the two-stage sequence \cite{vilenkin2000cosmic}
\begin{align}
    G \xrightarrow{\rm strings} H \times Z_2 \xrightarrow{\rm walls} H,
\end{align}
with $\pi_0(G/H) = I$. Strings form in the first transition when $G$ breaks to a subgroup containing a discrete symmetry (since 
$\pi_1(G/(H \times Z_2))\supseteq \pi_0(H \times Z_2) \neq I)$).  The walls form and connect to the strings when the \textit{same} discrete symmetry associated with the strings is broken. 

As indicated in Fig.~{\ref{fig:chains}}, many symmetry breaking paths from $SO(10)$ to the Standard Model yield hybrid defects. An example chain that produces all hybrid defects discussed in this paper is $SO(10) \rightarrow 5_{\rm F} 1 \rightarrow 3211\rightarrow 321Z_2 \rightarrow 321$, which we now go over as a concrete example of the different types of gastronomy signals discussed in this paper.

In the first breaking, $SO(10)\rightarrow 5_{\rm F}1$ generates 
monopoles which must be inflated away. The second breaking, $5_{\rm F}1 \rightarrow 3211$, also generates monopoles, but these lighter monopoles can get connected by the strings formed at the third breaking, $3211\rightarrow 321Z_2$.  Thus, this sequence can produce gravitational wave gastronomy signals discussed in Secs. \ref{sec:schwingerStringsMonopole} and \ref{sec:stringDestruction} with each section corresponding to when inflation occurs relative to monopole and string formation. Specifically, if inflation dilutes both heavy and light monopoles before the strings form, then the string network evolves as a pure string network until light monopoles nucleate and `eat' the string network (Sec. \ref{sec:schwingerStringsMonopole}). Note that for the nucleation to occur within cosmological timescales, the relative hierarchy between the second and third symmetry breaking chains cannot be too large.  However, if inflation occurs before the formation of the light monopoles, then the light monopoles connect to strings at the string formation scale and `eat' the monopole network (Sec. \ref{sec:stringDestruction}). 

At the third breaking, in addition to the previous strings, $Z_2$-strings appear and get filled by the domain wall formed at the fourth breaking, $321\times Z_2 \rightarrow 321$. This sequence can produce gravitational wave gastronomy signals discussed in Secs. \ref{sec:dwconsumedbystrings} and \ref{sec:stringboundedWalls}, with each section corresponding to when inflation occurs relative to string and wall formation. If inflation dilutes the $Z_2$ strings before the walls form, then the wall network evolves as a pure wall network until $Z_2$ strings nucleate and `eat' the wall (Sec. \ref{sec:dwconsumedbystrings}). For the wall to nucleate strings  before dominating the energy density of the Universe requires a relatively small hierarchy between the third and fourth symmetry breaking scales. However, if inflation occurs before the $Z_2$ strings form, then the strings get filled by the domain walls and the walls proceed to `eat' the string network (Sec. \ref{sec:stringboundedWalls}). In this gastronomy scenario, no degeneracy between scales is necessary.

\section{Gravitational wave detectors}
\label{sec:GWDetectors}

Topological defects leave a variety of gravitational wave signals that are in many cases detectable by proposed experiments. This means that gravitational wave detectors have a unique opportunity to probe the cosmological history of symmetry breaking. In the nHz to $\mu$Hz range, pulsar timing arrays including EPTA, PPTA, NANOGrav and SKA~\cite{Arzoumanian:2020vkk, Manchester:2006xj, Lentati:2015qwp,Janssen:2014dka} and astrometry including Gaia and Theia~ \cite{Book:2010pf,Moore:2017ity,Mihaylov:2018uqm,Mihaylov:2019lft,Garcia-Bellido:2021zgu} can reach impressive sensitivity over the next few years. Spaced based interferometry experiments including LISA \cite{Audley:2017drz} (Tianqin~\cite{TianQin:2015yph, TianQin:2020hid} and Taiji~\cite{Hu:2017mde} also cover similar regions), DECIGO \cite{Kawamura:2020pcg}, and the Big Bang Observer (BBO) \cite{Harry:2006fi}, all will probe mHZ to Hz frequencies. Atom interferometry experiments including AEDGE \cite{Bertoldi:2019tck}, AION \cite{Badurina:2019hst} and MAGIS \cite{Graham:2017pmn} will probe a similar range. Finally, ground based experiments including aLIGO and aVIRGO \cite{harry2010advanced,LIGOScientific:2014pky,VIRGO:2014yos,LIGOScientific:2019lzm}, KAGRA~\cite{KAGRA:2020cvd}, the Cosmic Explorer (CE) \cite{Reitze:2019iox}, and the Einstein Telescope (ET) \cite{Maggiore:2019uih} are in principle sensitive to the frequencies up to around a kHz. 

Many topological defects leave quite broad spectra which can lead to a boost in the naive sensitivity of a detector \cite{Thrane:2013oya}. The integrated sensitivity of a detector to a specific signal is given by the signal to noise ratio
\begin{equation}
    \text{SNR} = \sqrt{T \int _{f_{\rm min}}^{f_{\rm max}} df\left[ \frac{h^2 \Omega _{\rm GW}(f)}{ h^2 \Omega _{\rm sens}(f)}  \right]^2}
\end{equation}
where $T$ is the observation time of the detector, $\Omega _{\rm sens}(f)$ is the sensitivity to a monochromatic gravitational wave spectrum. To register a detection,  the SNR must be above $1$ as indicated by the sensitivity curves of \cite{Schmitz:2020syl}, which we use throughout this work. In some cases the defects are only visible at frequencies higher than the reach of the above experiments. This can occur either in the case of strings consuming a pre-existing monopole network or walls consuming a pre-existing string network. Unfortunately, the strongest projected sensitivity at present for frequencies above a few kHz is the bound arising from constraints on the expansion of the Universe during big bang nucleosynthesis and recombination \cite{Aggarwal:2020olq}. The current constraint on the expansion rate of the Universe is generally expressed in terms of the departure from the Standard Model prediction of the effective number of relativistic degrees of freedom,
\begin{equation}
    \label{eq:deltaNeff}
    \Delta N_{\rm eff} = \frac{8}{7} \left( \frac{11}{4} \right)^{4/3}\frac{\Omega _{\rm GW}^0}{
    \Omega_\gamma}
\end{equation}
where
\begin{equation}
    \Omega ^0 _{\rm GW} = \int  \frac{df}{f} \Omega _{\rm GW} (f) \ .
\end{equation}
$\Delta N_{\rm eff}$ constraints on the total energy density of gravitational waves can provide powerful bounds on defects which only leave a high frequency, but large amplitude, gravitational wave spectrum.
Current constraints on $\Delta N_{\rm eff }<0.284$ arise from the Planck 2018 dataset using TT,TE,EE+lowE+lensing \cite{Aghanim:2018eyx}. This is expected to improve significantly to $\Delta N_{\rm eff} < 0.03$ as a conservative estimate of the sensitivity of next generation experiments \cite{Abazajian:2016yjj}. A hypothetical experiment limited only by the cosmic variance limit was found to be sensitive to changes to the number of relativistic degrees of freedom as small as \cite{Ben-Dayan:2019gll}
\begin{equation}
    \Delta N_{\rm eff} ^{\rm CVL} < 
    3.1 \times 10^{-6}
\end{equation}
which is in principle sensitive to gravitational wave spectra at arbitrary frequency with an amplitude as small as ${\cal O}(10^{-12})$. 
Beyond cosmological limits, there are promising proposals using interferometers \cite{Akutsu:2008qv,Nishizawa:2008se,Chou:2016hbb,Martinez:2020cdh,Vermeulen:2020djm} ($10^3-10^7$ Hz), levitated sensors \cite{Aggarwal:2020umq} ($10^3-10^4$ Hz) and magnetic conversion \cite{Ringwald:2020ist} ($10^9-10^{10}$ Hz) which may probe high frequency gravitational wave cosmology as summarized in ref. \cite{Aggarwal:2020olq}. 

We now turn to calculating the gravitational wave gastronomy signal for strings bounded by monopoles and walls bounded by strings.

\section{Monopoles Eating Strings}
\label{sec:schwingerStringsMonopole}
\begin{figure}
    \centering
    \includegraphics[width=0.45\textwidth]{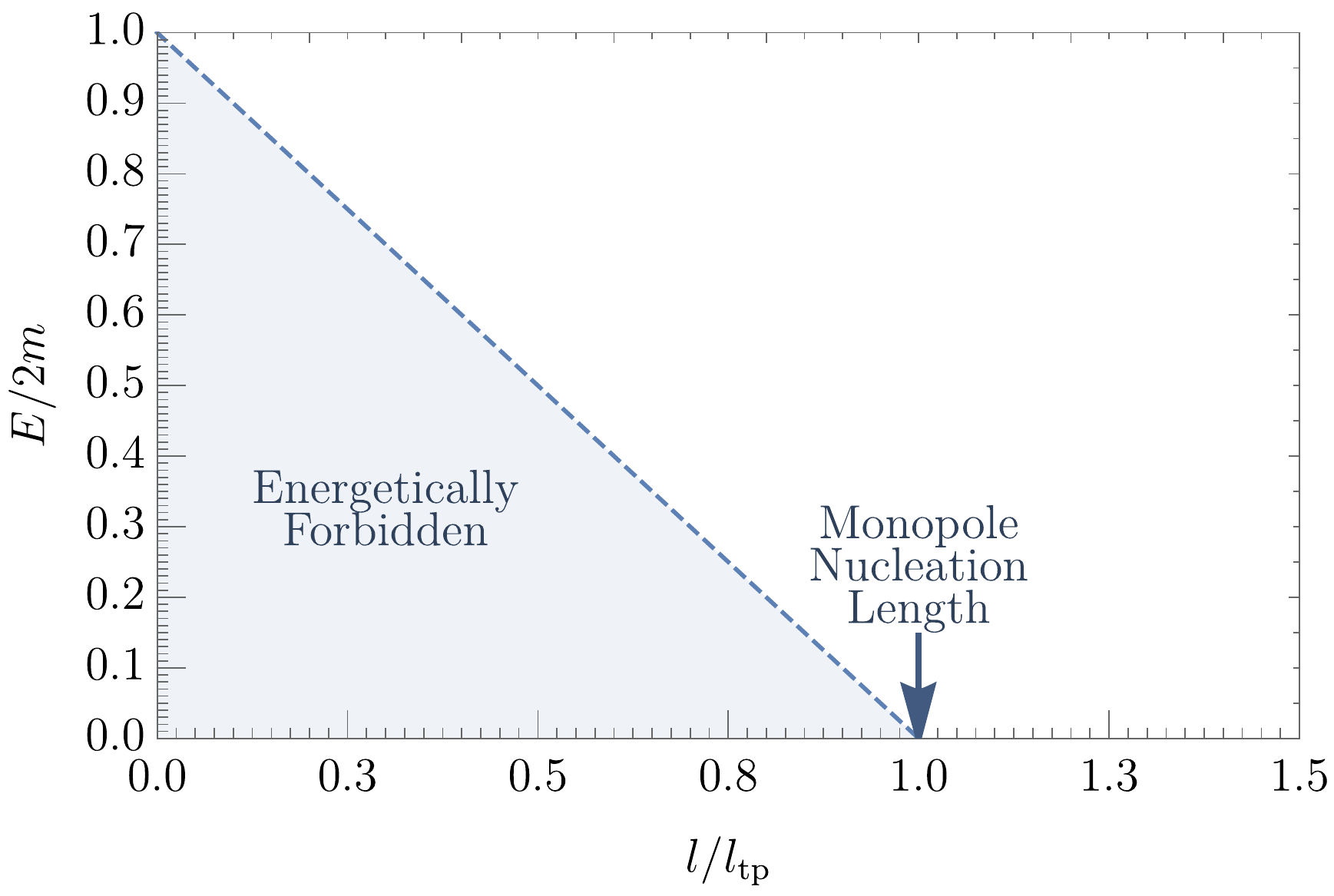}\vspace{.2cm}
    \includegraphics[width=0.45\textwidth]{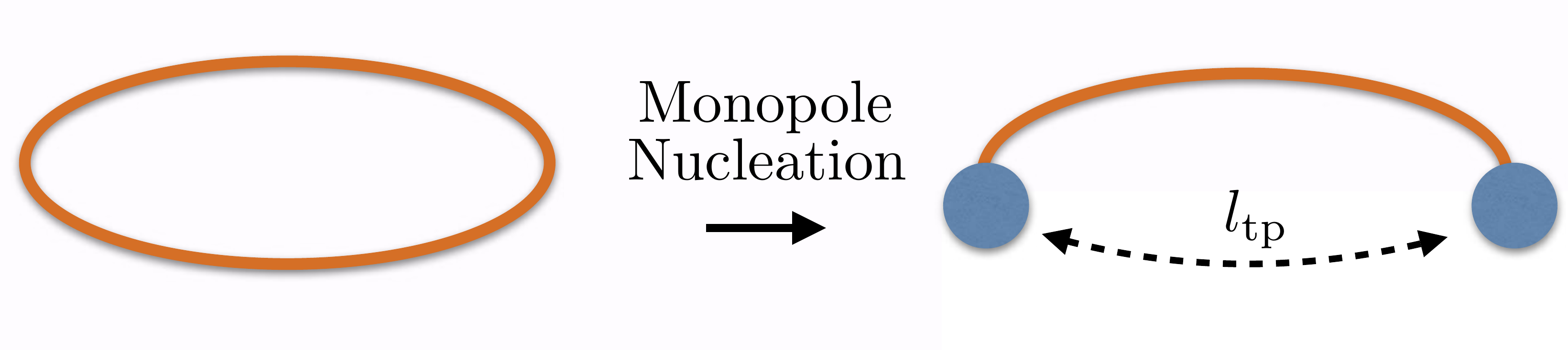}
    \caption{Top: Free energy diagram for a pair of monopoles nucleating on a string vs their nucleation separation, $l$. For $l > l_{\rm tp}$, the free energy of the system turns negative and it becomes energetically possible to nucleate a pair of monopoles in place of a string segment of length $l_{\rm tp}$. Bottom: Illustration of the nucleation process. For strings with length $l > l_{\rm tp}$ a string segment of length $l_{\rm tp}$ is `eaten' and replaced with a monopole-antimonopole pair which form the boundaries of the cut string piece.}
    \label{fig:WKBString}
\end{figure}

In this section, we consider the gastronomy signal of monopoles nucleating on strings.  As shown in Sec. \ref{sec:topologicalDefectsFromGUTs}, if they are related by the same $U(1)$, monopoles form first, (in the initial phase transition that leaves an unbroken $U(1)$ symmetry), and connect to strings in the second phase transition (when the $U(1)$ is broken). 
When inflation occurs after the formation of monopoles but before strings, the monopole abundance is heavily diluted by the time the strings form. The absence of monopoles initially prevents the formation of monopole-bounded strings at the second stage of symmetry breaking and the strings initially evolve as a normal string network.
Nevertheless, the strings can later become bounded by monopoles by the Schwinger nucleation of monopole-antimonopole pairs, which cuts the string into pieces bounded by monopoles as shown in Fig. \ref{fig:WKBString}. Conversion of string rest mass into monopole kinetic energy leads to relativistic oscillations of the monopoles before the system decays via gravitational radiation and monopole annihilation \cite{Martin:1996cp,Leblond:2009fq,Buchmuller:2021mbb}. 

Monopoles can only nucleate if it is energetically possible to. The energy cost of producing a monopole-antimonopole pair is $2 m$ where $m$ is the mass of each monopole, and the energy gained from reducing a string segment of length $l$  is $\mu l$ where $\mu$ is the string tension. The free energy of the monopole-string system is then
 \begin{equation}
    \label{eq:freeEnergyMonopoleString}
     E = 2m - \mu l \ .  
 \end{equation}
The energy balance between monopole creation and string length reduction leads to a critical string length, $l_{\rm tp}$, above which it is energetically favorable for the string to form a gap of length $l_{\rm tp}$ separating two monopole endpoints, as shown in Fig \ref{fig:WKBString}. $E < 0$ gives this turning point length
 \begin{equation}
     l_{\rm tp} = \frac{2m}{\mu} \ .
 \end{equation}
The probability for the monopoles to tunnel through the classically forbidden region out to radius $l_{\rm tp}$ can be estimated from the WKB approximation. The nucleation rate per unit string length is
\begin{align}
    \Gamma_m\propto \mu e^{-S_E} \ ,
\end{align}
where 
 \begin{eqnarray}
   S_{\rm E} &=& \int _0 ^{l_{\rm tp}}dl \sqrt{2 m E} 
   \propto \frac{m^2}{\mu} \ ,
 \end{eqnarray}
More precisely, the tunneling rate per unit string length can be estimated from the bounce action formalism ~\cite{Kibble:1982dd, Preskill:1992ck} and is found to be ~\cite{Leblond:2009fq}
\begin{equation}
    \label{eq:monopoleNucleationRate}
    \Gamma _{m} = \frac{\mu }{2 \pi } {\rm exp} (- \pi \kappa_m ) \ ,
\end{equation}
where $\kappa_m = m^2/\mu $. As we saw in section \ref{sec:topologicalDefectsFromGUTs}, typically $m\sim v_m$ and $\mu \sim v_\mu^2$  with little flexibility. 
Therefore, the exponential sensitivity of the decay rate \eqref{eq:monopoleNucleationRate} implies that if the hierarchy between the monopole and string breaking energy scales is large, $\kappa_m \gg 1$ and the string is stable against monopole nucleation on time scales greater than the age of the Universe. If this occurs, the gravitational wave spectrum is identical to the standard stochastic string spectrum and no gastronomy signal is observable. Consequently, monopole nucleation typically requires a moderate coincidence of string and monopole scales, $v_m \sim v_\mu$, so that $\kappa$ is not extremely large.


The remaining ingredients needed to determine the gravitational wave power spectrum for a stochastic background of metastable strings is the string number density spectrum as a function of the loop size and time as well as the gravitational power spectrum for an individual string. Here, we use the number density of string loops, formed by the intercommutation of long (`infinite') strings in the superhorizon string network, as derived by the velocity-dependent one-scale (VOS) model~\cite{Martins:1995tg,Martins:1996jp,Martins:2000cs,Sousa:2013aaa}. After their formation,
the infinite string network quickly approaches a scaling regime, with approximately  $\mathcal{O}(1)$ long strings per horizon with curvature radius $R \approx t$ for all time $t$ prior to nucleation. In the one-scale model, the typical curvature radius and separation between infinite strings is the same scale, $R$, so that the energy density of the infinite string network is
\begin{align}
    \rho_{\infty} \approx \frac{\mu R}{R^3} \approx \frac{\mu}{t^2}.
\end{align}
Prior to monopole nucleation, string loops break off from the infinite string network as intercommutation byproducts, with roughly one new loop formed every Hubble time. Loops that form at time $t_k$ typically are of length $l_k \approx \alpha t_k$, where $\alpha \approx 0.1$ is found in simulations \cite{Blanco-Pillado:2017oxo,Blanco-Pillado:2013qja}. 
If the probability a long string intersection produces a string loop is $p \sim 1$, and the number of string intersections per Hubble volume in a time interval $dt$ is $dN_{\rm int} \sim dt/t$ \cite{Vilenkin:1984ib}, then the rate of loop formation per volume at time $t_k$ is of the form
\begin{align}
    \frac{dn}{dt_k} \sim p \frac{\rho_{\infty}}{\mu l_k}\frac{dN_{\rm int}}{dt_k} = \frac{p}{\alpha t_k^4}.
\end{align}

Indeed, the loop number density production rate as calculated from the one-scale model and calibrated from simulations is \cite{Cui:2018rwi,Gouttenoire:2019kij,Sousa_2013}
\begin{align}
    \label{eq:stringLoopFormationRate}
    \frac{dn}{dt_k} &= \left(\frac{\mathcal{F} C_{\rm eff}(t_k)}{\alpha t_k^4} \frac{a(t_k)^3}{a(t)^3}\right)
    .
\end{align}
Here, $\mathcal{F}$ and $C_{\rm eff}$ are roughly constants refined from the one-scale model and simulations. $C_{\rm eff} \approx 5.4$ is the loop formation efficiency in a radiation dominated era \cite{Cui:2017ufi,Blasi:2020wpy}, and $\mathcal{F} \approx 0.1$ is the fraction of energy ultimately transferred by the infinite string network into loops of size $l_k$ \cite{Blanco-Pillado:2013qja}.

Since the loops are inside the horizon, they oscillate with roughly constant amplitude and hence redshift $\propto 1/a^3$, as shown by the rightmost term of Eq.~\eqref{eq:stringLoopFormationRate}, before decaying via gravitational radiation emission. Because the length of new string loops increases linearly with time, the nucleation probability of monopoles also grows with time, eventually cutting off loop production if $\kappa_m$ is sufficiently small. This results in a maximum string size
\begin{equation}
    \ell_{\rm max} \approx \frac{1}{t\Gamma_m} \approx \sqrt{\frac{\alpha}{\Gamma_m}}
\end{equation}
which is generally much greater than $l_{\rm tp}$.

The total power emitted in gravitational waves by strings loops prior to nucleation or by the relativistic monopoles post-nucleation can be estimated from the quadrupole formula, $ P_{\rm GW} \approx \frac{G}{45} \sum_{i,j} \langle \dddot{Q}_{ij} \dddot{Q}_{ij} \rangle \sim G (\mu l_k l_k^2  \, \omega^3)^2 \propto G \mu^2$. The power emitted by the string loops or monopole-bounded strings should be comparable since the kinetic energy of the relativistic monopoles originates from rest mass of the string. Indeed, more precise numerical computations and calibrations with simulations find the total power emitted \cite{Vilenkin:1981bx,Quashnock:1990wv,Blanco-Pillado:2017oxo} 
\begin{align}
    \label{eq:quadrupoleStrings}
    P_{\rm GW} = \Gamma G \mu^2
\end{align}
where $\Gamma \approx 50$ for string loops prior to nucleation and $\Gamma \approx 4 \ln \gamma_0^2$ for relativistic monopoles bounded to strings post-nucleation \cite{Leblond:2009fq}. Here, $\gamma_0 \approx 1 + \mu l /2m$ is the monopole Lorentz factor arising from the conversion of string rest mass energy to monopole kinetic energy.

The power emitted by gravitational waves reduces the string length, evolving in time as
\begin{align}
\label{eq:stringLength}
 l= \alpha t_k -\Gamma G \mu (t-t_k)\ . 
\end{align}
giving a loop lifetime of order $\alpha t_k/\Gamma G \mu$.
The string length and harmonic number $n$ is set by the emission frequency, $f' = n/T = 2 n/l$, where $T= l/2$ is the period of any string loop \cite{Weinberg:1972kfs,vilenkin2000cosmic}. The frequency observed today arises from redshift of $f'$ with the expansion of the Universe,
\begin{align}
    f = \frac{2n}{l} \frac{a(t_0)}{a(t)},
\end{align}
where $t_0$ the present time.

The number density spectrum of string loops then follows from Eqns. \eqref{eq:monopoleNucleationRate}, \eqref{eq:stringLoopFormationRate} and \eqref{eq:stringLength}, 
\begin{align}
\label{eq:stringNumberDensitySpectrum}
 \mathcal{N}(l,t)_{\rm Schwinger} &\equiv \frac{dn}{dl}(l,t) \approx \frac{dn}{dt_k}\frac{dt_k}{dl} e^{-\Gamma_m l (t-t_k)}  \nonumber \\
 &=
 \frac{\mathcal{F} C_{\rm eff}(t_k)}{t_k^4 \alpha(\alpha+\Gamma G\mu)}\Bigl(\frac{a(t_k)}{a(t)}\Bigr)^3 e^{-\Gamma_m l (t-t_k)}
, 
\end{align}
The exponential factor on the right side of \eqref{eq:stringNumberDensitySpectrum} is  the monopole nucleation probability which effectively cuts off loop production and destroys loops with lengths large enough to nucleate with significant probability. For $\Gamma_m l (t-t_k) \ll 1$, the probability of nucleation is negligible and the string network evolves like a standard, stable string network.
\footnote{Using a Heaviside function $\theta(\Gamma_m l (t - t_k) - 1)$ or $\theta(\Gamma_m l t - 1)$ to cutoff the loop production gives a nearly identical spectrum.}
Note that this cutoff is time-dependent, 
\begin{align}
\Gamma_m\, l (t-t_k) =\Gamma_m \frac{2n}{f}\frac{a(t)}{a(t_0)} (t-t_k)\ .
\end{align}

Although the number density of string loops decreases when nucleation occurs, as manifest by the exponential drop in the loop number density of Eq.~\eqref{eq:stringNumberDensitySpectrum}, the number density of string-bounded monopoles increases. Since $l_{\rm max} \gg l_{\rm tp}$, a loop that nucleates monopoles will continue to nucleate and fragment into many monopole-bounded strings, each with asymptotic size of order $l \sim l_{\rm tp} \ll l_{\rm max}$. While the total energy density in these pieces is comparable to the original energy density of the parent string loop, the net energy density eventually deposited into gravitational waves is much less. This is because the lifetime of the string-bounded monopoles $\sim \mu l_{\rm tp}/\Gamma G \mu^2$ is much smaller than the parent loop because their power emitted in gravational waves is similar to pure loops while their mass is much smaller. The net energy density that is transferred into gravitational waves is, to a good approximation, the energy density of the defect at the time of decay. Since these pieces decay quickly and do not redshift $\propto a^3$ for as long as pure string loops, their relative energy density compared to the background at their time of decay is much less than for pure string loops. Consequently, the net energy density that goes into gravitational radiation by monopole-bounded string pieces compared to string loops is small, and we do not consider their contribution to the spectrum.

The gravitational wave energy density spectrum generated from a network of metastable cosmic strings, including dilution and redshifting due to the expansion history of the Universe is 
\begin{gather}
    \label{eq:monopoleNucleationRho}
    \frac{d \rho_{\rm GW}(t)}{df} =
    \int_{t_{\rm scl}}^t dt' \frac{a(t')^4}{a(t)^4} \int dl \frac{dn(l,t')}{dl} \frac{dP(l,t')}{df'}\frac{df'}{df} 
    \\
    \frac{df'}{df} = \frac{a(t)}{a(t')}  \quad \quad \frac{dn}{dl}(l,t') =  \mathcal{N}(l,t')_{\rm Schwinger}
    \\
    \frac{dP(l,t')}{df'} = \Gamma G \mu^2 l \, g\left(f \frac{a(t)}{a(t')} l\right),
\end{gather}
where $t'$ is the emission time, $f' = a(t)/a(t')f$ is the emission frequency, and $f$ is the redshifted frequency observed at time $t$.  The normalized power spectrum for a discrete spectrum is \cite{vilenkin2000cosmic,Sousa_2013}
\begin{align}
    \label{eq:g(x)Strings}
    g(x) = \sum_n \mathcal{P}_n \delta(x - 2 n) \,
\end{align}
which ensures the emission frequency is $f' = 2n/l$. $\mathcal{P}_n = n^{-q}/\zeta(q)$ is the fractional power radiated by the $n$th mode of an oscillating string loop where the power spectral index, $q$, is found to be $4/3$ for string loops containing cusps \cite{Auclair:2019wcv,PhysRevD.31.3052}. Eqns. \eqref{eq:monopoleNucleationRho}-\eqref{eq:g(x)Strings} allow the stochastic gravitational wave spectrum from metastable strings to be written as
\begin{align}
   \Omega_{\rm GW}(f) \equiv& \frac{f}{\rho_c}\frac{d \rho_{\rm GW}}{d f} 
   \\
   =& \frac{8\pi}{3 H_0^2} (G \mu)^2 \sum_{n = 1}^\infty \frac{2n}{f} \int_{t_{\rm form}}^{t_0}
 dt \left(\frac{a(t)}{a(t_0)}\right)^5  
 \nonumber \\
 &\times {\cal N}_{\rm Schwinger} \Bigl(l=\frac{2n}{f}\frac{a(t)}{a(t_0)},t\Bigr) \mathcal{P}_n .
    \label{GW_schwinger_monopole}
\end{align}

\begin{figure}
    \centering
    \includegraphics[width=0.49\textwidth]{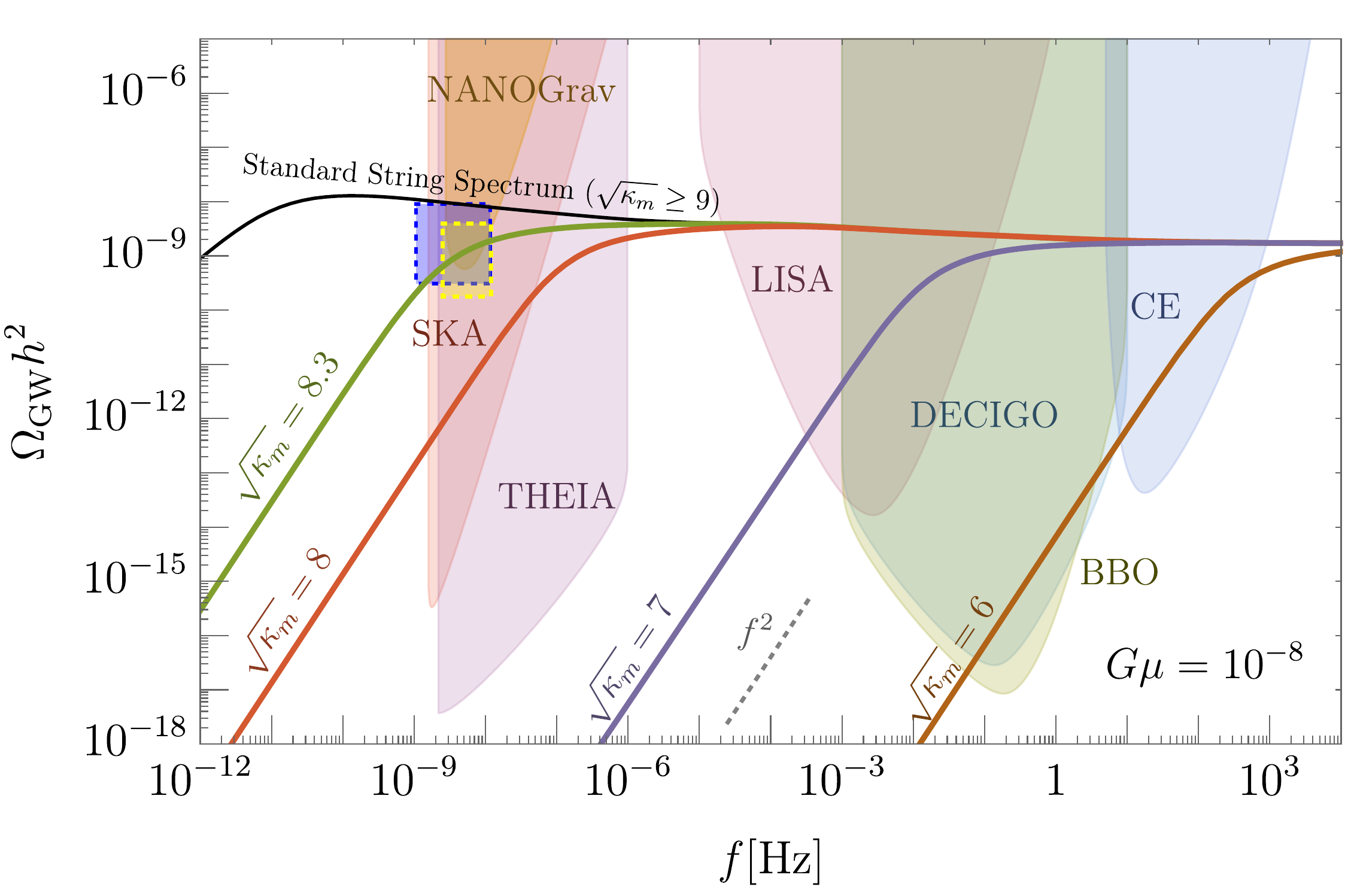}
    \caption{
   Representative spectra of gravitational waves emitted by cosmic strings that are eaten by the nucleation of monopoles for fixed $G\mu = 1 \times 10^{-8}$. Each colored contour corresponds to a different value of $\kappa_m = m^2/\mu$ which parameterizes the ratio between monopole and string symmetry breaking scales and sets the nucleation time of the monopoles on the string. Since nucleation is an exponentially suppressed process, the metastable string network is typically cosmologically long-lived and behaves as pure string network before nucleation.  At high frequencies, $\Omega_{\rm GW} \propto f^0$ like a pure string network while after nucleation, $\Omega_{\rm GW}$ decays as $f^2$. The black contour shows the pure string spectrum without monopoles. For $\kappa\geq 9$, the nucleation timescale of monopoles is greater than the age of the Universe and the metastable string network is indistinguishable from the pure string spectrum. The dotted-yellow and blue boxes highlight the potential signals of NANOGrav ~\cite{Arzoumanian:2020vkk} and PPTA ~\cite{Goncharov:2021oub}, respectively.
    } 
    \label{fig:benchmark}
\end{figure}

We numerically compute the gravitational wave spectrum, Eq.~\eqref{GW_schwinger_monopole}, over a range of string tensions $\mu$ and monopole masses $m$. Fig.~\ref{fig:benchmark} shows a benchmark plot of the gravitational wave spectrum from cosmic strings consumed by monopoles for fixed $G\mu = 1 \times 10^{-8}$ and
a variety of $\kappa_m = m^2/\mu$.
In computing the spectrum, we sum up $10^4$ normal modes and 
solve for the evolution of the scale factor from the Friedmann equations in a $\Lambda$CDM cosmology. 
The colored contours in Fig. \ref{fig:benchmark} show the effect of the nucleation rate parameter, $\kappa_m$, on the spectrum,  with larger $\kappa_m$ corresponding to a longer lived string network. In the limit $\sqrt{\kappa_m} \gtrsim 9$, the nucleation rate is so weak that the string network is stable on cosmological time scales, reducing to the standard stochastic string spectrum as shown by the black contour.
Larger loops, corresponding to lower frequencies and later times of formation, vanish because of Schwinger production of monopole-antimonopole pairs and hence the gravitational wave spectrum is suppressed at low frequencies, scaling as an $f^2$ power law in the infrared.  The slope is easily distinguishable from other signals such as strings without monopole pair production and strings consumed by domain walls, as discussed in Sections \ref{sec:dwconsumedbystrings} and \ref{sec:stringboundedWalls}. Importantly, from Fig. \ref{fig:benchmark}, we see that it is possible to detect the $f^2$ slope in the low frequency region of the power spectrum through many gravitational wave detectors, including NANOGrav, 
PPTA, SKA, THEIA, LISA, 
DECIGO, BBO, and CE. 

\begin{figure}
    \centering
    \includegraphics[width=0.48\textwidth]{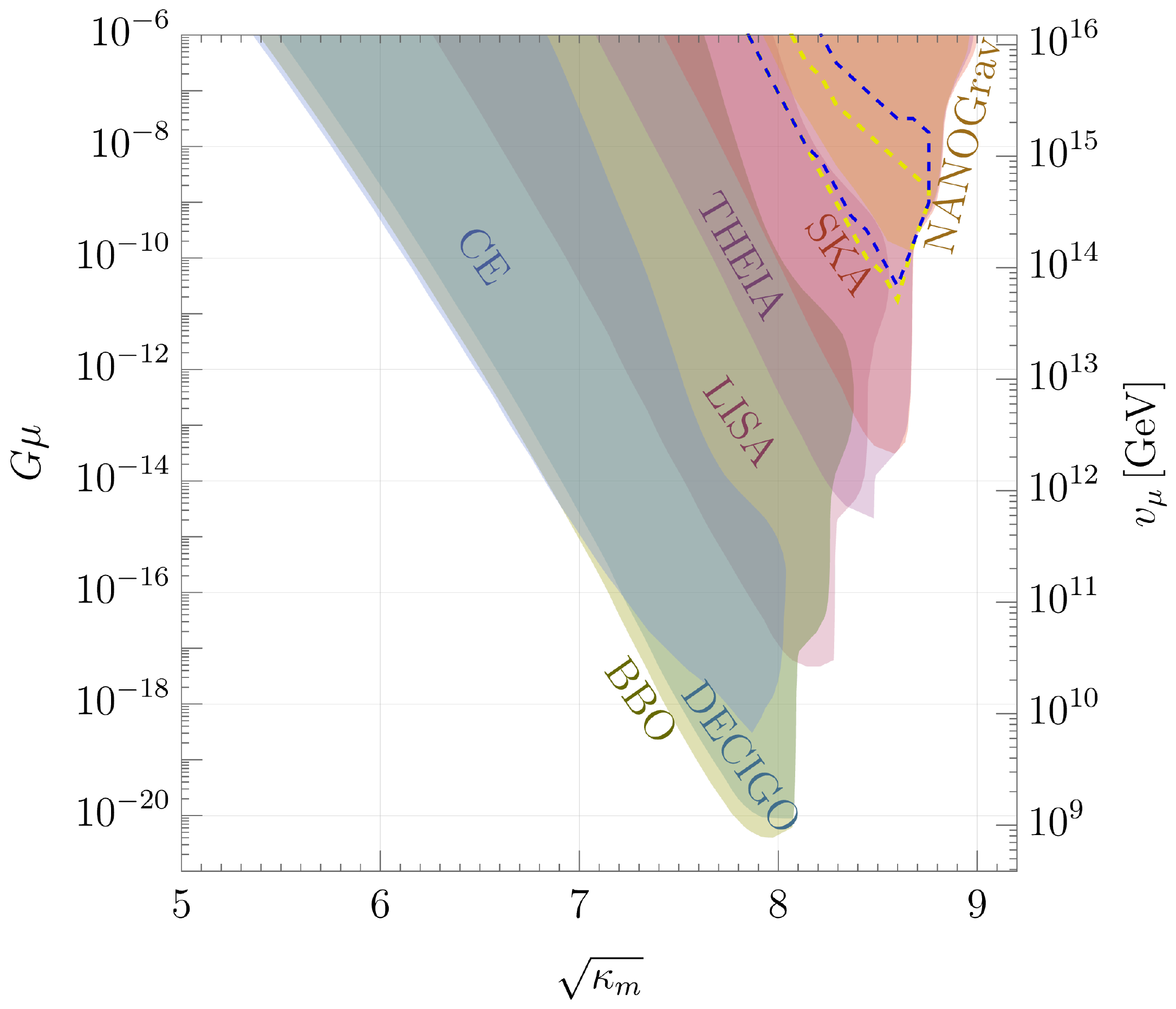}
    \caption{The parameter region in the $G\mu$--$\sqrt{\kappa_m}$ plane where the gravitational wave spectrum from cosmic strings eaten by the nucleation of monopoles  can be detected. For a given $(G\mu,\sqrt{\kappa_m})$, a detection is registered when $\Omega_{\rm GW}$ is greater than the sensitivity curve of the given detector \textit{and} the relative difference in spectra between cosmic strings eaten by monopoles and a pure string spectrum with the same $G\mu$ is greater than 10\%. The latter condition ensures the two signals are sufficiently distinguishable and the detection of the infrared $f^2$ slope shown, for example in Fig. \ref{fig:benchmark}, can be achieved. The yellow and blue dashed lines highlight the potential signals by NANOGrav and PPTA, respectively, as in Fig.~\ref{fig:benchmark}}.
    \label{fig:ps_graham}
\end{figure}

Fig. \ref{fig:ps_graham} shows the parameter space in the $G\mu$--$\sqrt{\kappa_m}$ plane where the $f^2$ decaying slope can be detected and distinguished from the standard string spectrum. For a given ($G\mu,\sqrt{\kappa_m})$, we register a detection of the monopole nucleation gastronomy in a similar manner to the “turning-point” recipe of \cite{Gouttenoire:2019kij}: First, $\Omega_{\rm GW}h^2$ must exceed the threshold of detection for a given experiment. Second, to actually distinguish between the monopole-nucleation gastronomy spectrum and the standard string spectrum, we require that their percent relative difference be greater than a certain threshold within the frequency domain of the experiment. Following \cite{Gouttenoire:2019kij}, we take this threshold at a conservative 10 \%. Fig. \ref{fig:ps_graham} demonstrates that a wide range of $\mu$ and $\kappa_m$ can be probed. String symmetry breaking scales $v_\mu \equiv \sqrt{\mu}$ between $10^9$ GeV and $10^{16}$ GeV and $\kappa_m$ between $30-80$ can be detected by current and near future gravitational wave detectors. Interestingly, the yellow and blue dashed boxes show the particular $\mu$ and $\sqrt{\kappa_m}$ that generate a spectrum that passes through the recent NANOGrav (yellow) ~\cite{Arzoumanian:2020vkk} and PPTA (blue) \cite{Goncharov:2021oub} signals.

Last, note that the benchmark spectra of Fig. \ref{fig:benchmark} are similar to the spectra found in a previous paper~\cite{Buchmuller:2019gfy}, but the slope in the low frequency region is not $f^{3/2}$ as found in ~\cite{Buchmuller:2019gfy}, but $f^2$. The difference comes from the authors of \cite{Buchmuller:2019gfy}  using a fixed time at which loop production ceases, corresponding to when the average length of the string loop network, $\langle l \rangle = l_{\rm max}$. However, the average length of string loops in the loop network is dominated by the smallest loops, even though there exists much larger loops up to $l \approx \alpha t$ in the network at any given time. We take into account the nucleation rate on individual loop basis. This is necessary due to the shorter nucleation lifetime of longer strings than shorter strings because the probability of pair production of monopoles on a string is proportional to the length of the string. Our results agree with the more recent work of \cite{Buchmuller:2021mbb}.

\section{Strings Eating Monopoles}
\label{sec:stringDestruction}
In this section, we consider the case where strings attach to, and consume, a pre-existing monopole network. The symmetry breaking chains that allow this are the same as in Sec. \ref{sec:schwingerStringsMonopole}, with the difference between the two scenarios arising from when inflation occurs relative to monopole formation. For the monopole nucleation gastronomy of Sec. \ref{sec:schwingerStringsMonopole}, inflation occurs after monopole formation but before string formation. For strings attaching to a pre-existing monopole network as considered in this section, inflation occurs before monopole and string formation. In this scenario, the monopole network is not diluted by inflation and at temperatures below the string symmetry breaking scale $v_\mu$, the magnetic field of the monopoles squeezes into flux tubes (cosmic strings) connecting each monopole and antimonopole pair \cite{Lazarides:1981fv,Vilenkin:1984ib}. 
Note that since this is not a tunneling process, there does not have to be a coincidence of scales between $v_m$ and $v_\mu$ as in the case of monopoles nucleating on strings as discussed in Sec. \ref{sec:schwingerStringsMonopole}. Moreover, since \textit{every} monopole and antimonopole get connected to a string which eventually shrinks and causes the monopoles to annihilate, the monopole problem is absent in such symmetry breaking chains. As shown in Fig. \ref{fig:chains}, an example chain where this gastronomy scenario occurs is  $3221 \rightarrow 3211 
\rightarrow 321$. The first breaking produces monopoles and the second breaking connects the monopoles to strings. Since there are no stable monopoles or domain walls that are also generated in this breaking pattern, inflation need not occur after the monopoles form when $3221$ breaks to $3211$.

The scenario where strings attach to a pre-existing monopole network has been considered before \cite{Lazarides:1981fv,Holman:1992xs,Martin:1996ea,vilenkin2000cosmic}, but only with an initial monopole abundance of roughly one monopole per horizon at formation as computed originally by Kibble \cite{Kibble:1976sj}, and with the conclusion that there is no gravitational wave amplitude.
\footnote{The case where monopoles are only partially inflated away so that eventually monopoles re-enter the horizon was considered in \cite{Martin:1996cp,Martin:1996ea}. We do not consider that scenario.} Here, we redo the calculation with the enhanced abundance of monopoles using the Kibble--Zurek mechanism \cite{Zurek:1985qw} and take into account monopole-antimonopole freeze-out that can occur between monopole and string formation \cite{Preskill:1979zi}. 
We find that after string formation, monopole-antimonopole pairs annihilate in generally less than a Hubble time with the typical monopole velocities being non-relativistic, often leading to no gravitational wave spectrum. However, for some monopole masses $m$ and string scales $v_\mu$, the monopole-bounded strings can be relativistic and emit a pulse of gravitational waves before decaying if friction is not severe. Moreover, the  greater number density of monopoles predicted in the Kibble--Zurek mechanism compared to Kibble's original estimate gives rise to significantly enhanced gravitational wave amplitude.

We begin with the Kibble--Zurek mechanism, where the initial number density of monopoles is set by the correlation length, $\xi_m$, of the Higgs field associated with the monopole symmetry breaking scale, $v_m$. For a Landau-Ginzburg free energy near the critical temperature $T_c$ of the phase transition of the form 
\begin{align}
    V(\phi) = (T - T_c)m\phi^2 + \frac{1}{4}\lambda \phi^4,
\end{align}
the initial number density of monopoles is approximately \cite{Murayama:2009nj}
\begin{align}
    \label{eq:kibbleZurekDensity}
    n_m(T_c) = \frac{1}{\xi_m^3} \approx \frac{\lambda}{2}H T_c^2
\end{align}
where $H$ is the Hubble scale. Note that the monopole formation density calculated by Zurek, \eqref{eq:kibbleZurekDensity}, is roughly a factor of $(M_{Pl}/T_c)^2  \approx (M_{ Pl}/v_m)^2 \gg 1$ greater than the original estimate by Kibble. $M_{Pl} = 1/\sqrt{G}$ is the Planck mass.

After formation, the monopole-antimonopole pairs annihilate, with a freeze-out abundance \cite{Preskill:1979zi}
\begin{align}
   \label{eq:monopoleNumDensity}
    \frac{n_m(T)}{T^3} = \left[\frac{T_c^3}{n_m(T_c)} + \frac{h^2}{\beta_m} \frac{C M_{Pl}}{m}\left(\frac{m}{T} - \frac{m}{T_c}\right)\right]^{-1}
\end{align}
where $C = (8\pi^3 g_*/90)^{-1/2}$ and 
\begin{align}
   \beta_m \simeq \frac{2 \pi}{9} \sum_i b_i \left(\frac{h e_i}{4\pi}\right)^2 \ln{\Lambda}
\end{align}
counts the particles of charge $e_i$ in the background plasma that the monopole scatters off of. The magnetic coupling is $h = 2\pi/e$ where $e$ is the $U(1)$ gauge coupling constant, $\Lambda \sim {1/(g_* e^4/16 \pi^2)}$ is the ratio of maximum to minimum scattering angles of charged particles in the plasma, and $b_i = 1/2$ for fermions and $1$ for bosons \cite{vilenkin2000cosmic,goldman1981gravitational}. With $e \sim 0.3$ and a comparable number of electromagnetic degrees of freedom as in the Standard Model, $\beta_m \sim 20$. For $T \ll v_m$, Eq. \ref{eq:monopoleNumDensity} asymptotes to a frozen-out abundance $n_m/T^3 \simeq \beta_m \text{Max}(T,T_*)/h^2 C M_p$, where $T_* = (4\pi/h^2)^2 m/\beta_m^2$ is approximately the temperature when the monopole mean free path becomes longer than the monopole-antimonopole capture distance \cite{Preskill:1979zi} 

Below the scale $v_\mu$, the magnetic fields of the monopoles squeeze into flux tubes, with the string length set by the typical separation distance between monopoles,
\begin{align}
    \label{eq:kibbleStringLength}
    l \approx \frac{1}{n_m(T = v_\mu)^{1/3}}. 
\end{align}
Eq.~\eqref{eq:kibbleStringLength} is valid when the correlation length of the string Higgs field, $\xi_\mu \geq l$ \cite{vilenkin2000cosmic}. If $\xi_\mu < l$, the monopole-bounded strings are straight on scales smaller than $\xi_\mu$ and Brownian on greater scales which gives the strings a length longer than \eqref{eq:kibbleStringLength}. For an initial abundance of strings set by the Kibble--Zurek mechanism, $\xi_\mu/l(T=v_\mu) \approx (2 \beta_m/\lambda_\mu h^2)^{1/3}$, which coincidentally, is usually of order or just marginally less than unity. Nevertheless, since the string correlation length grows quickly with time $\propto t^{5/4}$ \cite{Kibble:1976sj,vilenkin2000cosmic}, the string-bounded monopole becomes straightened out within roughly a Hubble time of string formation and ends up with a length close to Eq.~\eqref{eq:kibbleStringLength}. For $T_c = v_m \lesssim 10^{17} \, \rm GeV$, $l$ is far below the horizon scale. Consequently, $l$ is not conformally stretched by Hubble expansion and only can decrease with time by energy losses from friction and gravitational waves.

\begin{figure}
    \centering
    \includegraphics[width=0.45\textwidth]{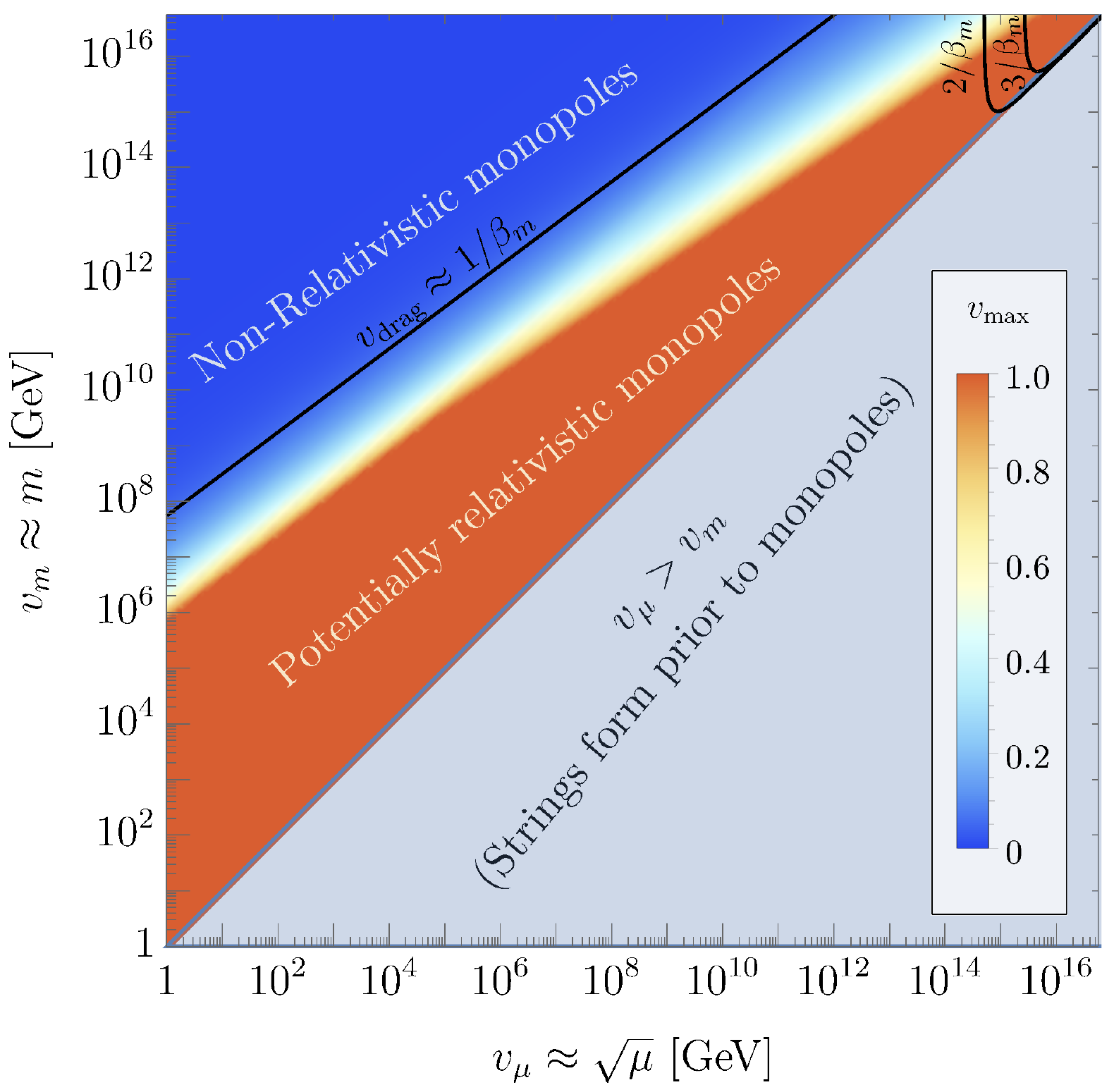}
    \caption{The $v_m - v_\mu$ parameter space where monopoles attached to strings can be relativistic. In the dark blue region at large $v_m$, the monopoles are sufficiently heavy that the conversion of string rest mass to monopole kinetic energy cannot accelerate the monopoles to relativistic speeds and any gravitational wave signal is heavily suppressed. In the red region, the monopoles are light enough that the string can accelerate them to relativistic speeds, neglecting friction. This region of parameter space can potentially generate a gravitational wave signal. The black contours shows the typical maximum drag speed of the monopoles from friction with the thermal bath. For sufficiently large $\beta_m$, a model dependent friction parameter, the drag speed prevents the monopoles from reaching relativistic speeds and the gravitational wave signal can be suppressed. In the light blue region, $v_\mu > v_m$, which is forbidden for composite monopole-bounded strings.}
    \label{fig:relativisticMonopoles}
\end{figure}

Because the string rest mass is converted to monopole kinetic energy, the initial string length \eqref{eq:kibbleStringLength} determines whether or not the monopoles can potentially move relativistically. Relativistic monopoles can emit a brief pulse of gravitational radiation before annihilating while non-relativistic monopoles will generally not. Energy conservation implies the maximum speed of the two monopoles on each string is roughly
\begin{align}
    \label{eq:relativisticMonopoles}
    v_{\rm max} \approx \sqrt{1 -\left(1 + \frac{\mu l}{2m}\right)^{-2}} \sim \text{Min}\left\{\sqrt{\frac{\mu l}{m}}\,, \, 1 \right\}.
\end{align}
The density plot of Fig. \ref{fig:relativisticMonopoles} shows the parameter space in the $v_\mu - v_m$ plane where $v_{\rm max} \sim 1$ (in red) and the monopoles can reach relativistic speeds according to Eq.~\eqref{eq:relativisticMonopoles}.
Initially, however, monopole friction can prevent the monopoles from reaching $v_{\rm max}$. This is because the relative velocity of the monopoles induced by the string produces an electromagnetic frictional force on the monopoles scattering with the background plasma. The force of friction between the monopole and plasma is \cite{vilenkin2000cosmic,vilenkin1982cosmological}
\begin{align}
    \label{eq:frictionForceMonopole}
    F_{f} \approx -\beta_m T^2 v e^{-M r(T)} \left(1 +M r(T)\right)
\end{align}
where $v$ is the relative speed between monopole and the bulk plasma flow.  We include the Yukawa exponential factor, $e^{-M r}(1+M r)$ to take into account the exchange of the now massive photon of mass $M \simeq e v_\mu /2$ at temperatures below $v_\mu$.  $r(T) \simeq \sqrt{4\pi T/e^2 n_e}$ is the inverse plasma mass associated with the screened magnetic field of the monopole. To a good approximation then, 
\begin{align}
   M r(T) = \frac{v_\mu}{T} \sqrt{\frac{30}{\zeta(3) \pi g_{*e}}} \approx \frac{v_\mu}{T}
\end{align}
where $g_{*e}$ is the charged relativistic degrees of freedom in the thermal bath.

The balance between the string tension and friction is described by the equation of motion of each monopole,
\begin{align}
    m \frac{dv}{dt} \simeq \mu + F_f(T) .
\end{align}
To an excellent approximation, the drag speed, or terminal velocity, of the monopoles satisfy the quasi-steady state solution $dv/dt \simeq 0$, which gives the monopole drag speed as a function of temperature
\begin{align}
   \label{eq:monopoleDragSpeed}
   v_{\rm drag} =\frac{\mu}{\beta_m T^2 e^{-v_\mu/T}(1 + \frac{v_\mu}{T})}.
\end{align}
The frictional damping of the monopole motion ends when $v_{\rm drag}$ equals $v_{\rm max}$, which occurs roughly a Hubble time after formation because of the decrease in $T$.
However, even in this brief period of damping, the friction force 
\eqref{eq:frictionForceMonopole} causes the string-monopole system to lose energy at a rate
\begin{align}
   \label{eq:powerfrictionMonopoles}
    P_{f} \approx - \beta_m T^2 v^2 e^{-v_\mu/T}\left(1 + \frac{v_\mu}{T}\right),
\end{align}
which can be considerable even in a Hubble time. Above, $v = \max{(v_{\rm drag},v_{\rm max})}$. For example, near string formation when $T^2 \sim \mu$, the power lost to friction is roughly $\beta/\Gamma G\mu v^4 \gg 1$ greater than the power lost to gravitational radiation, $P_{\rm GW} \approx \Gamma G\mu^2 v^6$.  Note that for the monopole nucleation gastronomy of Sec. \ref{sec:schwingerStringsMonopole}, the monopole nucleation occurs at a far lower temperature than the string formation time, and hence $P_{f} \ll P_{\rm GW}$ for that gastronomy scenario. In the gastronomy scenario of this section, where strings eat a pre-existing monopole network, $P_{f} \gg P_{\rm GW}$. Consequently, the power lost from friction determines the lifetime, $\tau$, of the string-bounded monopoles, with 
\begin{align}
   \tau \approx  -\left.\frac{E}{P_f}\right\vert_{T \simeq v_\mu} \approx \frac{\mu l}{\beta_m  \mu v^2} \approx 
   \begin{cases}
      \beta_m l \quad &v = v_{\rm drag}
      \\[5pt]
      \dfrac{m}{\beta_m \mu} \quad &v =v_{\rm max} \lesssim 1
      \\[10pt]
      \dfrac{l}{\beta_m} \quad &v =v_{\rm max} \sim 1 \, .
   \end{cases}
\end{align}
To more precisely determine the monopole-string lifetime,  we integrate Eq.~\eqref{eq:powerfrictionMonopoles} to determine the energy of the string-monopole system as a function of time and find that for $\beta_m \gtrsim 3$, the energy in the monopole-string system is entirely dissipated by friction before $v_{\rm drag}$ reaches $v_{\rm max}$ and hence relativistic speeds. The contours of Fig. \ref{fig:relativisticMonopoles} show the typical highest speed of the monopoles before losing energy via friction. Since the energy of the system is entirely dissipated in around a Hubble time, the largest monopole speed is typically set by the drag speed when $T^2 \sim \mu$; that is, $v_{\rm drag} \sim \beta_m^{-1}$ according to Eq.~\eqref{eq:monopoleDragSpeed}. Consequently, we see analytically that the terminal velocity of the monopoles is not relativistic unless $\beta_m \sim 1$. If the number of particles interacting with the monopole in the primordial thermal bath is comparable to the number of electrically charged particles in the Standard Model and with similar charge assignments, then $\beta_m \sim 20$ and thus the monopole-string system is never relativistic before decaying. In this scenario, the gravitational wave signal is heavily suppressed. 

If $\beta_m \sim 1$, however, which can occur in a dark sector with fewer charged particles in the thermal bath or with smaller $U(1)$ charges, then the monopoles reach the speed $v_{\rm max}$ before decaying via friction. In this case, the red region of Fig. \ref{fig:relativisticMonopoles} indicates where a gravitational wave signal can be efficiently emitted by the monopoles before annihilating. Unlike the monopole nucleation gastronomy of Sec. \ref{sec:schwingerStringsMonopole}, $v_\mu$ does not need to be as nearly degenerate with $v_m$ for gravitational waves to be produced.
Moreover, since the lifetime of the string pieces is shorter than Hubble, the pulse of energy density emitted by relativistic monopoles in gravitational waves is well approximated by
\begin{align}
    \rho_{\rm GW, burst} \approx  n_m(v_\mu)P_{\rm GW} \, \tau ,
\end{align}
where $P_{\rm GW} = \Gamma G\mu^2$ is the power emitted by oscillating monopoles connected to strings \eqref{eq:quadrupoleStrings}. The peak amplitude of the monopole gravitational wave burst is
\begin{align}
    \label{eq:monopoleBurstAmplitude}
    \Omega_{\rm GW, burst} &= \frac{ \rho_{\rm GW, burst}}{\rho_c(v_\mu)} \Omega_r \left(\frac{g_{*0}}{g_*(v_\mu)}\right)^{ \scalebox{1.01}{$\frac{1}{3}$}}  
    \nonumber
    \\
    &\approx \frac{30 \pi^2}{g_*(v_\mu)\beta}\Gamma G \mu \left(\delta \frac{m}{M_{\rm Pl}}\right)^{ \scalebox{1.01}{$\frac{2}{3}$}}  , 
\end{align}
where 
\begin{align}
   \label{eq:delta}
   \delta = \frac{1}{C\beta_m h^2}\left(\frac{4\pi}{h^2}\right)^2\text{Max}\left\{1, \, \frac{v_\mu}{m} \left(\frac{\beta_m h^2}{4\pi}\right)^2 \right\}.
\end{align}
and $\rho_c(v_\mu)$ is the critical energy density of the Universe at string formation, which is assumed to be in a radiation dominated era. 
The `Max' argument of \eqref{eq:delta} characterizes the amount of monopole-antimonopole annihilation that occurs prior to string formation at $T = v_\mu$. For sufficiently small $v_\mu/m$, the freeze-out annihilation completes before string formation and the max function of \eqref{eq:delta} is saturated at its lowest value of $1$. In this conservative scenario, $\delta \approx 10^{-4}\beta_m^{-1}(e/0.5)^4$.

Similarly, the peak frequency is
\begin{align}
    f_{\rm burst} \sim \frac{1}{l}\frac{a(t_\mu)}{a(t_0)}
    &\approx   10^8 \, {\rm Hz} \left(\frac{v_m}{10^{14} \, \rm GeV} \, \frac{\delta}{10^{-4}} \, \frac{106.75}{g_*(v_\mu)}\right)^{ \scalebox{1.01}{$\frac{1}{3}$}}
\end{align}
where $a(v_\mu)$ and $a(t_0)$ are the scale factors at string formation and today, respectively.  Note that redoing the analysis of this section but with Kibble's original estimate for the number density of monopoles yields a gravitational wave spectrum that is roughly $(v_m/M_{\rm Pl})^{4/3} \ll 1$ suppressed compared to Eq.~\eqref{eq:monopoleBurstAmplitude}.

With the qualitative features of the monopole burst spectrum understood, we can turn to a numerical computation of $\Omega_{\rm GW}$ in the case where $\beta_m \sim 1$. The gravitational wave energy density spectrum is 
\begin{gather}
    \label{eq:monopoleBurstRho}
    \frac{d \rho_{\rm GW}(t)}{df} =
    \int^t dt' \frac{a(t')^4}{a(t)^4} \int dl \frac{dn(l,t')}{dl} \frac{dP_{l}(l,t')}{df'}\frac{df'}{df} \ ,
    \\
    \label{eq:redshiftMonopole}
    \frac{df'}{df} = \frac{a(t)}{a(t')}\ ,  \quad \quad \frac{dn}{dl}(l,t') = \frac{dn}{dt_k}\frac{dt_k}{dl} \ ,
    \\
    \label{eq:dPdf}
    \frac{dP_l (l,t')}{df'} = \Gamma G \mu^2 l \, g\left(f \frac{a(t)}{a(t')} l\right),
\end{gather}
where primed coordinates refer to emission and unprimed refer to the present so that gravitational waves emitted from the monopoles at time $t'$ with frequency $f'$ will be observed today with frequency $f = f' a(t')/a(t)$. $t_k$ is the formation time of monopole-bounded strings of length $l(t_k)$,
\begin{align}
    \frac{dn}{dt_k} \simeq n_m(t_k)\delta(t_k - t_\mu) \left(\frac{a(t_k)}{a(t)} \right)^3
\end{align}
is the string-bounded monopole production rate, which is localized in time to the string formation time, $t_\mu \simeq C M_{Pl}/v_\mu^2 $ . $dt_k/dl$ is found by noting that the energy lost by relativistic monopoles separated by a string of length $l$ is
\begin{align}
   \frac{dE}{dt} = \frac{d}{dt}(\mu l + 2m) \approx -\beta_m v^2 \mu \,.
\end{align}
In the red region of Fig. \ref{fig:relativisticMonopoles} where a gravitational wave signal can be generated, $\mu l \gg 2m$ (otherwise the monopoles would not be relativistic). As a result,  monopole-bounded strings that form at time $t_k$ with initial size $l(t_k)$ decrease in length according to 
\begin{align}
   l(t) \simeq l(t_k) - \beta_m v^2 (t - t_k)
\end{align}
so that
\begin{align}
   \frac{dt_k}{dl} \simeq \frac{1}{\beta_m v^2} \approx 1 .
\end{align}
The normalized power spectrum for a discrete spectrum is
\begin{align}
   \label{eq:g(x)Monopoles}
    g(x) = \sum_n \mathcal{P}_n \delta(x - n \xi) \qquad \xi \equiv \frac{l}{T}
\end{align}
ensures the emission frequency of the $n$th harmonic is $f' = n/T$, where $T$ is the oscillation period of the monopoles. For pure string loops, $T = l/2$ ($\xi = 2$, reducing to Eq.~\eqref{eq:g(x)Strings}), whereas for monopoles connected to strings, $T = 2m/\mu + l \simeq l$ \cite{Leblond:2009fq,Martin:1996cp} ($\xi \approx 1$).  $\mathcal{P}_n \approx n^{-1}$ is found \cite{Leblond:2009fq,Martin:1996cp} for harmonics up to $n \approx \gamma_0^2$, where $\gamma_0 \simeq (1 + \mu l/2m)$, is the Lorentz factor of the monopoles. For $n > \gamma_0^2$, $P_n \propto n^{-2}$. $\Gamma \approx 4 \ln \gamma_0^2$. 

Integrating the energy density spectrum \eqref{eq:monopoleBurstRho} and normalizing by the present day critical density, $\rho_{\rm c} = 3H_0^2/8\pi G$,
yields the present day gravitational wave spectrum from monopoles eaten by strings
\begin{align}
    \Omega_{\rm GW} 
    &= \sum_n \frac{8 \pi (G \mu)^2}{3 H_0^2} \left(\frac{a(t_\mu - l*)}{a(t_0)}\right)^5 \left(\frac{a(t_\mu)}{a(t_\mu - l*)}\right)^3
    \nonumber \\
    &\times \Gamma \mathcal{P}_n \frac{\xi n}{f}  \frac{n_m(t_\mu)}{\beta_m v^2}
\end{align}
where 
\begin{align}
   l_* = \frac{\frac{\xi n}{f}\frac{a(t_\mu)}{a(t_0)} - n_m(t_\mu)^{ \scalebox{1.01}{$-\frac{1}{3}$}}}{\beta_m v^2} .
\end{align}
 \begin{figure}
    \centering
    \includegraphics[width=0.48\textwidth]{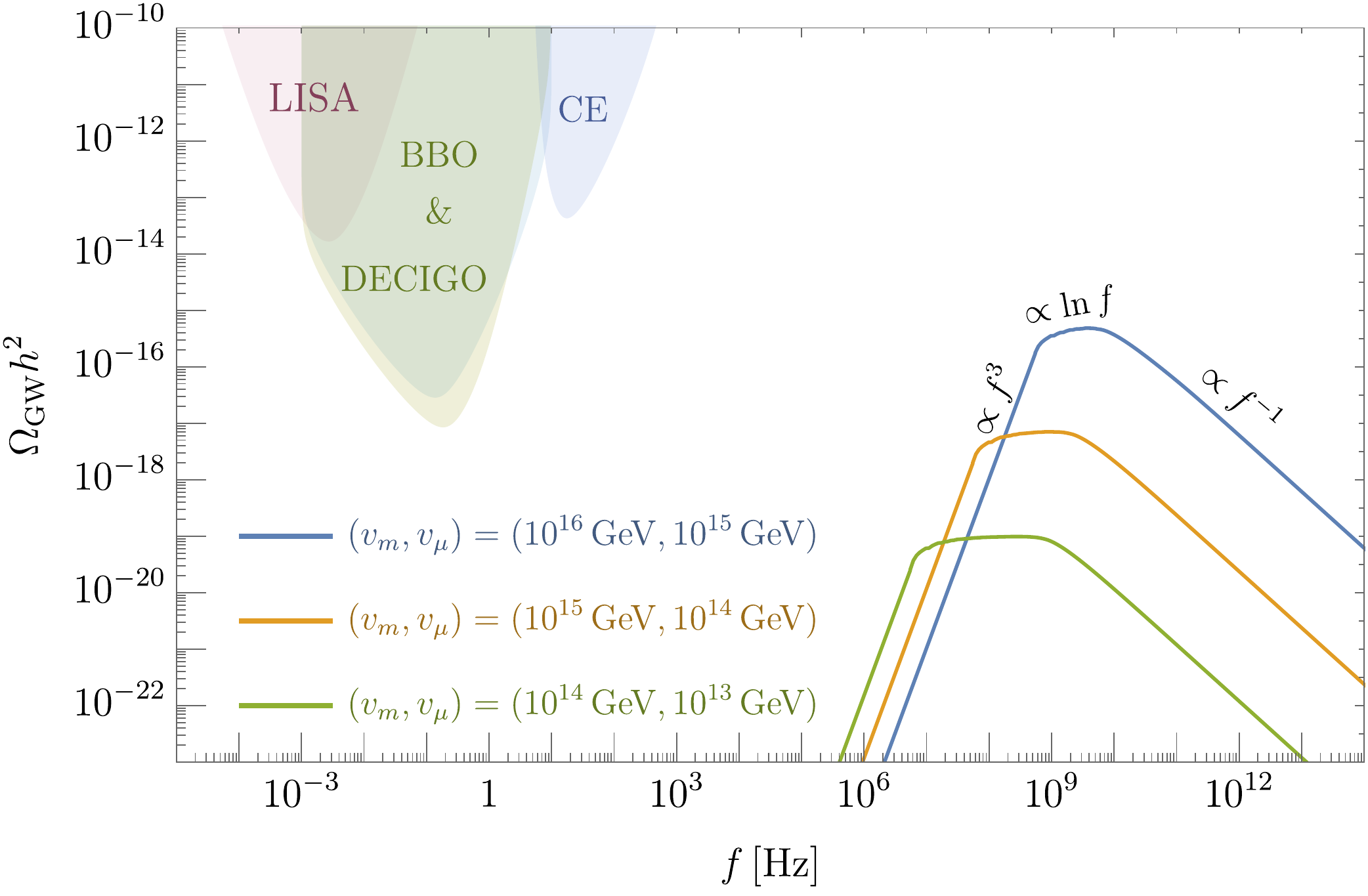}
    \caption{Representative spectra of gravitational waves emitted by monopoles that are eaten by strings. Each colored contour corresponds to a different value of symmetry breaking scales $(v_m,v_\mu)$. In all cases, we fix $\beta_m$ and the monopole speed $v$ near unity. The dominant energy loss by the monopoles is from friction which causes the monopole-bounded string to decay within a Hubble time. The emission of gravitational waves thus occurs in a `burst' and is peaked at high frequencies corresponding to the monopole-antimonopole separation distance when $T \approx v_\mu$. At high frequencies, $\Omega_{\rm GW} \propto f^{-1}$ while at low frequencies $\Omega_{\rm GW} \propto f^3$ by causality. The frequency dependence near the peak of the spectrum interpolates scales as $\Omega_{\rm GW} \propto \ln f$.}
    \label{fig:monopoleBurstSpectrum}
\end{figure}
The contours of Fig. \ref{fig:monopoleBurstSpectrum} show $\Omega_{\rm GW}h^2$ for range of a $v_\mu$ and $v_m$ where monopoles can oscillate relativistically before decaying via friction, assuming $\beta_m v^2 \sim 1$. For frequencies much lower than the inverse string length, we take the causality limited spectrum $f^3$ \cite{Caprini:2009fx}. Fig. \ref{fig:monopoleBurstSpectrum} shows that the spectral shape  goes as $f^{-1}$ at high frequencies, plateaus logarithmically for a brief period, and decays as $f^3$ at low frequencies. The duration of the logarithmic plateau corresponds to the number of modes where $P_n \propto 1/n$, which is set by $\gamma_0$ and hence $v_{\rm max}$.  As suggested by the estimate $f_{\rm burst}$, the frequency at which the spectrum decays typically occurs at very high frequencies because the separation length of the monopoles is small when eaten by strings at $T \simeq v_\mu$. Consequently, to observe the monopole burst gastronomy signal, future gravitational wave detectors near megahertz frequencies are needed.
 
Finally, we comment that string loops or open strings without monopoles also form at the string symmetry breaking scale $v_\mu$. For $\xi_s \sim l$, as is generally the case, both simulations and free-energy arguments \cite{Mitchell:1987th,Copeland:1987ht,vilenkin2000cosmic} suggest that these pure strings are clustered around the monopole separation scale $l$, with the distribution of strings of length greater than $l$ exponentially suppressed and only making a subdominant $\lesssim 10\%$ of all strings \cite{Copeland:1987ht}. Essentially, it becomes exponentially unlikely for a string with length greater than $l$ to not terminate on two monopoles. 

Like the monopole string segments, the dominant energy loss mechanism for these loops is friction with the plasma. Here, the friction is mainly due to Aharonov-Bohm scattering, which exerts a force
\begin{align}
   \label{eq:aharonovBohmForce}
   F_{\rm AB} \simeq -\beta_s T^3 v l 
\end{align}
where 
\begin{align}
   \beta_s \simeq \frac{2\zeta(3)}{\pi^2} \sum_i a_i \sin^2(\pi \nu_i)
\end{align}
counts the particles in the background plasma that experience a phase change $2 \pi \nu_i = e_i \Phi$ when moving around the string of magnetic flux $\Phi$, thereby scattering off the string via the Aharonov-Bohm mechanism \cite{Alford:1988sj,vilenkin2000cosmic}. $a_i = 3/4$ for fermions and $1$ for bosons. $v$ is the relative perpendicular motion of the string with respect to the plasma.

Just like the monopoles, the frictional force on the strings initially prevents the string loops, which are subhorizon, from freely oscillating relativistically \cite{Garriga:1993gj}. Balancing the string curvature tension, $\mu$, and friction force gives the string drag speed as a function of temperature 
\begin{align}
   v_{\rm drag} \approx \frac{\mu}{\beta_s T^3 l} \,.
\end{align}
For $T^2 \sim \mu$, $\beta_s \geq 1$, and for string lengths of order the monopole separation distance, \eqref{eq:kibbleStringLength}, the string drag velocity is initially non-relativistic for all $v_\mu, v_m \lesssim 10^{17}$ GeV. The frictional damping of the string motion causes the string loops to be conformally stretched, $l(t) \propto a(t)$, until $v_{\rm drag}$ becomes relativistic, or equivalently, their conformally stretched size drops below the friction scale $L_f \approx \mu l/|F_{\rm AB}|$ (see Sec. \ref{subsec:friction} for a further discussion). This occurs at time $t_f \approx t_0 \max(\beta_s l_0 v_\mu,1)$ and final string size $l_f \approx l_0 \max((\beta_sl_0 v_\mu)^{1/2},1)$, where $l_0 = l(T=v_\mu)$ is the typical monopole separation at string formation.
However, even after this brief period of damping, the Aharonov-Bohm friction force, \eqref{eq:aharonovBohmForce} still causes the string to lose energy at a rate $\mu dl/dt = -P_{\rm AB}$, where 
\begin{align}
   \label{eq:powerAB}
   P_{\rm AB} = - \beta_s T^3 v^2 l.
\end{align}
with $v \sim 1$. The power lost via Ahronov-Bohm friction causes the string length to exponential decrease in size. These small loops will then completely and quickly decay via gravitational radiation that, depending on the fraction of stings in loops, can generate a comparable $\Omega_{\rm GW}$ to the monopole burst spectrum of Fig. \ref{fig:monopoleBurstSpectrum}. Unlike the monopole bursts, the ultraviolet frequency dependence of the string burst spectrum will scale approximately as $f^{1-q}$, where $q = 4/3$ is the power spectral index of string loops with cusps. This is because for $\mathcal{P}_n \propto n^{-q}$, the contribution of higher harmonics, and hence higher frequencies, becomes more important for smaller $q$, as discussed in \cite{Co:2021lkc}.

\section{Strings Eating Domain Walls}
\label{sec:dwconsumedbystrings}
\begin{figure}
    \centering
    \includegraphics[width=0.45\textwidth]{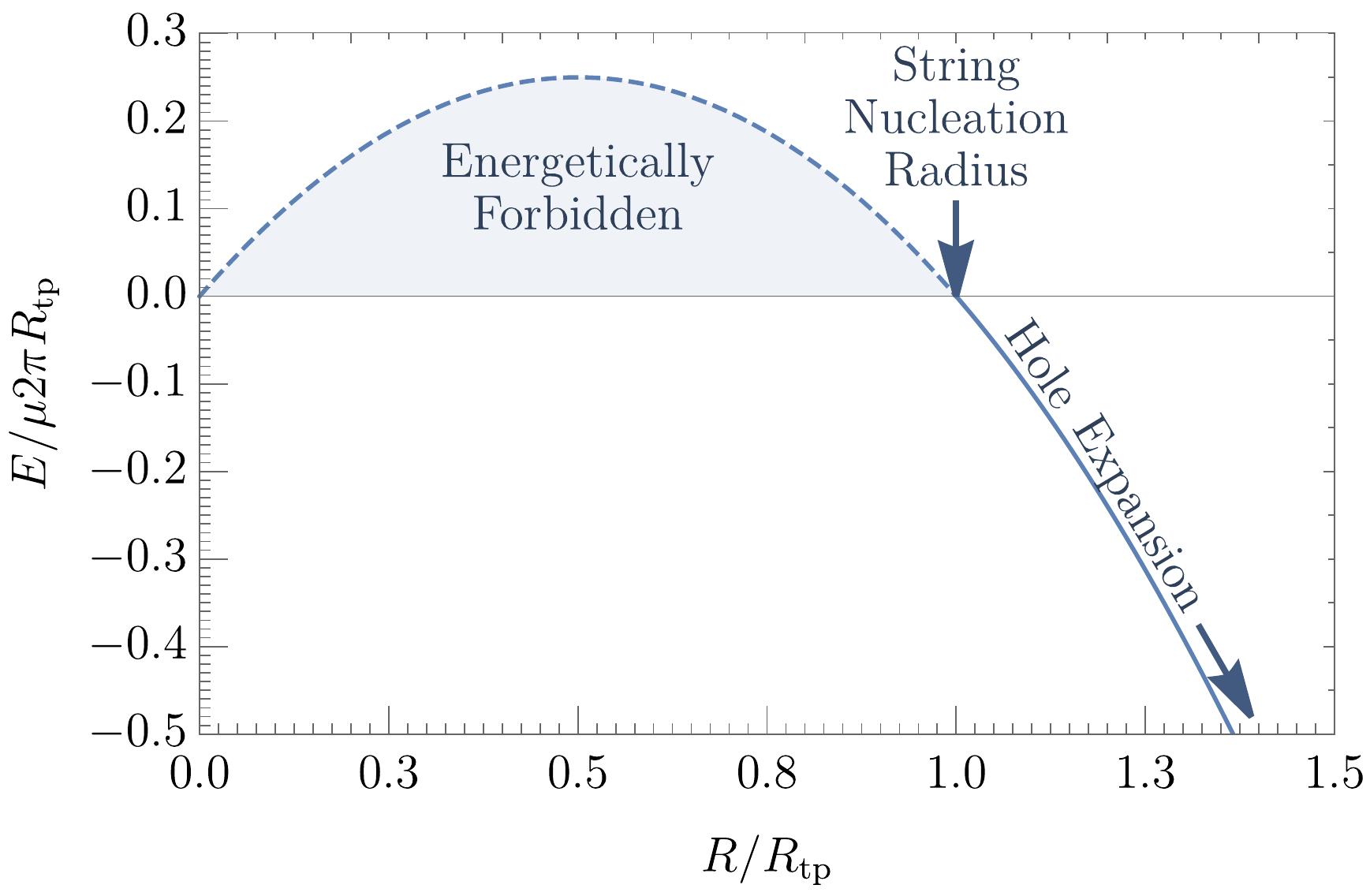}
     \includegraphics[width=0.45\textwidth]{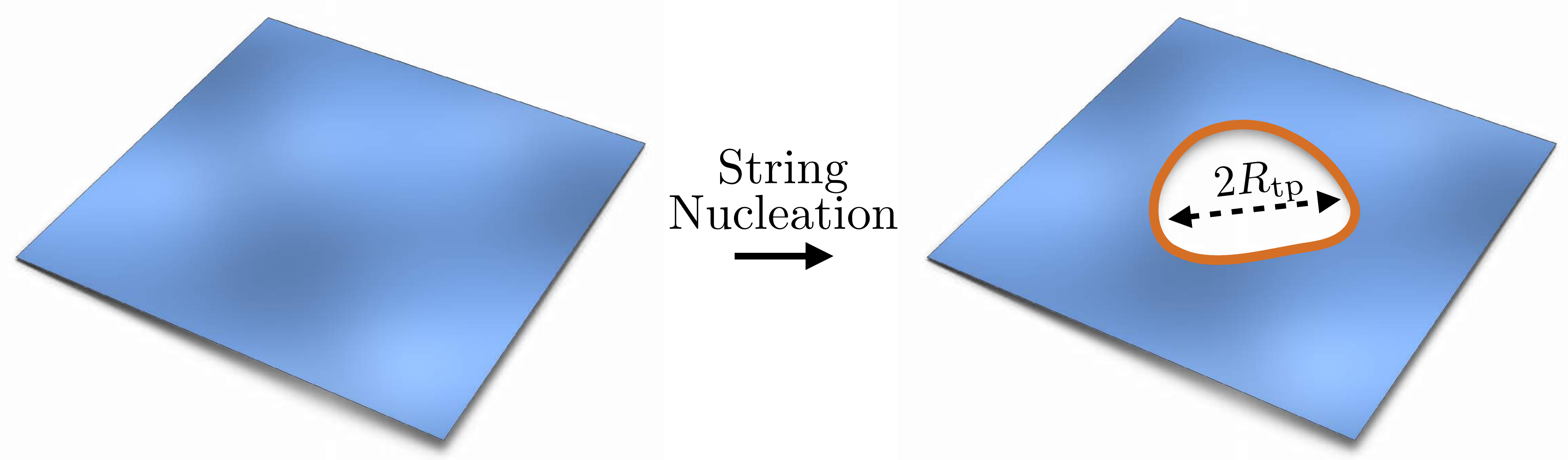}
    \caption{Top: Free energy diagram for a circular string-bounded hole nucleating on a domain wall vs the string nucleation radius, $R$. For $R > R_{\rm tp}$, the free energy of the system turns negative and it becomes energetically possible to nucleate a string in place of a wall element of area $\pi R_{\rm tp}^2$. Bottom: Illustration of the nucleation process. For walls with radii $R > R_{\rm tp}$ a piece of wall with area $\pi R_{\rm tp}^2$ is `eaten' and replaced with a string which forms the boundary of the punctured hole.}
    \label{fig:WKBWalls}
\end{figure}

In this section, we consider the case of strings nucleating on domain walls. As discussed in Sec. \ref{sec:topologicalDefectsFromGUTs}, if they are related by the same discrete symmetry, strings form first (in the initial phase transition that leaves an unbroken discrete symmetry), and connect to domain walls in the second phase transition (when the discrete symmetry is broken). When inflation occurs after the formation of strings but before domain walls, the string abundance is heavy diluted by the time the walls form. The absence of strings initially prevents the formation of string-bounded walls at the second stage of symmetry breaking and the walls initially evolve as a normal wall network.  Nevertheless, the walls can become bounded by strings later by the Schwinger nucleation of string holes as shown in Fig. \ref{fig:WKBWalls}. Conversion of wall rest mass into string kinetic energy causes the string to rapidly expand and `eat' the wall, causing the wall network to decay.

Strings can only nucleate on the wall if it is energetically possible to. The energy cost of producing a circular string loop is $\mu l$ where $l = 2\pi R$ is the length of the string, and the energy gained from destroying the interior wall is $\sigma A$ where $A = \pi R^2$ is the area of the eaten wall. The free energy of the string-wall system is then
 \begin{equation}
    \label{eq:freeEnergy}
     E = \mu 2 \pi R - \sigma \pi R^2 \ .  
 \end{equation}
The balance between string creation and domain wall destruction leads to a critical string radius, $R_{\rm tp}$, above which it is energetically favorable for the string to nucleate and continue expanding and consuming the wall as shown in Fig. \ref{fig:WKBWalls}. $E < 0$ gives this turning point radius
 \begin{equation}
     R_{\rm tp} = 2 R_c  \qquad R_c \equiv \frac{\mu}{\sigma} \ .
 \end{equation}
The probability for the string to tunnel through the classically forbidden region out to radius $R_{\rm tp}$ can be estimated from the WKB approximation. The nucleation rate per unit area is
\begin{align}
    \Gamma_s \propto \sigma e^{-S_E} \ ,
\end{align}
where 
 \begin{eqnarray}
   S_{\rm E} &=& \int _0 ^{R_{\rm tp}}dr \sqrt{\mu r E} 
   \propto \frac{\mu^3}{\sigma^2} \ ,
 \end{eqnarray}

More precisely, the tunneling rate can be estimated from the bounce action formalism and is found to be ~\cite{Kibble:1982dd, Preskill:1992ck}
\begin{align}
    \Gamma_s \sim \sigma \exp(-\frac{16 \pi}{3} \kappa_s)
\end{align}

where $\kappa_s = \mu^3/\sigma^2$. As a result, the string nucleation rate on the domain wall is typically exponential suppressed and the domain wall can be cosmologically long-lived if $\mu^3$ and $\sigma^2$ are disparate, similar to the string and monopole scales in Sec. \ref{sec:schwingerStringsMonopole}. For the coincidence of scales $\mu^3 \sim \sigma^2$, the domain wall network is metastable and may decay before dominating the energy density of the Universe. 

In terms of the symmetry breaking scale,  Eqs.~\eqref{eq:derivation_sigma} and \eqref{eq:derivation_mu} suggest
\begin{equation}
    \kappa_s =
    \frac{9 \pi^3}{\lambda_\sigma} \left(B\left(\frac{2\lambda}{e^2}\right)\right)^3 \left(\frac{v_\mu }{v_\sigma}\right)^6  \ . 
\end{equation}
for the fiducial models of Sec. \ref{sec:topologicalDefectsFromGUTs}. Since the homotopy selection rules require $v_\mu \geq v_\sigma$, nucleation of strings within cosmological timescales requires $B(2\lambda/e^2) \ll 1$, which can occur for $\lambda \ll e^2$.

Before decaying via string nucleation, the evolution of the metastable domain wall network is that of a pure domain wall network. The dynamics of a pure domain wall is well-described by the wall Nambu-Goto action \cite{vilenkin2000cosmic}
\begin{align}
    S=  - \sigma \int d^3 \zeta \sqrt{\gamma} \,,
    \label{eq:action_for_dw}
\end{align}
where $d^3 \zeta$ is the infinitesimal worldvolume swept out by the domain wall of tension $\sigma$, $\gamma \equiv |\det(\gamma_{ab})|$ is the determinant of the induced metric on the wall with $\gamma_{ab} = g_{\mu \nu} \frac{\partial X^\mu}{\partial \zeta^a} \frac{\partial X^\nu}{\partial \zeta^b}$. $X^\mu(\xi^a)$ are the spacetime coordinates of the wall with $\xi^a (a = 1,2,3)$ parameterizing the wall hypersurface, and $g_{\mu\nu} = a^2(\eta)(d\eta^2 - d \mathbf{x}^2)$ is the Friedmann-Robertson-Walker metric in conformal gauge. For large, roughly planar walls with a typical curvature radius, $R$, the  Euler-Lagrange equation of motion of \eqref{eq:action_for_dw} is \cite{Kawano:1989mw,Avelino:2005kn}
\begin{align}
\label{eq:vosDW}
 \frac{dv_{w}}{dt} &= (1 - v_w^2)\left(\frac{k_w}{R} - 3H v_{w} \right),
\end{align}
where $v_w$ is the average wall velocity perpendicular to the wall surface, $H$ is Hubble, and $k_w$ is an $\mathcal{O}(1)$ velocity-dependent function that parameterizes the effect of the wall curvature on the wall dynamics. Conservation of energy implies
\begin{align}
\label{eq:rhoDW}
\frac{d\rho_{w}}{dt} + H(1 + 3 v_w^2)\rho_{w} = - \frac{ c_w v_w}{R} \rho_{w},
\end{align}
which is coupled to Eq.~\eqref{eq:vosDW} via the `one scale' ansatz
\begin{align}
\label{eq:VOSansatzDW}
\rho_{w} \equiv \frac{\sigma R^2}{R^3} = \frac{\sigma}{R} .
\end{align}
Eq.~\eqref{eq:VOSansatzDW} states that the typical curvature and separation between infinite walls is the same scale, $R$. $c_w$ is an $\mathcal{O}(1)$ constant parameterizing the chopping efficiency of the infinite wall network into enclosed domain walls 
\footnote{These enclosed walls, known as `vacuum bags', are analogous to string loops forming from the intercommutation of a infinite string network. However, unlike string loops which can be long-lived, the vacuum bags collapse under their own tension and decay quickly. This is because the wall velocity becomes highly relativistic during collapse causing length contraction of the wall thickness and hence efficient particle emission of the scalar field associated with the wall \cite{Widrow:1989vj}.}. Note that Eq.~\eqref{eq:rhoDW} does not include gravitational wave losses which are small as long as the walls do not dominate the Universe.

Generally, the tunneling rate is sufficiently suppressed so that the domain walls reach the steady-state solution of Eqns. \eqref{eq:vosDW}-\eqref{eq:VOSansatzDW} before decaying, which is the scaling-regime such that $R/t \sim 1$ \cite{Avelino:2005kn}. In the scaling regime, the energy lost by the infinite wall network from self-intercommutation balances with the energy gained from conformal stretching by Hubble expansion so that the network maintains roughly one domain wall per horizon, similar to the scaling regime of the infinite string network in Sec. \ref{sec:schwingerStringsMonopole}. As a result, the energy density in the domain wall network before decay evolves with time as
\begin{align}
   \label{eq:rhoDWScaling}
    \rho_{w} = \mathcal{A}\frac{\sigma t^2}{t^3} = \mathcal{A}\frac{\sigma}{t},
\end{align}
where $\mathcal{A}$ is found to be $\mathcal{O}(1)$ from simulations \cite{Hiramatsu:2013qaa}.
For domain walls that are not highly relativistic, the total power emitted as gravitational radiation for a wall of mass $M_{w}$ and curvature radius $R_{}$ follows from the quadrupole formula \cite{Vilenkin:1984ib},
\begin{align}
    \label{eq:PowerGWDW}
    P_{\rm GW} \approx \frac{G}{45} \sum_{i,j} \langle \dddot{Q}_{ij} \dddot{Q}_{ij} \rangle \sim G (M_{w} R_{}^2  \, \omega^3)^2 =  \mathcal{B}\, G \sigma M_{w}.
\end{align}
In the last equation, we take the typical oscillation frequency $\omega$ and curvature $R_{}^{-1}$ to be comparable.
Numerical simulations of domain walls in the scaling regime confirm Eq.~\eqref{eq:PowerGWDW} with $\mathcal{B} \approx \mathcal{O}(1)$ \cite{Hiramatsu:2013qaa,Saikawa:2017hiv} .

In the scaling regime and prior to nucleation, the energy density rate lost into gravitational waves by the domain walls at time $t$ is then
\begin{align}
    \label{eq:GWBoltzmann}
   \frac{d\rho_{w}}{dt}^{(\rm GW)} = -n_{w} P_{\rm GW} \simeq -\mathcal{A}{\mathcal{B}}\frac{G \sigma^2}{t}.
\end{align}
In writing the right hand side of \eqref{eq:GWBoltzmann}, we use $\rho_{w} \simeq n_{w} M_{w}$ and insert Eq.~\eqref{eq:rhoDWScaling}. The energy density injected into gravitational waves is subsequently diluted with the expansion of the Universe.
The total energy density, $\rho_{\rm GW}$ in the gravitational wave background is thus described by the Boltzmann equation,
\begin{align}
    \label{eq:GWNucleation}
    \frac{d\rho_{\rm GW}}{dt} + 4H \rho_{\rm GW}  = \mathcal{A}\mathcal{B}\frac{G \sigma^2}{t}\theta(t_\Gamma - t) - x \frac{d \rho_{\rm DW}}{dt} \theta(t - t_\Gamma) ,
\end{align}
where $x \in [0,1]$ is an efficiency parameter characterizing the fraction of the energy density of the wall transferred into gravitational waves after strings begin nucleating and eating the wall, which occurs at time 
\begin{align}
    t_\Gamma \sim \frac{1}{\sigma A} e^{S_E} \sim \frac{1}{\sigma^{1/3}}\exp{\frac{16 \pi \kappa_s}{9}}.
\end{align}
Here, we take the wall area, $A$ at time $t_{\Gamma}$ to be $\sim t_{\Gamma}^2$ in accordance with the scaling regime. When the strings begin nucleating at $t_\Gamma$, they quickly expand from an initial radius $R_{\rm tp} = 2 R_c$ according to
\begin{align}
   \label{eq:stringExpansionWall}
    R_s(t) = \sqrt{4 R_c^2 + (t-t_{\Gamma})^2} ,
\end{align}
as shown in Appendix \ref{ap:DWstring} for circular string-bounded holes. Consequently, the strings rapidly accelerate to near the speed of light as they `eat' the wall. The increase in string kinetic energy arises from the devoured wall mass.
Thus, shortly after $t_\Gamma$, most of the energy density of the wall is transferred to strings and string kinetic energy. Numerical simulations outside the scope of this work are required to accurately determine the gravitational waves emitted from the typical relativistic collisions of the string bounded holes which mark the end of the domain wall network and hence the determination of $x$. As a result, we conservative take $x = 0$ when computing the resulting gravitational wave spectrum. Nevertheless, we can estimate the potential effect of non-zero $x$ by taking the sudden decay approximation for the wall. That is, assuming the destruction of the wall following nucleation occurs shortly after $t_\Gamma$, we may take $d\rho_{\rm DW}/dt \approx -\rho_{\rm DW}\delta(t - t_\Gamma)$.

The solution to \eqref{eq:GWNucleation} during an era with scale factor expansion $a(t) \propto t^\nu$ is then
\begin{align}
    \label{eq:gwEnergyDensityNucleation}
    \rho_{\rm GW}(t) &=
    \begin{cases}
    \mathcal{A}\mathcal{B}\dfrac{ G \sigma^2}{4 \nu}\left(1 - \left(\dfrac{t_{\rm scl}}{t}\right)^{4\nu}\right) &\; t \leq t_{\Gamma} 
    \\ 
    \left(\rho_{\rm GW}(t_\Gamma) +  x  \mathcal{A}\dfrac{\sigma}{{t_\Gamma}}\right) \left(\dfrac{a(t_{\Gamma})}{a(t)}\right)^{4} &\; t > t_{\Gamma} .
    \\
    \end{cases}
\end{align}
Eq.~\eqref{eq:gwEnergyDensityNucleation} demonstrates that the gravitational wave energy density background quickly asymptotes to a constant value after reaching scaling at time $t_{\rm scl}$ and to a maximum at the nucleation time $t_\Gamma$. We thus expect a peak in the gravitational wave amplitude of approximately 
\begin{align}
    \label{eq:peakAmplitudeNucleation}
    \Omega_{\rm GW, max} &\approx \frac{\rho_{\rm GW}(t_\Gamma)}{\rho_c(t_\Gamma)}\left(\frac{g_{*0}}{g_*(t_{\Gamma})}\right)^{ \scalebox{1.01}{$\frac{1}{3}$}}
    \\
    &=\frac{16\pi}{3}\left[
    (G \sigma t_\Gamma)^2 + 2 x 
    G \sigma t_\Gamma \right] \Omega_r \left(\frac{g_{*0}}{g_*(t_{\Gamma})}\right)^{ \scalebox{1.01}{$\frac{1}{3}$}}
\end{align}
where we take $t_\Gamma > t_{\rm scl}$, $\mathcal{A} = \mathcal{B} = 1$, and a radiation dominated background at the time of decay with $\nu = \frac{1}{2}$. $\Omega_r = 9.038 \times 10^{-5}$ is the critical energy in radiation today \cite{Aghanim:2018eyx}.

The first term in the second line of \eqref{eq:peakAmplitudeNucleation}, the contribution to the peak amplitude from gravitational waves emitted prior to nucleation, agrees well with the numerical results of \cite{Hiramatsu:2013qaa} if $t_\Gamma$ maps to the decay time of unstable walls in the authors' simulations. Note that in \cite{Hiramatsu:2013qaa}, the domain walls are global domain walls and are unstable due to a vacuum pressure difference arising from the insertion of a $Z_2$ breaking term in the domain wall potential. In this work, we consider gauged domain walls in which such a discrete breaking term is forbidden.

The second term in \eqref{eq:peakAmplitudeNucleation}, the contribution to the peak amplitude from gravitational waves emitted after nucleation, has not been considered in numerical simulations. The post-nucleation contribution dominates the pre-nucleation contribution if $x \gtrsim G \sigma t_{\Gamma}$, which may be important for short-lived walls. The complex dynamics of string collisions during the nucleation phase motivates further numerical simulations. 

The frequency dependence on the gravitational wave amplitude may be extracted from numerical simulations of domain walls in the scaling regime. The form of the spectrum was found in \cite{Hiramatsu:2013qaa} to scale as
\begin{align}
    \Omega_{\rm GW}(f) &=\frac{f}{\rho_c}\frac{d \rho_{\rm GW}(t_0,f)}{df} 
    \nonumber 
    \\ 
    \label{eq:nucleationSpectrumSimulation}
    &\approx \Omega_{\rm GW,max} 
    \begin{cases}
        \left(\dfrac{f}{f_{\rm peak}}\right)^{-1} \quad & f > f_{\rm peak}
        \\
        \left(\dfrac{f}{f_{\rm peak}}\right)^{3} \quad & f \leq f_{\rm peak}
    \end{cases}
\end{align}
where 
\begin{align}
    f_{\rm peak} \sim \frac{1}{t_{\Gamma}}\frac{a(t_\Gamma)}{a(t_0)}
\end{align}
is the fundamental mode of oscillation at the time of decay. 
The infrared $f^3$ dependence for $f < f_{\rm peak}$ arises from causality arguments for an instantly decaying source \cite{Caprini:2009fx}.

\begin{figure}
    \centering
    \includegraphics[width=0.48\textwidth]{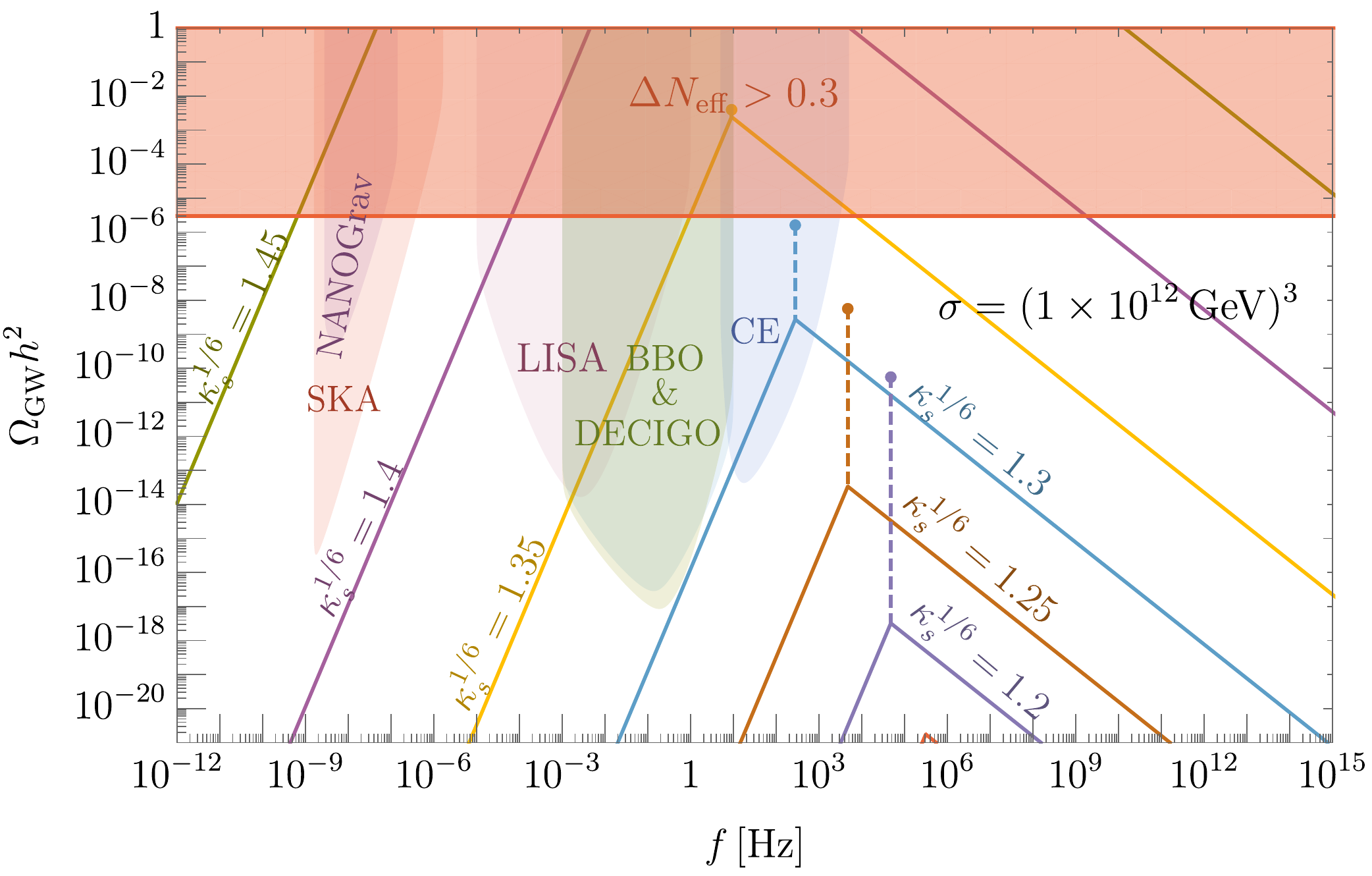}
    \caption{Representative spectra of gravitational waves emitted by domain walls that are eaten by nucleation of strings for fixed $\sigma^{1/3} = 10^{12}\, {\rm GeV}$. Each colored contour corresponds to a different value of $\kappa_s = \mu^3/\sigma^2$ which parameterizes the ratio between string and wall symmetry breaking scales and sets the nucleation time of the strings on the wall. Since nucleation is an exponentially suppressed process, the metastable wall network is typically cosmologically long-lived and behaves as pure wall network before nucleation.  At high frequencies, $\Omega_{\rm GW}$ scales as $f^{-1}$ while after nucleation, $\Omega_{\rm GW}$ decays as $f^3$ by causality \cite{Hiramatsu:2013qaa}. For sufficiently large $\kappa_s$, the domain wall network is long-lived enough to dominate the energy density of the Universe at decay and emits enough gravitational radiation to violate measurements of $\Delta N_{\rm eff}$, as shown by the red region. Consequently, $\kappa_s$ must be close to unity so that walls decay by string nucleation before wall domination.}
    \label{fig:nucleationSpectrum}
\end{figure}

Fig. \ref{fig:nucleationSpectrum} shows a benchmark plot of the gravitational wave spectrum from domain walls consumed by string nucleation for fixed $\sigma = (10^{12} \,{\rm GeV)^3}$ and a variety of $\kappa_s = \mu^3/\sigma^2$. In computing the spectrum, we evaluate  \eqref{eq:nucleationSpectrumSimulation}, in the conservative limit of $x = 0$. The corresponding dots above each triangular vertex shows the potential peak of the spectrum in the $x \rightarrow 1$ limit which corresponds to the assumption that all of the wall energy at nucleation goes into gravitational waves. For sufficiently large $\kappa_s$, the domain wall energy density grows relative to the background and can come to dominate the critical density of the Universe at the time of decay. This can lead to gravitational radiation producing too large $\Delta N_{\rm eff}$, \eqref{eq:deltaNeff}, as shown by the red region. For relatively long-lived walls nucleating prior to wall domination, it is possible for many gravitational wave detectors to observe the $\Omega_{\rm GW, peak}$ and the characteristic $f^{-1}$ ultraviolet slope and $f^3$ infrared slope.

\begin{figure}
    \centering
    \includegraphics[width=0.48\textwidth]{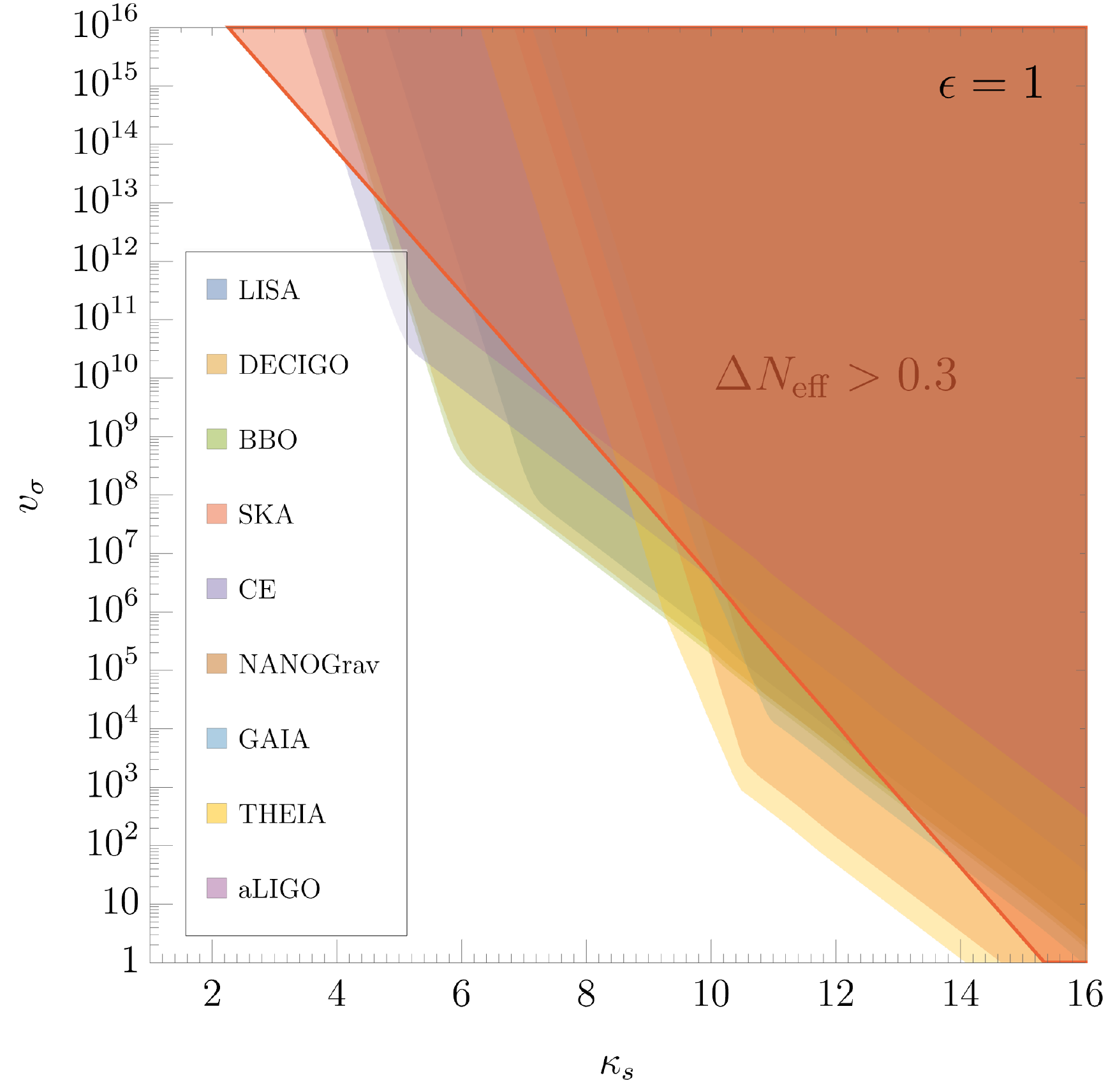}
    \caption{The parameter region in the $v_\sigma-\kappa_s$ plane where the gravitational wave spectrum from domain walls eaten by the nucleation of strings can be detected. We take the fiducial value $\epsilon = 1$ so that $\sigma = v_\sigma^3$. For a given $(v_\sigma,\kappa_s)$, a detection is registered when $\Omega_{\rm GW}$ is greater than the sensitivity curve of the given detector. In the red region, the energy density emitted by walls into gravitational radiation is large enough to be excluded by $\Delta N_{\rm eff}$ bounds. Deep in the red region, $\kappa_s$ is sufficiently large that the walls are so long-lived that they dominate the energy density of the Universe.}
    \label{fig:nucleationParameterSpace}
\end{figure}

Fig. \ref{fig:nucleationParameterSpace}, shows the detector reach of $\Omega_{\rm GW}$ in the $v_\sigma - \kappa_s$ plane. Here we take $\epsilon = 1$ so that $v_\sigma = \sigma^{1/3}$. Since the triangular shaped spectrum from a domain wall eaten by strings is sufficiently different compared to a flat, stochastic string background, we register a detection of the string nucleation gastronomy as long as $\Omega_{\rm GW}h^2$ exceeds the threshold of detection for a given experiment.  Fig. \ref{fig:nucleationParameterSpace} demonstrates that a wide range of $\sigma$ and $\kappa_s$ can be probed. Note that most detection occurs when the walls decay shortly before coming to dominate the Universe as shown by the diagonal red $\Delta N_{\rm eff}$ region. In general, wall symmetry breaking scales $v_{\sigma}$ between $1$ and $10^{13}$ GeV and $\kappa_s$ between $4-15$ can be detected by current and near future gravitational wave detectors.

In addition, while the infrared ($f^{-3}$) and ultraviolet ($f^{-1}$) wall spectrum is similar to the monopole burst spectrum of Sec. \ref{sec:stringDestruction}, there is a logarithmic plateau at the peak of the monopole burst spectrum that is absent for the walls and hence can be used to distinguish both gastronomy signals. Moreover, in first order phase transitions where the bulk of the energy goes into the scalar shells, the envelope approximation predicts a similar spectrum ($f^3$ in the infrared, $f^{-1}$ in the ultraviolet) \cite{Kosowsky:1992vn}. However, more sophisticated analyses of this type of phase transition appear to predict a UV spectrum that scales as $f^{-1.5}$ \cite{Cutting:2018tjt} making it unlikely that a wall or monopole network eaten by strings can be mimicked by a first order phase transition.

\section{Domain Walls Eating Strings}
\label{sec:stringboundedWalls}
In this section, we consider the gastronomy case where domain walls attach to, and consume, a pre-existing string network. The symmetry breaking chains that allow this are the same as in the previous section, with the difference between the two scenarios arising from when inflation occurs relative to string formation. For the string nucleation gastronomy of Sec. \ref{sec:dwconsumedbystrings}, inflation occurs after string formation but before wall formation. For walls attaching to a pre-existing string network as considered in this section, inflation occurs before string and wall formation. In this scenario, the string network is not diluted by inflation and at temperatures below the wall symmetry breaking scale, $v_\sigma$, walls fill in the space between strings. Note that since the attachment of walls to a pre-existing string network is not a nucleation process, there does not have to be a coincidence of scales between $v_\mu$ and $v_{\sigma}$ as in the case of strings nucleating on walls as discussed in Sec. \ref{sec:dwconsumedbystrings}. 

The outline of this section is as follows: First, we derive the equation of motion for the string boundary of a circular wall and quantitatively show how the wall tension dominates the string dynamics when the radius, $R$, of the hybrid defect is greater than $R_c \equiv \mu/\sigma$, and how the string dynamics reduce to pure string loop motion for $R \ll R_c$. We then run a velocity one-scale model on an infinite string-wall network, and show how the walls pull their attached strings into the horizon when the curvature radius of the hybrid network grows above $R_c$. Once inside the horizon, the domain wall bounded string pieces oscillate and emit gravitational radiation, which we compute numerically. We find that power emitted in gravitational waves asymptotes to the pure string limit, $P_{\rm GW} \propto G\mu^2$ for pieces of string-bounded bounded walls with radii $R \ll R_c$, and to the expected power emitted by domain walls from the quadrupole approximation, $P_{\rm GW} \propto G\sigma^2 R^2$, for $R \gg R_c$. We use the numerically computed gravitational wave power to derive the energy density evolution and the gravitational wave spectrum of a network of circular string-bounded wall pieces. We discuss the features of this gastronomy signal and its experimental detectability with current and future gravitational wave detectors. Last, we discuss how model dependent effects such as friction on the string or wall can affect the spectrum.

\subsection{The String-Wall Equation of Motion}
Let us begin with the total action of a wall bounded by a string with wall tension $\sigma$ and string tension $\mu$,
\begin{equation}
    S=  - \sigma \int d^3 \zeta \sqrt{\gamma}- \mu \int d^2 \zeta \sqrt{\Upsilon} .
    \label{eq:action_for_dw-str}
\end{equation}
The parameters of the wall action (left term) are the same as in Eq.~\eqref{eq:action_for_dw}. For the string action (right term), $d^2 \zeta$ is the infinitesimal wordsheet swept out by the string, $\Upsilon \equiv |\det(\Upsilon_{ab})|$ is the determinant of the induced metric on the string, and $\Upsilon_{ab} = g_{\mu \nu} \frac{\partial Y^\mu}{\partial \zeta^a} \frac{\partial Y^\nu}{\partial \zeta^b}$, where $Y^\mu(\zeta^0, \zeta^1)$ are the spacetime coordinates of the string which is fixed to lie at on the boundary of the wall.

Assuming the wall velocities are not ultra-relativistic and the string boundary on the wall is approximately circular, one can derive the the Lagrangian for the string boundary of the wall to be
\begin{align}
    L&= -2\pi \mu |\mathbf{r}_s(\eta)| a^2(\eta) \sqrt{1 - \left({\frac{d\mathbf{r}_s}{d\eta}}\right)^2} - \sigma \pi \mathbf{r}_s(\eta)^2 a^3(\eta),
    \label{eq:action_subs_dw-str}
\end{align}
where $\mathbf{r}_s$ is the comoving position vector of the string boundary, $\eta$ is conformal time, and $a$ the scale factor of the Universe. See Appendix \ref{ap:DWstring} for details, including a justification of the assumptions. The Lagrangian \eqref{eq:action_subs_dw-str} generates the following Euler-Lagrange equation of motion
\begin{align}
    \label{eq:EulerLagrangeEOM}
    \frac{d^2\mathbf{r}_s}{d\eta^2} 
    &=- \frac{\sigma}{\mu}\left(1- \left(\frac{d\mathbf{r}_s}{d\eta}\right)^2\right)^{3/2} a(\eta) \hat{\mathbf{r}}_s 
    \nonumber \\
    &-\left(1- \left(\frac{d\mathbf{r}_s}{d\eta}\right)^2\right)\left(\frac{\hat{\mathbf{r}}_s}{|\mathbf{r}_s|} + 2 \mathcal{H}\frac{d\mathbf{r}_s}{d\eta} \right),
\end{align}
where $\mathcal{H} = d\ln a/d\eta = H a$ is the conformal Hubble rate.

In the limit that the physical size of the wall, $\mathbf{R}_s = \mathbf{r}_s a$ is much smaller than the critical radius $ R_c \equiv \mu/\sigma$, the equation of motion for the string bounded wall reduces to the standard result of a pure circular string loop \cite{Garriga:1993gj,vilenkin2000cosmic}.  However, for $|\mathbf{R}_s| \geq R_c $, the domain wall tension dominates the string tension and the string motion becomes more relativistic. This can also be simply understood by noting that a wall-bounded string of curvature radius $R$ experiences a wall tension force $F \sim \sigma R$ and a string tension force  $F \sim \mu$, which become comparable at $R = R_c$ \cite{Everett:1982nm,Vilenkin:1984ib}. 

\begin{figure}
    \centering
    \includegraphics[width=0.45\textwidth]{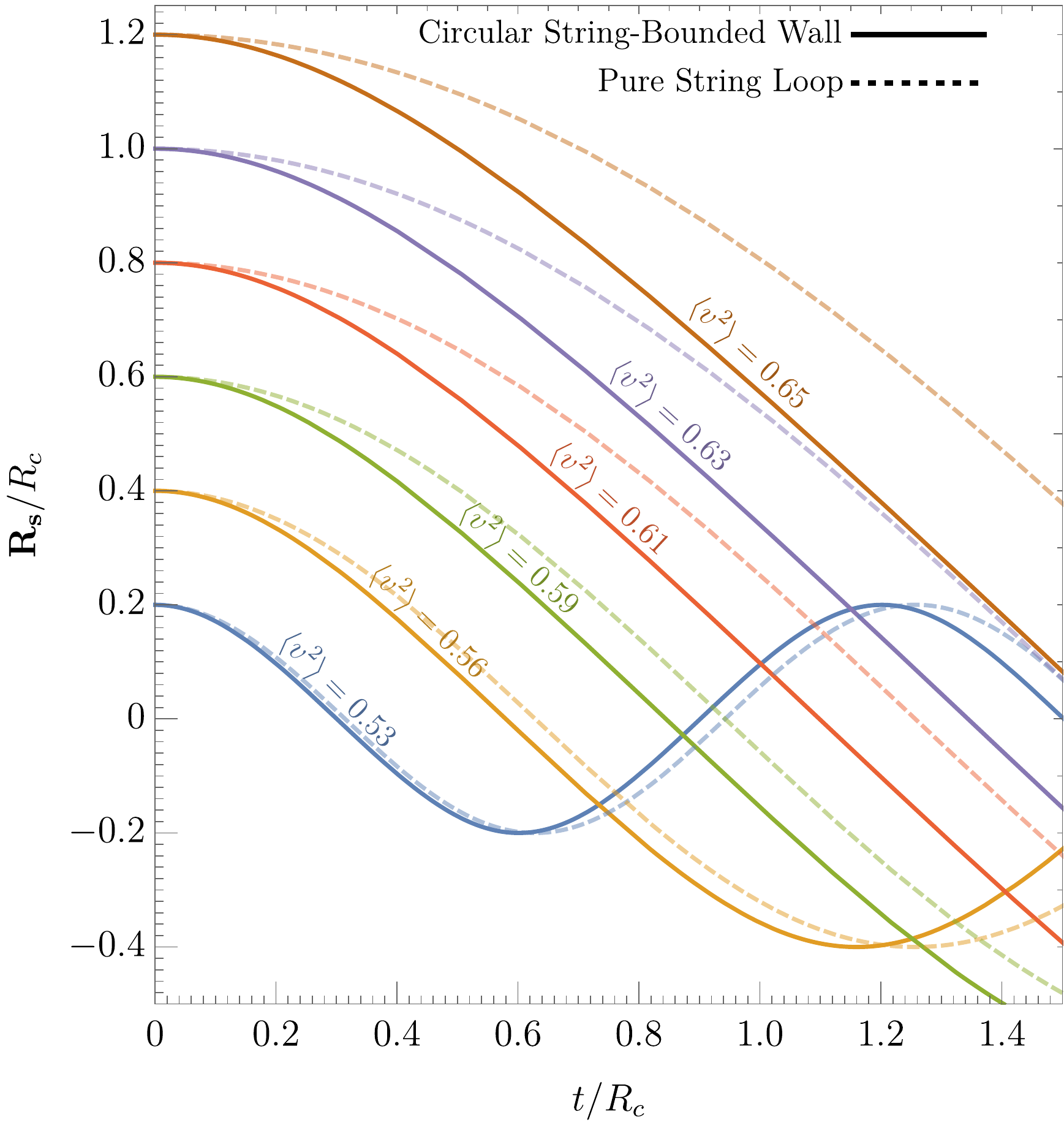}
    \caption{Evolution of a circular string radius $\mathbf{R}_s$ as a function of time in the flat spacetime limit (ie subhorizon strings) when the string is the boundary of domain wall (solid) and when it is a pure string loop (dashed). The colored contours show the evolution for a variety of different string sizes. When the string is small compared to $R_c = \mu/\sigma$, the string dominates the dynamics and circular string-bounded walls oscillate similarly to pure string loops of the same size. However, when the string size becomes of order or greater than $R_c$, the wall dominates the dynamics of the string and causes the string to oscillate highly relativistically compared to pure string loops of the same size. This can be seen by the increase of the period-averaged velocity squared, $\langle v^2 \rangle$, which increases from approximately $0.5$ in the pure string loop limit to more relativistic values as the size of the string-bounded wall grows above $R_c$.}
    \label{fig:GWBoundedStringRadius}
\end{figure}

Fig. \ref{fig:GWBoundedStringRadius} shows the numerical solution of Eq.~\eqref{eq:EulerLagrangeEOM} for the string boundary  as a function of the initial string size in the flat spacetime limit, ($a \rightarrow 1, \eta \rightarrow t$), or equivalently, after the loops have entered the horizon. For $|\mathbf{R}_s| \ll R_c$, the evolution of $\mathbf{R}_s$ for the string-bounded wall is identical to the pure string loop motion (dashed lines) \cite{Garriga:1993gj}. For string-bounded walls with $|\mathbf{R}_s| \gtrsim R_c$, the evolution deviates from the pure string loop, with the domain wall accelerating its string boundary to highly relativistic speeds for most of its oscillation period. The highly relativistic string boundaries are responsible for the gravitational wave emission of string-bounded walls as discussed later in this section. 
 \begin{figure}
    \centering
    \includegraphics[width=0.48\textwidth]{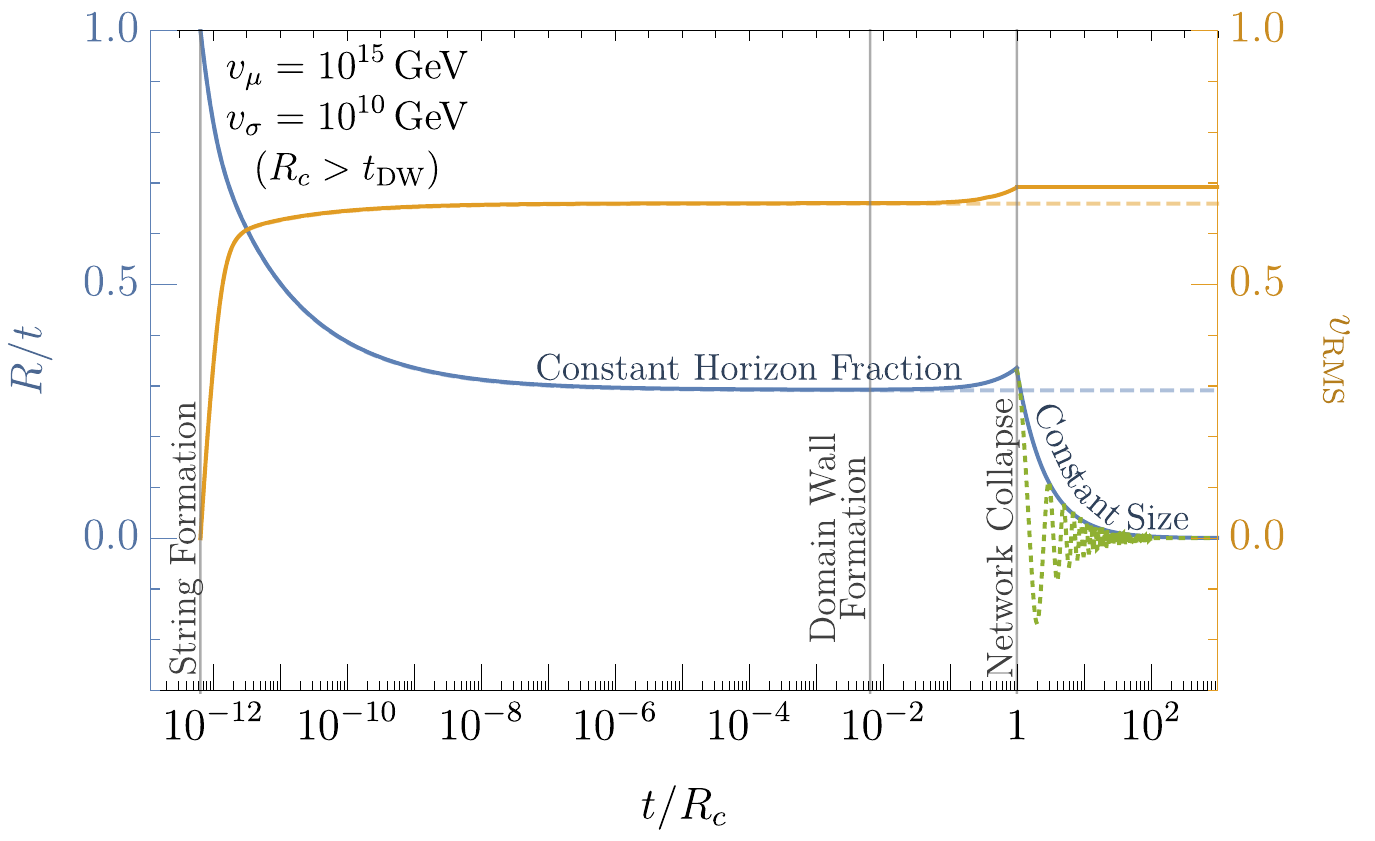}
    \includegraphics[width=0.48\textwidth]{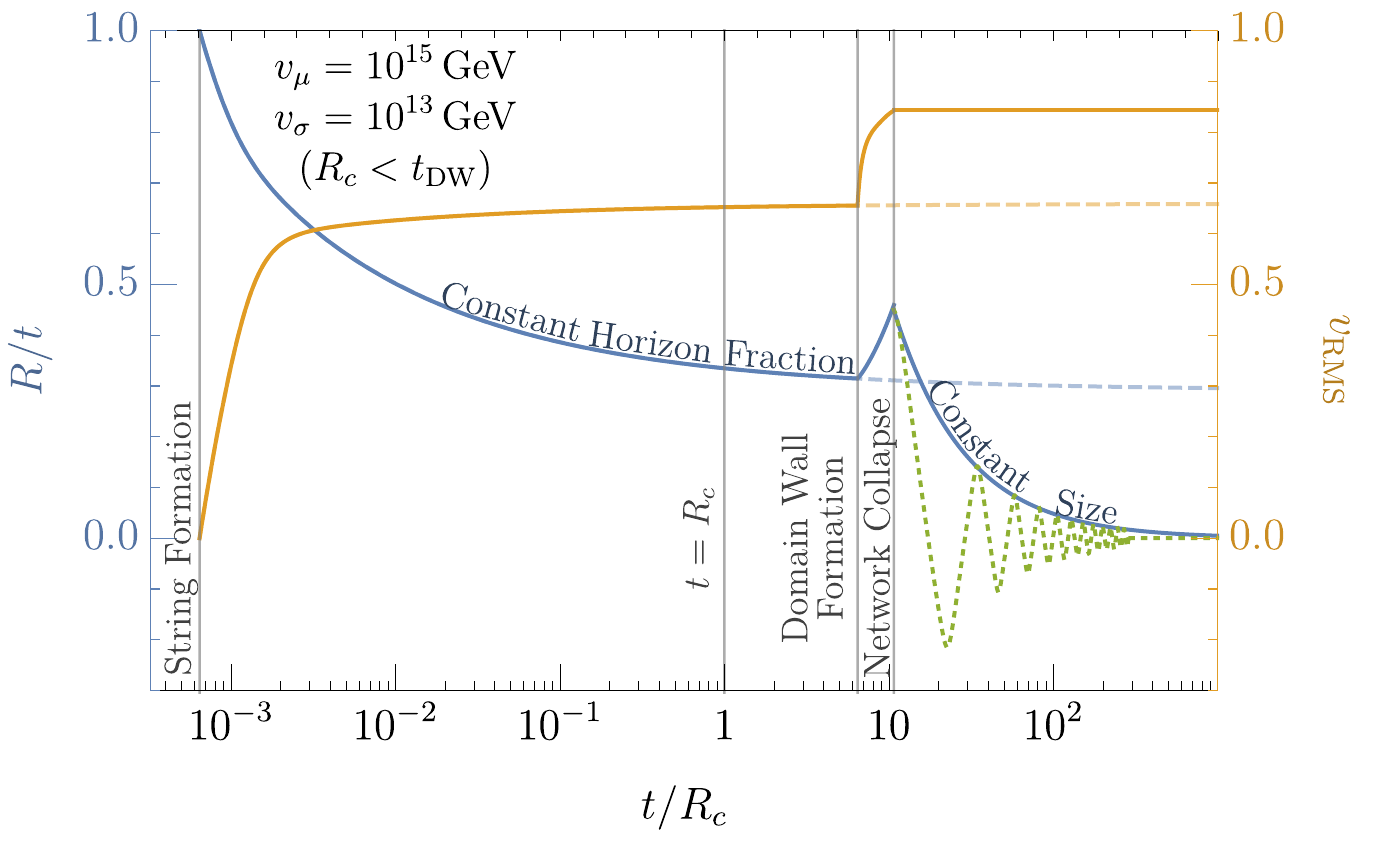}
    \caption{Evolution of the infinite string-wall network. The blue curve shows the curvature radius of the string-bounded walls over time, $R/t$, while the orange curve shows the string RMS velocity, $v_{\rm RMS}$. Top: Representative case where $t_{\rm DW} < R_c$ so that walls form before dominating the string dynamics. For $t < R_c$, we numerically compute the modified one-scale model equations. The string-wall network reaches a scaling regime where $R$ maintains a constant fraction of the horizon. As $t$ approaches $R_c$, the walls begin dominating the dynamics and the strings move more relativistically. At $t = R_c$, infinite string-bounded walls with curvature radius $R$  behave like wall-bounded strings of curvature radius $R$. We approximate this transition by piecewise connecting the one-scale model solution to the numerical solution of the Euler-Lagrange equation of motion for a circular string-bounded wall. For $t > R_c$, the infinite network collapses and the pieces oscillate at constant physical size before decaying via gravitational waves.
    Bottom: Same as the top but representative of the case where $t_{\rm DW} > R_c$ so that walls form already dominating the string dynamics. In this scenario, $v_{\rm RMS}$ of the infinite wall network abruptly increases at wall formation. We transition from the one-scale to the Euler-Lagrange solution when $v_{\rm RMS}$ of the infinite strings approximately reaches $v_{\rm RMS}$ of a string-bounded wall piece of the same curvature radius.}
    \label{fig:scalingCases}
\end{figure}
\subsection{Collapse of the Infinite String-Wall Network}
\label{subsection:collapseOfInfiniteNetwork}
For subhorizon loops, $|\mathbf R_s| \lesssim t$, the Hubble term in Eq.~\eqref{eq:EulerLagrangeEOM} is subdominant compared to the string curvature term and hence the motion of the domain wall bounded string loops approaches the flat spacetime limit. However, for superhorizon or `infinite' strings, the effect of the expansion of the Universe is critical. To understand the evolution and collapse of the infinite string-wall network, we implement a `one-scale' model ~\cite{Martins:1995tg,Martins:1996jp,Martins:2000cs} by rewriting Eq.~\eqref{eq:EulerLagrangeEOM} in terms of the RMS comoving velocity, $v_{s} \equiv -\sqrt{\langle \mathbf{v}_s \cdot \mathbf{v}_s \rangle} = -\sqrt{\langle d\mathbf{r}/d\eta \cdot d\mathbf{r}/d\eta \rangle}$ of the typical long string,
\begin{align}
    \label{eq:vos1}
    \frac{d v_{s}}{dt} &= (1 - v_{s}^2)\frac{k(R,v_{s})}{R} -2 H v_{\infty} 
\end{align}
where
\begin{align}
    k(v_{s},R)&= \frac{\langle(1 - \mathbf{v}_s^2 + \frac{R}{R_c}(1 - \mathbf{v}_s^2)^{3/2})\mathbf{v}_s \cdot \hat{\mathbf{r}}_s \rangle}{v_{s}(1-v_{s}^2)}
\end{align}
is the wall-modified curvature parameter. Similarly, the energy density of the infinite network, $\rho_\infty$, can be decomposed into infinite string, $\rho_s = \beta \rho_\infty$, and wall, $\rho_w = (1 - \beta)\rho_\infty$, contributions. That is, $0 \leq \beta \leq 1$ parameterizes the relative energy density between strings and walls with the entire energy density in strings when $\beta = 1$ and the entire energy density in walls when $\beta = 0$.
\footnote{A similar analysis for a string-monopole network with $Z_{N \geq 3}$ strings was considered in \cite{vachaspati1987evolution}. In \cite{vachaspati1987evolution}, monopoles are connected to multiple strings which allows the monopole-string `web' to be long-lived and reach a steady-state scaling regime.}
The energy density evolution of the infinite string-wall network is then
\begin{align}
    \label{eq:vos2} 
&    \frac{d \rho_\infty}{dt} + 3H(1 + w)\rho_\infty = -\frac{c v_\infty}{R}\rho_{\infty},
\end{align}
where $c$ is a chopping efficiency parameter and
\begin{align}
   \label{eq:EOS}
   w = \frac{2}{3}(1 + v_s^2) \beta + (\frac{1}{3} + v_w^2) (1-\beta) - 1
\end{align}
is the equation of state of the infinite wall-string network \cite{Boehm:2002bm,Sousa_2011}, with $v_s$ and $v_w$ the average string and wall speeds, respectively. Note the wall speed is unimportant to the wall-string evolution for the following reason: For $R \lesssim R_c$, the strings dominate the energy density and $\beta \simeq 1$. For $R \gtrsim R_c$, the energy density is initially mostly in the walls, but is quickly converted to string kinetic energy with $v_s$ and then $\beta$ quickly becoming approximately $1$. Thus, for any $R$, we expect the wall contribution in Eq.~\eqref{eq:EOS} (second term) to be subdominant to the string contribution (first term) and set $\beta \simeq 1$ for all time which eliminates $v_w$ from the wall-string dynamics. 
\par The chopping efficiency, $c$, of the infinite network into loops is expected to be an $\mathcal{O}(1)$ number \cite{Vilenkin:1984ib}. For definiteness, we take the pure-string result $c \approx 0.23$ inferred from simulations \cite{Martins:2000cs}. Last, the `momentum parameter' $k$, is an $\mathcal{O}(1)$ number which parameterizes the effect of the string curvature and wall tension on the infinite string dynamics and vanishes when $v_{\infty}$ matches the RMS velocity, $v_0$, of the string loops in flat space \cite{Martins:2000cs}.  $v_0 = 1/\sqrt{2}$ for any pure string loop \cite{vilenkin2000cosmic}, but is an increasing function of $R/R_c$ for string-bounded walls as shown graphically by Fig. \ref{fig:GWBoundedStringRadius}. As a result, we approximate $k(v,R)$ by the pure-string momentum parameter \cite{Martins:2000cs}
\begin{align}
    \label{eq:momentumParameter}
    k(v_s,R) \approx \frac{2\sqrt{2}}{\pi}(1-v_s^2)(1+2\sqrt{2}v_s^3)\frac{v_0(R)^6- v_s^6}{v_0(R)^6 + v_s^6},
\end{align}
but with $v_0$ now the $R/R_c$ dependent RMS velocity of the string bounded walls as computed numerically from Eq.~\eqref{eq:EulerLagrangeEOM}. In the pure string limit, $R_c \rightarrow \infty$, equations \eqref{eq:vos1}-\eqref{eq:momentumParameter} reduce to the standard one scale model.

The two equations \eqref{eq:vos1}, \eqref{eq:vos2}, are coupled via the `one scale' ansatz
\begin{align}
\label{eq:VOSansatz}
\rho_{\infty} \equiv \frac{\mu R + \sigma R^2 \theta(t - t_{\rm DW})}{R^3} =
\frac{\mu}{R^2}\left(1 + \frac{R}{R_c}\theta(t - t_{\rm DW})\right),
\end{align}
where $t_{\rm DW} \approx M_{Pl}C/v_{\sigma}^{2}$ is the wall formation time. The ansatz \eqref{eq:VOSansatz} amounts to assuming the typical curvature and separation between infinite string-bounded walls is the same scale, $R$. Note that while $\rho_{\infty}$ is the total rest mass energy density of the combined string-wall network, the allocation of the total energy density is shared among the two defects.

We evaluate the coupled system of equations \eqref{eq:vos1}-\eqref{eq:momentumParameter} in time up until the one-scale ansatz breaks down. This occurs when the curvature radius $R$ of the infinite strings approaches $R_c$, at which point the wall tension dominates the string tension and the walls pull the infinite strings with curvature radius $R$ effectively into string bounded domain walls of radius $R$. At this point, we evaluate Eq.~\eqref{eq:EulerLagrangeEOM} with the initial conditions taken from the one-scale solution and piecewise connect the two solutions so that each solution is valid in their respective regimes. 

For a given string tension $\mu$ and wall tension $\sigma$, two general collapse scenarios arise. One, when the walls form before $R \sim R_c$ and the other when they form after, as represented by the top and bottom panels of Fig. \ref{fig:scalingCases} , respectively.  If the wall formation time $t_{\rm DW}  < R_c$, the walls gradually come to dominate the infinite string dynamics with $v_{s}$ and $R$ rising slightly before $t = R_c$ as shown by the orange and blue curves, respectively. Here, we define the right-axis $v_{\rm RMS}$ as the RMS velocity for the infinite strings $v_{s}$ prior to network collapse, and to the RMS velocity of the wall-bounded string pieces, $v_0$, after network collapse.
\footnote{For the one-scale model, the energy density \textit{decreases} as $R$ \textit{increases}. $R$ increases slightly before $t = t_*$ because $\rho_{\infty}$ redshifts faster. This is because the equation of state of the wall-string network briefly behaves more like radiation due to the sudden increase in $v_{s}$ caused by the walls.}
In this scenario, we define the network collapse time as $t_* = R_c$ from which point on we evaluate Eq.~\eqref{eq:EulerLagrangeEOM} to determine the dynamics of the string system. If the wall formation time $t_{\rm DW} > R_c$, the walls dominate the strings upon formation, and $v_{s}$ increases abruptly as shown in the bottom panel of Fig. \ref{fig:scalingCases}. In this scenario, we define the network collapse time as the time when $v_{s}$ approximately matches $v_0$ as determined from Eq.~\eqref{eq:EulerLagrangeEOM}, from which point on we evaluate Eq.~\eqref{eq:EulerLagrangeEOM} to determine the dynamics of the system. Since the collapse proceeds shortly after domain wall formation, the collapse time of the infinite network is effectively at $t_* = t_{\rm DW}$. \par In summary, we take the time of collapse of the infinite string-wall network and hence the end of loop production, to be 
\begin{align}
    \label{eq:t*}
    t_* \equiv \text{Max}(R_c, t_{\rm DW}),
\end{align}
as first proposed by \cite{Martin:1996ea}. More realistic simulations beyond our 
one-scale analysis and piecewise approximations are required to more precisely determine $t_*$. Nevertheless, the sudden increase in $v_{s}$ and $R$ around $t_*$ according to the one-scale analysis or comparing each term in the string equation of motion to determine at what time each term dominates as done in subsection \ref{subsec:friction} when we consider friction, indicate that the walls begin dominating the infinite string dynamics near a time of order Eq.~\eqref{eq:t*}. Moreover, the gravitational wave spectrum from wall-bounded strings is fairly weakly dependent on the precise value of $t_*$, and knowing $t_*$ to within a factor of a few is sufficient to accurately compute the gravitational wave spectrum as discussed later in this section. 

\subsection{Gravitational Wave Emission from String-Bounded Walls}
When a string-bounded domain wall piece enters the horizon, it oscillates at constant amplitude as shown by the dotted green curves of Fig. \ref{fig:scalingCases} since they are subhorizon and do not experience the conformal expansion with the horizon.
 As they oscillate, the loops emit gravitational waves with a total power \cite{Weinberg:1972kfs}
\begin{gather}
    \label{eq:gwPower}
    P_{\rm GW} = \sum_n \int d\Omega \frac{dP_n}{d\Omega} 
    \\
    \frac{dP_n}{d\Omega} = \frac{G \omega_n^2}{\pi}\left[T^{\mu \nu*}(\mathbf{k}, \omega_n) T_{\mu \nu}(\mathbf{k},\omega_n) - \frac{1}{2}|T^{\mu}_{\; \; \mu} (\mathbf{k},\omega_n)|^2 \right]
\end{gather}
where $\omega_n = |\mathbf{k}| = 2 \pi n/T$ is the frequency of the $n$th harmonic of the string-bounded wall oscillating with period $T$. The stress tensor of the string-wall system is
\begin{align}
    T^{\mu \nu}(\mathbf{k}, \omega_n) &= \frac{1}{T} \int_0^T dt\,  e^{i \omega_n t} \int d^3\mathbf{x} e^{- i \mathbf{k} \cdot \mathbf{x}}T^{\mu \nu}(\mathbf{x},t) 
    \\
   \label{eq:stressTensor}
    T^{\mu \nu}(\mathbf{x},t) &= \int_{\rm string} \mu \, |\mathbf{R}_s| d\theta \, \gamma \frac{d Y^\mu}{dt} \frac{d Y^\nu}{dt} \delta^3(\mathbf{x} - \mathbf{R}_s(t))
\end{align}
where $dY/dt = (1, \mathbf{V}_s)$, $\gamma = (1 - \mathbf{V}_s \cdot \mathbf{V}_s)^{-1/2}$, and $\mathbf{V}_s = d\mathbf{R}_s/dt$ is the physical velocity of the string.
 \begin{figure}
    \centering
    \includegraphics[width=0.48\textwidth]{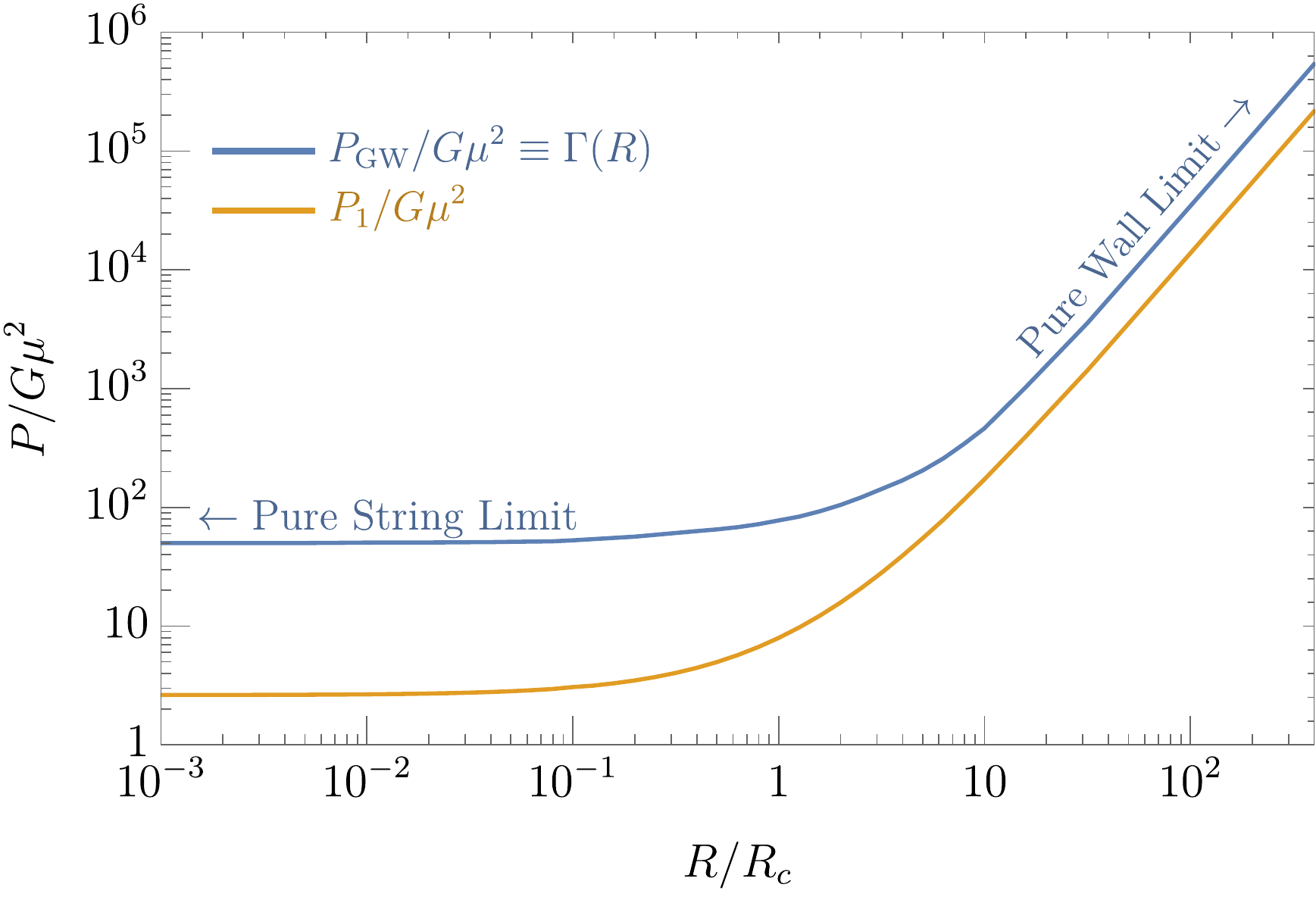}
   \includegraphics[width=0.48\textwidth]{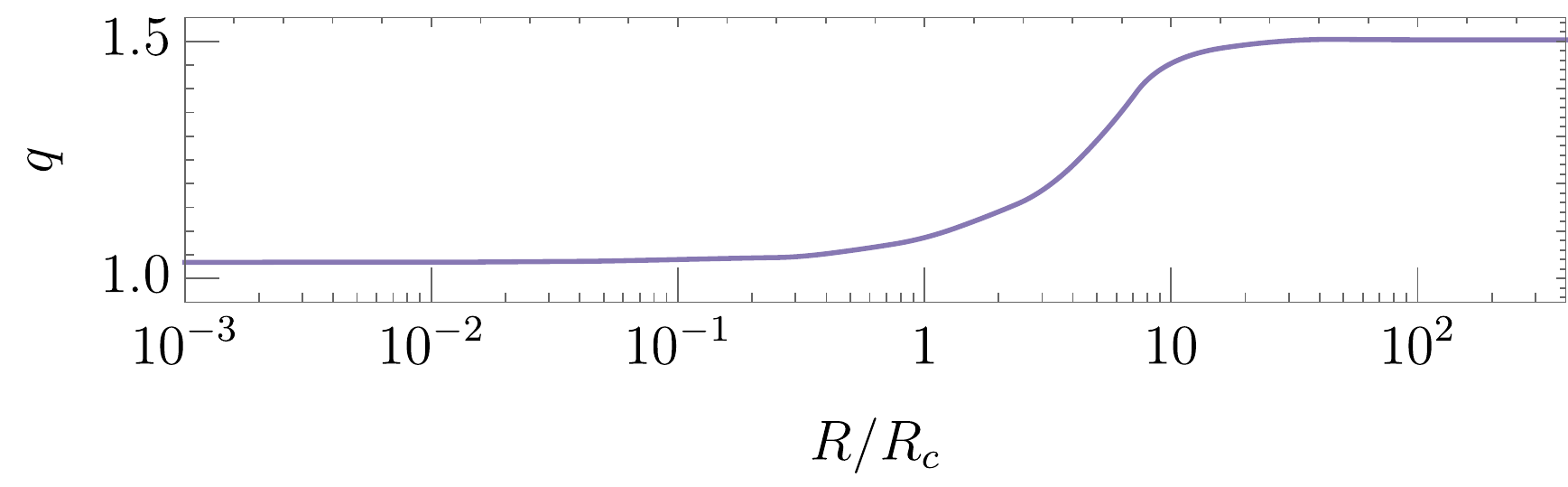}
    \caption{Top: The gravitational power, $P_{\rm GW}$, emitted by string-bounded walls as a function of $R/R_c$.  The orange contour shows the power in the first harmonic $P_1$ while the blue contour shows the total power. For $R \ll R_c$, the string dominates the dynamics and we recover the pure string loop limit, namely $P_{\rm GW}/G\mu^2 = \Gamma_s$, where $\Gamma_s \approx 50$ is a constant and is independent of string size. For $R \gg R_c$, the wall dominates the dynamics and we recover the pure domain wall limit, namely $P_{\rm GW} \approx G \sigma M_{\rm DW}$. Bottom: The power spectral index as a function of $R/R_c$, defined by $P_n \propto n^{-q}$. In the pure string limit, $q \rightarrow 1$ and in the pure wall limit, $q \rightarrow 3/2$. }
    \label{fig:GWBoundedStringPower}
\end{figure}

We calculate the gravitational wave power of the string-wall system by numerically computing Eqns. \eqref{eq:gwPower} - \eqref{eq:stressTensor} for circular string-bounded walls using the numerically computed time evolution of $\mathbf{R}_s$ from the Euler-Lagrange equation of motion \eqref{eq:EulerLagrangeEOM}. The orange contour of Fig. \ref{fig:GWBoundedStringPower} shows the ratio of the gravitational wave power in the first harmonic, $P_{1}$, to $G \mu^2$ as a function of $R/R_c$, where $R$ is the string oscillation radius. For $R \ll R_c$, the string dominates the dynamics and the power is independent of loop size, in agreement with the pure string case. However, for $R \gg R_c$, the domain wall dominates the dynamics and the power deviates from the pure string case, increasing quadratically
with $R/R_c$. Since $R_c \equiv \mu/\sigma$, this is equivalent to $P_{\rm GW} \propto G \sigma^2 R^2 \propto G \sigma M_{\rm DW}$, in agreement with the quadrupole formula expectation for gravitational wave emission from domain walls.

The bottom panel of Fig. \ref{fig:GWBoundedStringPower} shows the power spectral index, $q$, as a function of $R/R_c$ where $q$ is defined by the index $P_n \propto n^{-q}$. We numerically determine $q$ by examining the asymptotic dependence of $P_n$ for $n$ up to $\sim 300$. In the string dominated regime ($R \ll R_c$), $q \simeq 1$ which agrees with the pure string result of a \textit{perfectly circular} string loop \cite{Battye:1997ji}. In the domain wall dominated regime ($R \gg R_c$) we find $q \simeq 3/2$.  

Note the mild (logarithmic) divergence in the total power for $R \ll R_c$ is an artifact of perfectly circular loops \cite{PhysRevD.31.3052,Battye:1997ji} and more realistic loops, which will not be perfectly circular but have cusps, will moderate the divergence such that $P_n \propto n^{-4/3}$ for large $n$. Although realistic loops are not perfectly circular, nearly all loop configurations emit similar total power in gravitational waves \cite{PhysRevD.31.3052,Burden:1985md,vilenkin2000cosmic}, including nearly circular, but not completely symmetric loops. Indeed, numerically calculations of nearly circular pure string loops have $P_1$ nearly identical to our numerical result in the $R \ll R_c$ limit, but have finite total power similar to most string loop geometries, $P_{\rm tot} \approx (50-100) G \mu^2$ \cite{PhysRevD.31.3052}. As a result, to match with a realistic ensemble of loops which are not perfectly circular and contain cusps, we cut-off the artificial logarithmic divergence in the $R \ll R_c$ regime by normalizing $P_{\rm tot}$ to the typical string loop such that $P_{\rm tot}/G\mu^2 \equiv \Gamma_s \simeq 50$. For $R > R_c$ when $q < 1$, we take the total power $P_{\rm tot} \simeq P_1/\zeta(q)$ which is the total power for $P_n = P_1 n^{-q}$.  For convenience in computing the gravitational wave spectrum in the following subsection, we define the function $\Gamma(R) \equiv P_{\rm GW}(R)/G\mu^2$ for string-bounded walls, where $\Gamma(R)$ is now a function of $R/R_c$. The blue contour of Fig. \ref{fig:GWBoundedStringPower} shows $\Gamma(R)$ as a function of $R/R_c$. For $R/R_c \ll 1$, $\Gamma \rightarrow \Gamma_s$ while for $R/R_c \gg 1$, $\Gamma \rightarrow 3.7 (R/R_c)^2$. Note the power in the large $R/R_c$ regime is equivalent to $P_{\rm GW} \simeq 1.2 G \sigma M_{\rm DW}$ for a circular string-bounded wall, which agrees well with the numerical power inferred from simulations of domain walls in a scaling regime \cite{Hiramatsu:2013qaa}.

\subsection{Gravitational Wave Spectrum from String-Bounded Walls}
Now that the gravitational wave power emitted by a string-bounded domain wall is known, we may calculate the gravitational wave spectrum from a network of circular string-bounded walls. First, we analytically estimate the expected amplitude and frequency of the spectrum to gain intuition before computing it numerically.

Consider first a pure string loop without walls that forms at time $t_k$ with initial length $l_k = \alpha t_k$, where $\alpha \simeq 0.1$ is the typical fixed ratio between loop formation length and horizon size found in simulations \cite{Blanco-Pillado:2017oxo,Blanco-Pillado:2013qja}. Once inside the horizon, these loops oscillate and their energy density redshifts $\propto a^{-3}$ because their energy $E = \mu l$ is constant in the flatspace limit. The loops emit gravitational radiation with power $P_{\rm GW} = \Gamma_s G \mu^2$, where $\Gamma_s \approx 50$, and eventually decay from gravitational radiation at time
\begin{align}
    t_{\Gamma} \approx \frac{\mu l_k}{\Gamma_s G \mu^2} \quad (\text{Pure string loop lifetime}).
\end{align}
When the pure string loops form and decay in a radiation dominated era, their energy density at decay is
\begin{flalign}
    \rho(t_\Gamma) \approx \mu l_k n(t_k)\left(\frac{t_k}{t_\Gamma}\right)^{3/2} \quad \Big(\parbox{2.2cm}{\centering
    Pure string \\ decay density}\Big)
\end{flalign}
where $n(t_k) \approx  \frac{1}{3}\frac{\mathcal{F}C_{\rm eff}}{\alpha t_k^3}$ is the initial number density of loops of size $l_k$ that break off from the infinite string network in a scaling regime \cite{Cui:2018rwi,Gouttenoire:2019kij,Sousa_2013}. As found by simulations, $\mathcal{F} \approx 0.1$ \cite{Blanco-Pillado:2013qja} is the fraction of energy ultimately transferred by the infinite string network into loops of size $l_k$ and $C_{\rm eff} \approx 5.4$ is the loop formation efficiency in a radiation dominated era \cite{Cui:2017ufi,Blasi:2020wpy}.

As a result, the gravitational wave amplitude arising from these pure string loops is approximately
\begin{align}
    \label{eq:pureStringEstimation}
    \Omega_{\rm GW}^{(\rm str)} &\approx \frac{\rho(t_\Gamma)}{\rho_c (t_{\Gamma})} \Omega_{\rm r} \left(\frac{g_{*0}}{g_*(t_{\Gamma})}\right)^{ \scalebox{1.01}{$\frac{1}{3}$}}  \\
    &= \frac{32 \pi}{9}\mathcal{F}C_{\rm eff}\sqrt{\frac{\alpha G \mu}{\Gamma_s}}\Omega_{\rm r} \left(\frac{g_{*0}}{g_*(t_{\Gamma})}\right)^{ \scalebox{1.01}{$\frac{1}{3}$}}  \quad \Big(\parbox{1.75cm}{\centering
    Pure-string \\ amplitude}\Big) \nonumber
\end{align}
where $\rho_c(t_\Gamma)$ is the critical energy density of the Universe at $t_\Gamma$. 

Until $t = t_*$, the strings dominate the dynamics of any string-bounded walls, and the spectrum must be approximately that of a pure string spectrum with $\Omega_{\rm GW}$ given approximately by Eq.~\eqref{eq:pureStringEstimation}, independent of frequency. Now, consider a near circular string-bounded wall that forms at time $t_k = t_*$ with initial circumference $l_k = \alpha t_k$. If $l_k \lesssim 2 \pi R_c$, the power emitted and total mass of the system is effectively identical to the pure string case so that the $\Omega_{\rm GW}$ is again the same as Eq. \ref{eq:pureStringEstimation}. However, if $l_k \gtrsim 2\pi R_c$, the power emitted and mass of the system is dominated by the wall contribution of the wall-string piece. In this case, the wall bounded string decays from gravitational radiation at time
\begin{align}
    t_{\Gamma} \approx \frac{\sigma l_k^2/4\pi}{\Gamma(l_k) G \mu^2} \approx \frac{1}{G \sigma} \quad
    \Big(\parbox{2.6cm}{\centering
    String-bounded \\ wall lifetime}\Big)
\end{align}
When the wall bounded strings form and decay in a radiation dominated era, their energy density at decay is 
\begin{flalign}
    \rho(t_\Gamma) \approx \frac{\sigma  l_k^2}{4\pi} n(t_k)\left(\frac{t_k}{t_\Gamma}\right)^{3/2} \quad \Big(\parbox{2.8cm}{\centering
    String-bounded \\ wall decay density}\Big)
\end{flalign}
where $n(t_k) \approx  \frac{1}{3}\frac{\mathcal{F}C_{\rm eff}}{\alpha t_k^3}$ follows from the infinite string-wall network being in the scaling regime with $\mathcal{F}$ and $C_{\rm eff}$ expected to be similar to the pure string values right before the infinite network collapses at $t_*$.  

As a result, the gravitational wave amplitude arising from these string-bounded wall pieces is approximately
\begin{align}
    \label{eq:wallStringEstimation}
    \Omega_{\rm GW} &\approx \frac{\rho(t_\Gamma)}{\rho_c (t_{\Gamma})} \Omega_{r} \left(\frac{g_{*0}}{g_*(t_{\Gamma})}\right)^{1/3} \\
    &= \frac{8}{9}\mathcal{F}C_{\rm eff} \alpha \sqrt{G \sigma t_k}\Omega_{r}\Big(\frac{g_{*0}}{g_*(t_{\Gamma})}\Big)^{ \scalebox{1.01}{$\frac{1}{3}$}} \, 
    \Big(\parbox{2.5cm}{\centering
    String-bounded \\ wall amplitude}\Big). \nonumber
\end{align}
The largest amplitude of \eqref{eq:wallStringEstimation} occurs at the latest formation time $t_k$, which is $t_*$, the collapse time of the infinite network. Consequently, a `bump' relative to the flat string amplitude occurs if
\begin{align}
    \label{eq:omegaRatio}
    \frac{\Omega_{\rm GW}}{\Omega_{\rm GW}^{(\rm str)}} \approx \frac{1}{4\pi}\sqrt{\frac{\Gamma_s \alpha t_*}{R_c}} \approx 0.2 \left(\frac{\alpha}{0.1}\right)^{ \scalebox{1.01}{$\frac{1}{2}$}} \left(\frac{\Gamma_s}{50}\right)^{ \scalebox{1.01}{$\frac{1}{2}$}} \left(\frac{t_*}{R_c}\right)^{ \scalebox{1.01}{$\frac{1}{2}$}}
\end{align}
is greater than $1$ and at a frequency 
\begin{align}
    \label{eq:peakFreqEstimation}
    f_{\rm peak} \sim \frac{1}{l_k} \frac{a(t_\Gamma)}{a(t_0)}.
\end{align}
since the walls remain the same size once inside the horizon and dominantly emit at the frequency of the harmonic, $f_{\rm emit} \sim l_k^{-1}$. Here, $l_k \approx \alpha t_*$.
 \begin{figure}
    \centering
    \includegraphics[width=0.45\textwidth]{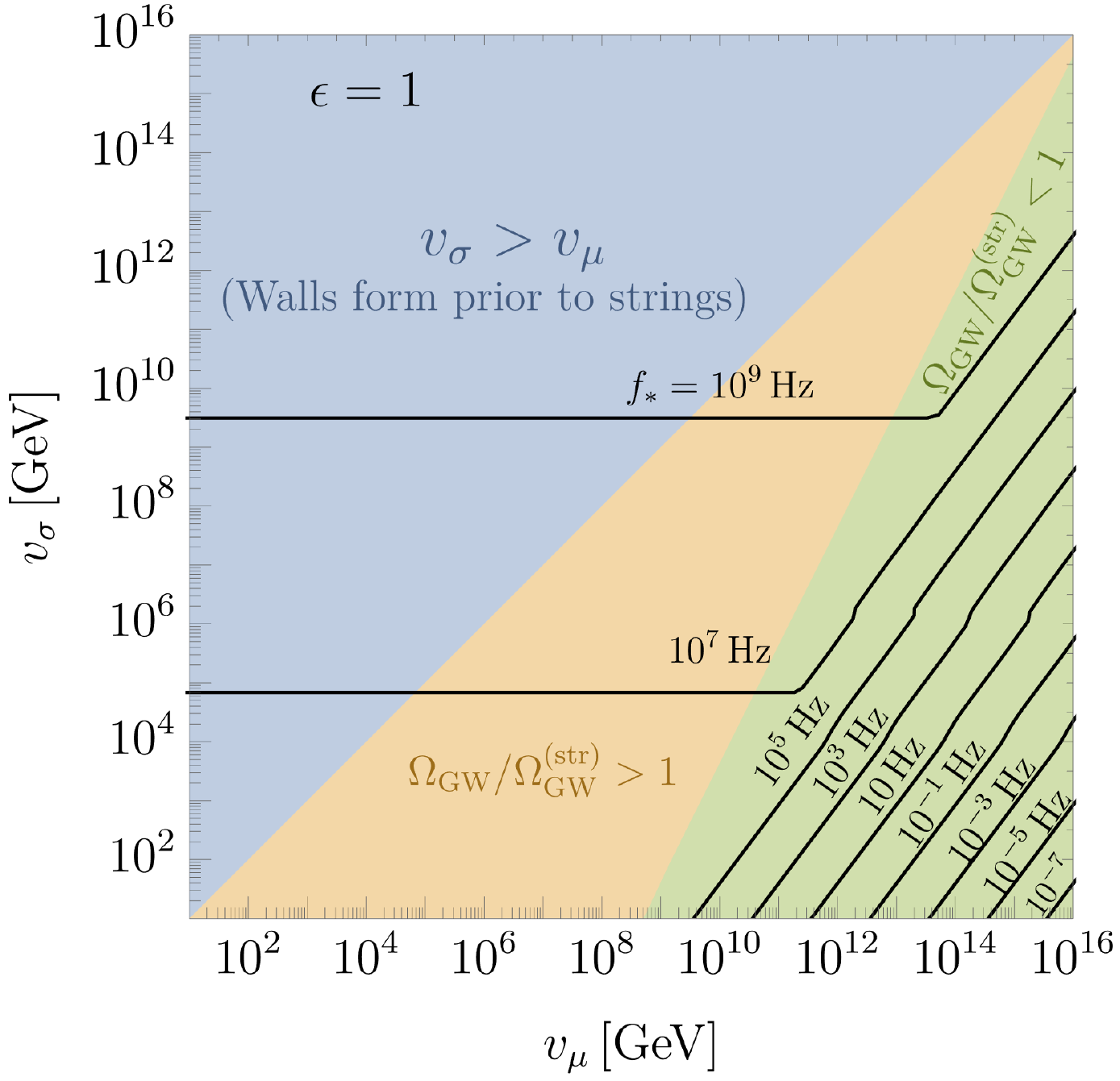}
    \includegraphics[width=0.45\textwidth]{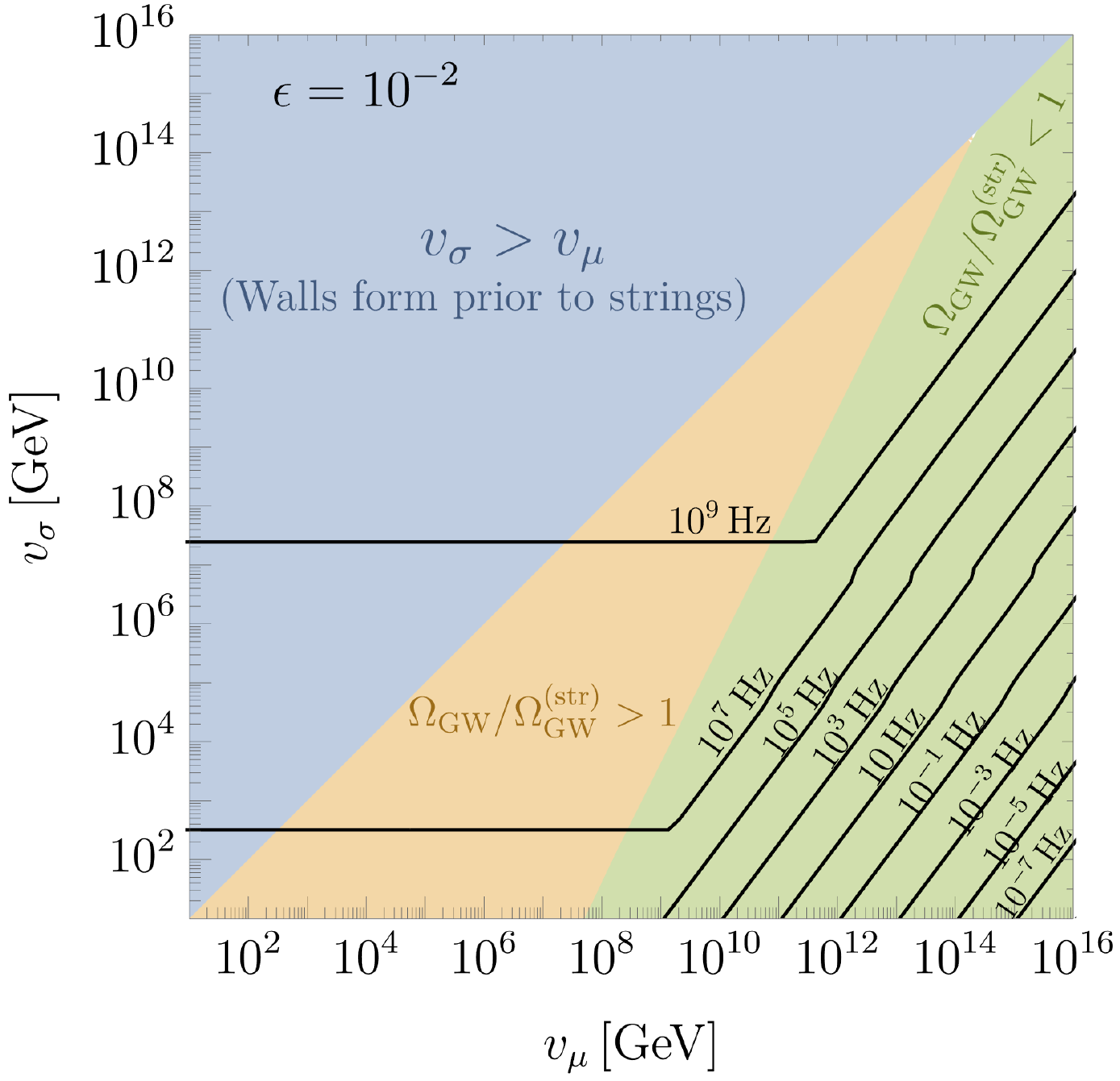}
    \caption{The $v_\sigma - v_\mu$ parameter space where wall-bounded strings can generate a gravitational wave signal. In the green region, the largest string-bounded walls at the network collapse time, $t_*$ have a lifetime comparable to pure-string loops of the same size. The energy density they deposit into gravitational waves when they decay is comparable to pure string loops and hence they do not produce a `bump' in $\Omega_{\rm GW}$ relative to the flat pure-string spectrum at high frequencies. In the yellow region, the largest string-bounded walls at the network collapse time, $t_*$ are sufficiently large that their lifetime is long compared to a pure string loop of the same size. The energy density they deposit into gravitational waves when they decay is greater than pure string loops and a `bump' in $\Omega_{\rm GW}$ can be observed relative to the flat string spectrum. In the blue region, $v_\mu < v_\sigma$ which is forbidden for composite string-bounded walls. The black contours show the approximate frequency, $f_*$, where $\Omega_{\rm GW}$ decays from the pure string spectrum. The top and bottom panels show the same regions for $\epsilon \equiv \sigma/v_\sigma^3 = 1$ and $10^{-2}$, respectively.}
    \label{fig:wallStringSpectrumParameterSpace}
\end{figure}

The estimation of Eq.~\eqref{eq:omegaRatio} indicates that if $t_* \gg R_c$, then $\Omega_{\rm GW}$ features a `bump' relative to the flat string spectrum before decaying. Qualitatively, in this limit, the walls are large enough and hence massive enough to live much longer than the pure string loops of the same size. As a result, their energy density before decaying from gravitational radiation is enhanced relative to shorter-lived pure string loops.  For $t_* \approx R_c$, the spectrum does not feature an enhancement over the pure string spectrum because the string-bounded walls are small in size and decay quickly. Nevertheless, as we will show numerically, the spectrum still decays as $f^3$ which can still be distinguished from the $f^2$ decay signal from monopoles eating strings as discussed in Sec. \ref{sec:stringDestruction}. As a result, for any $t_*$, we expect a unique gravitational wave gastronomy signature from gauge groups that produce string-bounded walls. 

Fig. \ref{fig:wallStringSpectrumParameterSpace} shows the parameter space in the $v_\mu - v_\sigma$ plane where we can expect certain gravitational wave signatures from cosmic gastronomy. Here, $v_\mu \equiv \mu^{1/2}$ and $v_{\sigma} \equiv (\sigma/\epsilon)^{1/3}$ where $\epsilon \lesssim 1$ is parameterizes the coupling constant of the scalar field which breaks the discrete symmetry associated with the domain wall. 
 
With the qualitative features of the spectrum understood, we turn to a numerical computation of $\Omega_{\rm GW}$.

The energy lost by oscillating circular loops of length $l = 2\pi R$ is 
\begin{align}
    \frac{dE}{dt} = \frac{d}{dt}\left(\mu l  + \frac{\sigma l^2}{4\pi}\right) = - \Gamma(l) G \mu^2,
\end{align}
As a result, loops that form at time $t_k$ with initial size $l_k = \alpha t_k$ slowly decrease in size according to 
\begin{align}
    \label{eq:lengthLoss}
    G\mu(t - t_k) = \int_l^{\alpha t_k}  dl' \frac{1 + \frac{l'}{2 \pi R_c}}{\Gamma(l')}.
\end{align}
As before, the stochastic gravitational wave energy density spectrum is 
\begin{gather}
    \label{eq:DWPiecesSpectrum}
    \frac{d \rho_{\rm GW}(t)}{df} =
    \int_{t_{\rm sc}}^t dt' \frac{a(t')^4}{a(t)^4} \int dl \frac{dn(l,t')}{dl} \frac{dP(l,t')}{df'}\frac{df'}{df} 
    \\
    \label{eq:redshift}
    \frac{df'}{df} = \frac{a(t)}{a(t')}  \quad \quad \frac{dn}{dl}(l,t') = \frac{dn}{dt_k}\frac{dt_k}{dl} 
    \\
    \frac{dP (l,t')}{df'} = \Gamma(l) G \mu^2 l \, g\left(f \frac{a(t)}{a(t')} l\right)
\end{gather}
where
\begin{align}
    \frac{dn}{dt_k} &= \left(\frac{\mathcal{F} C_{\rm eff}(t_k)}{\alpha t_k^4} \frac{a(t_k)^3}{a(t')^3}\right)\theta(t_* - t_k) 
\end{align}
is the loop number density production rate which follows from roughly one loop of size $\alpha t_k$ breaking off from the infinite wall-string network every Hubble time and then redshifting $\propto a^{-3}$.
\begin{align}
    \frac{dt_k}{dl} &= \frac{1 + \frac{l}{2 \pi R_c}}{\Gamma(l) G \mu}\left(1 + \frac{\alpha(1 + \frac{\alpha t_k}{2 \pi R_c})}{\Gamma(\alpha t_k) G\mu} \right)^{-1}
\end{align}
follows from differentiating Eq.~\eqref{eq:lengthLoss} with respect to $t_k$, and
\begin{align}
    \label{eq:g(x)}
    g(x) = \sum_n \mathcal{P}_n \delta(x - \xi n) \qquad \xi \equiv \frac{l}{T}
\end{align}
is the normalized power spectrum for a discrete spectrum where $2 \leq \xi \leq \pi$ with $\xi = 2$ corresponding to the pure string limit ($l \ll 2\pi R_c$) and $\xi = \pi$  corresponding to the ultrarelativistic limit ($l \gg 2\pi R_c$). As discussed in the previous subsection, we take $\mathcal{P}_n = n^{-q}/\zeta(q)$ with $q = 4/3$ to match on to more realistic non-circular strings with cusps. Above, primed coordinates refer to emission and unprimed refer to the present so that gravitational waves emitted from the string-bounded wall at time $t'$ with frequency $f'$ will be observed today with frequency $f = f' a(t')/a(t)$. Last, $t_k$ is solved for numerically from Eq.~\eqref{eq:lengthLoss}.  

Integrating the energy density spectrum, \eqref{eq:DWPiecesSpectrum} over loop length $l$ and normalizing by the present day 
energy density, $\rho_{\rm c} = 3H_0^2/8\pi G$,
yields the present day gravitational wave spectrum from domain wall bounded strings
\begin{align}
   \label{eq:dwPieceSpectrum}
    \Omega_{\rm GW} 
    &= \sum_n \frac{8 \pi (G \mu)^2}{3 H_0^2} \int_{t_{\rm sc}}^{t_0} dt' \frac{a(t')^5}{a(t_0)^5}\left(\frac{\mathcal{F} C_{\rm eff}(t_k)}{\alpha t_k^4} \frac{a(t_k)^3}{a(t')^3}\right) 
    \nonumber \\
    & \mathcal{P}_n \frac{\xi n}{f}\left(1 + \frac{\xi n}{2\pi R_c f}\frac{a(t')}{a(t_0)} \right)\frac{\Gamma(\alpha t_k)\theta(t_* - t_k) }{\Gamma(\alpha t_k) G \mu + \alpha(1 + \frac{\alpha t_k}{2\pi R_c})}.
\end{align}
 \begin{figure}
    \centering
    \includegraphics[width=0.48\textwidth]{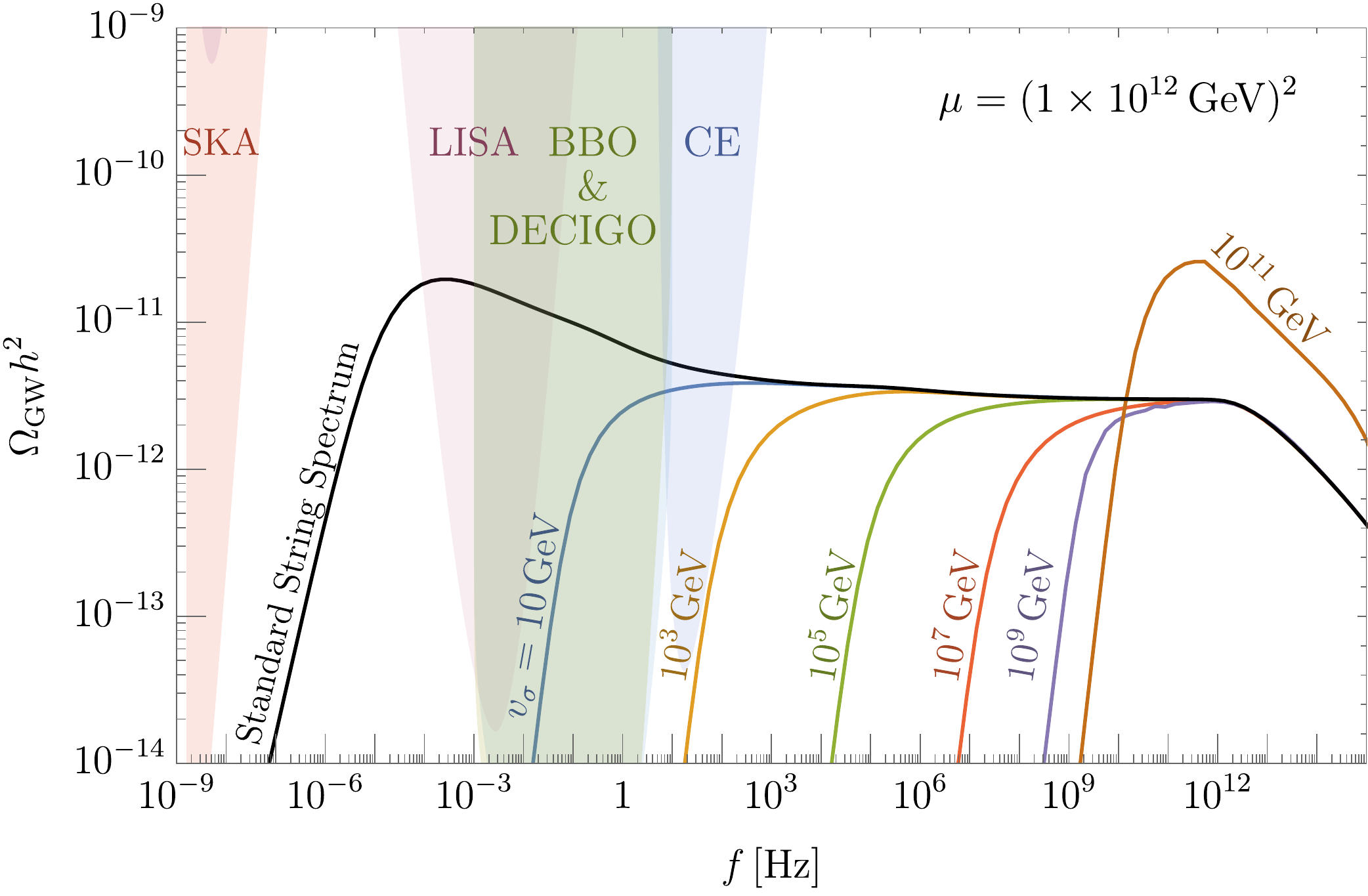}
    \caption{Representative spectra of gravitational waves emitted by strings that are eaten by domain walls for fixed $\sqrt{\mu} = 10^{12} \, \rm GeV$. Each colored contour corresponds to a different value of the wall symmetry breaking scales $v_\sigma$. Prior to wall domination at $t_*$, the wall-string network behaves similarly to a pure string network and $\Omega_{\rm GW} \propto f^0$ at high frequencies. After the network collapses and the largest string-bounded walls decay, $\Omega_{\rm GW}$ drops as $f^3$ at low frequencies. For $t_{\rm DW} < R_c$, the largest wall-bounded string pieces at decay do not live longer compared to pure string loops of the same size and hence do not deposit significantly more energy density into gravitational waves compared to pure string loops. There is no `bump' in $\Omega_{\rm GW}$ in this case. For $t_{\rm DW} \gg R_c$, the largest wall-bounded strings pieces at decay have size $R \gg R_c$ and are long-lived compared to pure string loops of the same size. These pieces deposit significant energy into gravitational waves at decay and generate a `bump' in $\Omega_{\rm GW}$ as shown by the $v_\sigma = 10^{11} \, \rm GeV$ contour.}
    \label{fig:GWPieceSpectrum}
\end{figure}
Fig. \ref{fig:GWPieceSpectrum} shows a benchmark plot of the gravitational wave spectrum from cosmic strings consumed by domain walls for fixed $v_\mu \equiv \sqrt{\mu} = 10^{12}$ GeV and a variety of $v_\sigma \equiv (\sigma/\epsilon)^{1/3}$, where we take $\epsilon = 1$. In computing the spectrum, we sum up $10^4$ normal modes and solve for the evolution of the scale factor in a $\Lambda$CDM cosmology. The colored contours in Fig. \ref{fig:GWPieceSpectrum} show the effect of $t_*$ on the spectrum while the  black contour shows the pure string spectrum, equivalent to the limit $t_* \rightarrow \infty$. When $v_\sigma \ll v_\mu$, the
walls form before dominating the strings and the network collapses at $t_* = R_c$, with the largest wall bounded strings approximately of size $\alpha R_c$. These wall bounded string pieces decay approximately with the same lifetime as pure strings of the same size, implying the spectrum is not enhanced over the pure string spectrum before decay. The smaller $R_c$ is, the longer the string network evolves as a pure string network, which is why the $f^3$ decaying spectrum in Fig. \ref{fig:GWPieceSpectrum} occurs at lower frequencies the lower $v_{\sigma}$ is. Conversely, when $v_\sigma \sim v_\mu$, as shown for instance, by the brown $v_{\sigma} = 10^{11}$ GeV contour, the walls form already dominating the strings. The network collapses at $t_* = t_{\rm DW} \gg R_c$ with the largest string-bounded walls approximately of size $\alpha t_{\rm DW}$. These string-bounded wall pieces decay much later than pure string loops of the same size, causing the spectrum to be enhanced over the pure string spectrum before decay, as shown by the bump near $10^{11}$ Hz. Because $v_{\sigma}$ must be near $v_\mu$ in this scenario, the frequency of the bump generally occurs at very high frequencies, as shown, for instance, by the yellow region of Fig. \ref{fig:wallStringSpectrumParameterSpace}. 

Finally, note that the spectrum is identical to the monopole nucleation spectrum of Sec. \ref{sec:schwingerStringsMonopole} at high frequencies, namely a pure string spectrum, but at low frequencies, the two gastronomy spectra are distinguishable by the slope of their infrared tails, which goes as $f^3$ and $f^2$, respectively.
\subsection{Frictional Losses and Chopping}.
\label{subsec:friction}
Until now, we have ignored the effect of string friction and wall friction on the gastronomy signal from walls eating strings. In this subsection, we investigate how friction can affect the evolution of the wall-string network and hence the gastronomy signal.

First, we consider friction on the strings due to the Aharonov-Bohm force, Eq.~\eqref{eq:aharonovBohmForce}. It can be shown \cite{Alford:1988sj,Vilenkin:1991zk,vilenkin2000cosmic} that the effect of this frictional force on the string equation of motion, \eqref{eq:EulerLagrangeEOM}, is to replace $\mathcal{H} \rightarrow \mathcal{H} + a(t)/L_f$, where $L_f = \mu/\beta_s T^3$ is known as friction length, which is effectively the reciprocal of the friction force per unit string mass. There are then four relevant scales (forces per unit mass) in the string equation of motion, with each dominating at a different stage in the evolution of the wall-string network:
\begin{align}
   &(a) \qquad 2 H v \quad &\text{(Hubble)} \nonumber
   \\
   &(b) \qquad \frac{\beta_s T^3v}{\mu} \quad &\text{(String Friction)}\nonumber
   \\
   &(c) \qquad\frac{1}{R}  \quad &\text{(String Tension)}\nonumber
   \\
   &(d) \qquad \frac{\sigma}{\mu}= \frac{1}{R_c}  \quad &\text{(Wall Tension)}\nonumber
\end{align}
Consider first the network evolution when $R < R_c$, which is the pure string limit. In this case, strings will be damped by friction until the Hubble (a) and friction terms (b) are equal. For a radiation-dominated era, this occurs at time
\begin{align}
   t_{f} = \frac{M_{Pl}^3 \beta_s^2 C^3}{8 \mu^2}.
\end{align}
where $C = (8\pi^3 g_*/90)^{-1/2}$ as before. After $t_{f}$,  the Hubble (a) and string curvature (c) terms dominate; the strings oscillate freely and the network reaches the standard scaling regime. If $R_c > t_{f}$, the walls do not dominate the string network until after the strings reach scaling and the results of this section are unchanged. The condition for the wall to dominate the string dynamics only after $t_f$ then occurs when 
\begin{align}
   \label{eq:stringFricCondition}
   \frac{R_c}{t_{\rm DW}} \geq \frac{\beta_s^{2/3}}{\epsilon}
\end{align}
is satisfied. For nearly all $(\mu, \sigma)$ with $t_* = R_c$, $R_c \gg t_{\rm DW}$ and hence Eq.~\eqref{eq:stringFricCondition} is easily satisfied and the gastronomy signal discussed in the previous subsection remain unchanged. 

However, for $t_* = t_{\rm DW}$, $R_c < t_{\rm DW}$ and Eq.~\eqref{eq:stringFricCondition} is generally not satisfied. In this case, the walls dominate the string dynamics during the initial string friction era. In this scenario, the two largest terms in the string equation of motion around the time of domain wall formation are the string friction term (b) and the wall tension term (d). Balancing the two terms gives the string terminal velocity
\begin{align}
   v = \frac{\sigma}{\beta_s T^3} \simeq \frac{\epsilon}{\beta_s}\left(\frac{t}{t_{\rm DW}}\right)^{3/2} ,
\end{align}
valid until $v$ becomes relativistic. Friction prevents the string-wall system from initially collapsing since the friction scale of the system, \footnote{We find a more rigorous derivation of the evolution of the string curvature from the Euler-Lagrange equation of motion gives the same scaling. Also note the different scaling compared to pure strings when the string curvature and friction balance, which gives $R \propto t^{5/4}$, known as the Kibble regime \cite{Kibble:1976sj,Hindmarsh:1994re}.}
\begin{align}
   R_f \sim v t = t\left(\frac{t}{t_{\rm DW}}\right)^{3/2}\frac{\epsilon}{\beta_s}
\end{align}
can be smaller than $R$. Specifically, perturbuations on the string larger than $R_f$ remain stuck by friction while those smaller than $R$ have been smoothed out by friction and can move freely. The wall-bounded strings cease expanding conformally when $R_f$ equals the string radius, $R = R(t_{\rm DW})(t/t_{\rm DW})^{1/2}$, which occurs at radius
\begin{align}
   \label{eq:frictionStringRadius}
   R \approx R(t_{\rm DW}) \times {\rm Max} \left(1, \, \left(\frac{R(t_{\rm DW}) \beta_s}{t_{\rm DW}\epsilon} \right)^{1/4} \right).
\end{align}
%
Unless $\epsilon \ll 1$ or $\beta_s \gg 1$, Eq.~\eqref{eq:frictionStringRadius} occurs at or very close to the string curvature radius at wall formation, $R(t_{\rm DW})$, since $R(t_{\rm DW})/t_{\rm DW} < 1$ in the friction regime. Thus, when $t_* = t_{\rm DW}$, the strings oscillate highly relativistically nearly immediately after wall formation, even with string friction. Nevertheless, there are still frictional energy losses after the strings move freely shortly after $t_{\rm DW}$. The power lost to Aharonov-Bohm friction for these pieces is given by equation \eqref{eq:powerAB}. Since the energy of the wall-bounded string piece is dominated by walls in this case, $E \sim \sigma R^2$, Eq.~\eqref{eq:powerAB} can be integrated to obtain the string-wall size vs time,
\begin{align}
   R(t) \approx R(t_{\rm DW}) - \frac{\beta_s C^{3/2}}{\sigma G^{3/4}}(t_{\rm DW}^{-1/2} - t^{-1/2}).
\end{align}
where we take $v\sim 1$. For $t > t_{\rm DW}$, the curvature radius quickly decreases to its asymptotic size
\begin{align}
   \label{eq:frictionAsymptoticR}
   R_{\rm final} \simeq R(t_{\rm DW})\left(1 - \frac{\beta_s}{\epsilon} \frac{t_{\rm DW}}{R(t_{\rm DW})}\right).
\end{align}
If the term in parenthesis remains of $\mathcal{O}(1)$, the wall bounded string pieces do not appreciably shrink due to friction and will decay via gravitational radiation. In such a scenario, the domain wall induced bump in $\Omega_{\rm GW}$ right before decay, a feature of the $t_* = t_{\rm DW}$ regime, still occurs but without the flat $f^0$ part of the spectrum to the right because the string network is frozen prior to $t_{\rm DW}$ and does not significantly emit gravitational waves. According to Eq.~\eqref{eq:frictionAsymptoticR}, the condition for the wall-bounded string pieces to remain long-lived is then
\begin{align}
   \label{eq:frictionAsymptoticCondition}
   R(t_{\rm DW}) \gtrsim \frac{\beta_s}{\epsilon} t_{\rm DW} .
\end{align}
In the friction regime, it is generally the case that $R(t_{\rm DW}) \ll t_{\rm DW}$, in contrast to the frictionless scaling regime when $R \sim t$ at wall formation.
\footnote{If the number of strings in the horizon at string formation time $t_\mu$ is sufficiently dilute such that  $( G \mu /\beta_s^2 C^2)^{1/4} \lesssim R(t_\mu)/t_\mu \lesssim 1$ \cite{Martins:1995tg,Aulakh:1998ua}, the strings are stretched with the scale factor $R(t)/t_\mu = \xi (t/t_\mu)^{1/2}$ where $\xi \equiv R(t_\mu)/t_{\mu}$. Since the horizon grows with $t$, the abundance of strings in the horizon increases with time. For sufficiently large string densities within the horizon, whether initially at $t_\mu$ or after increasing in the stretching regime, the curvature radius of strings enters the Kibble regime with $R(t)/t_\mu \sim (G \mu/\beta_s^2 C^2)^{1/4} (t/t_\mu)^{5/4}$, independent of whether the strings start in the stretching regime or Kibble regime. Only at the very beginning of the stretching regime if $\xi_0 \sim 1$ or at the end of the Kibble regime does $R \sim t$, (the latter of which anyway marks the end of the friction era), can Eq.~\eqref{eq:frictionAsymptoticCondition} be satisfied and a gastronomy signal be observed.}
As a result, if $t_* = t_{\rm DW} < t_f$, the wall-string sytem decays quickly to friction unless $\beta_s \ll 1$. If $\beta_s \sim 1$ at wall formation, the friction dominates and the wall-string system decays via friction in around a Hubble time, and the gastronomy signal is suppressed. This may eliminate the `bump' feature that occurs in $t_* = t_{\rm DW}$ cosmologies, as shown, for example, by the rightmost contour of Fig. \ref{fig:GWPieceSpectrum}. Nevertheless, there can still be an appreciable gravitational wave pulse from walls bounded by strings in this scenario. This is because the number of string-bounded walls in the horizon in the friction era can be significant, giving rise to a brief, but significant pulse of graviational waves similar to the monopole burst of Sec. \ref{sec:stringDestruction}. Moreover, the gastronomy signal for walls eating strings for the case of $t_* = R_c$ is still observable and distinguishable from other gastronomy signals even without its bump due to its $f^3$ infrared spectrum. 

In addition, after the string friction era, there can be friction on the walls from scattering with the bulk motion of the plasma \cite{Kibble:1976sj,Everett:1982nm}. Like string friction, wall friction is model dependent, and gives rise to a temperature dependent retarding force \cite{Everett:1982nm,vilenkin2000cosmic}
\begin{align}
   \label{eq:wallfriction}
   F_w \sim - \beta_w T^4 v R^2 ,
\end{align}
where $v$ is the velocity of the wall relative to the plasma, $R$ the wall curvature radius, and 
\begin{align}
   \beta_w \sim \sum_i w_i \frac{30 \zeta(3)}{\pi^4}
\end{align}
characterizes the number of relativistic particles that scatter with the scalar field composing the wall and where $w_i = 1$ for bosons and $6/7$ for fermions. If there are no particles with mass $m \ll T$ in the bath that stongly scatter with the scalar field of the wall, then $\beta_w = 0$, and the following discussion is inapplicable. Likewise, if the only particles that strongly scatter with the wall are of order $v_\mu$, such as scalar field of the wall itself, $\beta_w$ quickly becomes exponentially suppressed and the following discussion is inapplicable. If there exists such particles and $\beta_w \gtrsim 1$, the balance of the friction force \eqref{eq:wallfriction}, with the wall tension force, $F \sim \sigma R$, gives the terminal velocity of the walls,
\begin{align}
   \label{eq:wallTerminalSpeed}
   v \sim \frac{4 G \sigma t^2}{\beta_w c^2 R}
\end{align}
in a radiation dominated era. We now follow the discussion of \cite{Everett:1982nm}, but further generalize the authors' results to the case when $t_* = t_{\rm DW}$ which was not considered previously. The wall friction scale is
\begin{align}
   R_f \sim v t \approx \sqrt{\frac{4G \sigma t^3}{\beta_w c^2}}
\end{align}
whereas the string curvature of the infinite string-wall network scales as $R \sim t$ in the scaling regime. Perturbations on the wall larger than $R_f$ remain stuck by friction while those smaller than $R$ have been smoothed out by friction and can move freely. At $t = t_*$, the wall dominates the string dynamics, and normally, this would cause the walls to pull the strings into the horizon and oscillate at constant amplitude as discussed in Sec. \ref{subsection:collapseOfInfiniteNetwork}. However, the $R > R_f$ periphery of the wall and hence string boundary (which goes along for the ride) is conformally stretched  until $R_f$ equals the string radius, $R = R(t_*)(t/t_*)^{1/2}$, which occurs at time $t_1$ and curvature radius
\begin{align}
   \label{eq:endFrictiontime}
   t_1 \sim \frac{t_*}{\delta} \qquad R(t_1) \sim \frac{t_*}{\sqrt{\delta}}
\end{align}
where
\begin{align}
   \delta = \sqrt{\frac{G \sigma t_*}{\beta_w C^2}}
\end{align}
valid for $t_* = R_c$ or $t_* = t_{\rm DW}$. At time $t_1$, the wall-bounded string pieces cease being conformally stretched and oscillate at constant size. Nevertheless, the walls lose energy via friction. The power lost to friction by the walls is
\begin{align}
   \label{eq:wallPowerFriction}
   P_f = F_w v \sim - \frac{\beta_w c^2}{G t^2} R(t_1)^2 v^2
\end{align}

where $v = \delta^{1/2}(t/t_1)^2$ for $t > t_1$ using Eq.~\eqref{eq:wallTerminalSpeed} and \eqref{eq:endFrictiontime}. The integral of \eqref{eq:wallPowerFriction} gives the energy of the system as a function of time,
\begin{align}
   \label{eq:wallEnergyVsTime}
   E(t) = E(t_1)- \frac{1}{3}E(t_1)\left(\frac{t^3}{t_1^3}  - 1\right)
\end{align}
where $E(t_1) \sim \sigma R(t_1)^2 = G \sigma^2 t_1^3/\beta_w C^2$ is the initial energy of the wall at $t_1$. Eq.~\eqref{eq:wallEnergyVsTime} demonstrates that the walls lose most of their energy in a Hubble time after $t_1$. The energy loss causes the size of the string-bounded walls to shrink until they become relativistic, which occurs at $t_2 \sim t_1$, and, according to \eqref{eq:wallTerminalSpeed}, at $R(t_2) \sim t_*$. At this point, Eq.~\eqref{eq:wallEnergyVsTime} is invalid and we must return to Eq.~\eqref{eq:wallPowerFriction} to describe the power lost to friction by the relativistic wall-string piece. If $t_* = R_c$, the time at which the walls become relativistic coincides with the moment the strings return to dominating the dynamics of the shrinking wall-bounded string piece, that is, $R \approx R_c$ when $v$ becomes $1$. If $t_* = t_{\rm DW}$, $R > R_c$ when $v$ becomes $1$, and the walls still dominate the dynamics. However, it is easy to see that if the wall still dominates the dynamics for $v \sim 1$, the curvature radius exponentially drops in time so that even for the case $t_* = t_{\rm DW}$, the wall-bounded string pieces will shrink to $R_c$ at $t \sim t_1$. 

But the shrinking can continue further. Once the string dominates the dynamics and $v \sim 1$, the integration of the power loss, Eq.~\eqref{eq:wallPowerFriction}, gives the curvature radius of the wall-bounded string piece as
\begin{align}
   R(t) = \left[\frac{1}{R(t_2)} - \frac{\beta_w c^2}{G \mu}\left(\frac{1}{t} - \frac{1}{t_2} \right)\right]^{-1}
\end{align}
which asymptotically shrinks to 
\begin{align}
   R_{\rm final} = \frac{G \mu t_2}{\beta_w c^2} = R_c \delta .
\end{align}
For $\beta_w \gtrsim 1$, $\delta \ll 1$ and the wall-bounded string pieces shrink far below $R_c$ and subsequently decay quickly via gravitational radiation. In this scenario, the gastronomy signal is again suppressed, even for $t_* = R_c$, unlike the case for string friction. However, this is highly model dependent and requires relativistic particles in the thermal bath to scatter off the domain wall far past wall formation so that $\beta_w \gtrsim 1$ still at $t_1$. If the only particles that scatter with the wall have mass compared to $v_\sigma$, then $\beta_w \ll 1$ by $t_1$ so that the wall friction becomes negligible and the gastronomy discussion of the previous subsections are unchanged.

Last, we mention that it is possible that the wall-bounded string pieces can potentially lose energy from self-intercommutation when they oscillate, thereby chopping themselves into finer pieces. If this occurs, the chopping is likely to stop becoming important once the pieces slice and dice themselves below $R < R_c$ at which point the strings dominate the dynamics and the wall-bounded string pieces dynamically behave similar to pure string loops. If this occurs, it only effects the $t_* = t_{\rm DW}$ parameter space where the wall-bounded string pieces can have curvature radii $R > R_c$. Moreover, the final number of chopped pieces of size $R_c$ will be greater than the usual $t_* = R_c$ cosmology because the total energy in the wall-bounded string pieces post chopping is similar to pre-chopping due to energy conservation
\footnote{The total energy density of the system pre and post chopping may be somewhat smaller if the chopped pieces inherit a large translation kinetic energy which can be redshifted away by the expansion of the Universe.}.
Furthermore, for an asymptotic chopped radius of $R \sim R_c$, the lifetime of the chopped wall-bounded string pieces is comparable to larger pieces with $R > R_c$ because the gravitational wave power at this radius is approximately proportional to the wall mass so that the lifetime is the same for string-bounded walls for any size $R \gtrsim R_c$. Thus, because the total energy density and lifetime of the chopped pieces remains similar to the pre-chopped pieces, the `bump' in the spectrum for $t_* = t_{\rm DW}$ cosmologies should stay roughly the same height if there was no chopping, but may be shifted to slightly higher frequencies because the pieces are smaller than before.

\section{Topological defects washed out by inflation}
\label{sec:pre-inflation}

Inflation exponentially dilutes all topological defects. This is useful for removing monopoles, for which even small relic abundances are in tension with present-day cosmology. However other topological defects such as superhorizon strings and domain walls dilute slower than the background radiation and hence can replenish even after enduring many $e-$folds of inflation. Examples of symmetry breaking chains in Fig. \ref{fig:chains} where this can occur are
\begin{eqnarray}
  {\rm SO}(10) &\to & G_{\rm SM} \times \mathbb{Z}_2 \nonumber \\
  {\rm SO} (10) & \to & {3221D} \nonumber 
\end{eqnarray}
since these chains \textit{simultaneously} produce stable monopoles and strings and thus require the strings to be diluted by inflation too.

Recent work \cite{Cui:2019kkd} found that if a string network forms early in inflation, the strings can replenish enough such that bursts emanating from ultrarelativistic cusps can give an observable signal at frequencies around pulsar timing arrays. However, there is a limit to how many e-folds strings can be diluted and still leave an observable signal.

Limits on monopole flux are most stringent for monopoles that catalyze baryon number violation. Such bounds on the flux, $\Phi$, are at no stricter than \cite{Freese:1998es} 
\begin{equation}
    \Phi \lesssim 10^{-28} {\rm cm} ^{-2}{\rm sr}^{-1} {\rm sec}^{-1}
\end{equation}
which requires at least 30 $e-$foldings of inflation to dilute, whereas strings can replenish after many more $e-$foldings \cite{Cui:2019kkd} - up to about 54. 

Domain walls in principle can also replenish after being diluted by inflation. The evolution of a domain wall network can be estimated by taking a conservative initial number density to be $H_I^3$ (the Kibble or scaling limit) and the initial mass of a domain wall to be $\sigma /H_I^2$, where $H_I$ is the value of Hubble during inflation. After formation, the domain wall is stretched by $N$ e-foldings and the number density is diluted by a factor $e^{-3N}$. Due to the superhorizon size, the walls are conformally stretched with the evolution of the curvature radius $R$ stretching with the scale factor until horizon re-entry when $HR=1$.
After horizon re-entry, that is when the domain wall size is the Hubble size, the domain walls reach a scaling regime and $\rho_{\rm DW} \approx \sigma/R \propto 1/t$, which is slower than all other energy densities bar the vacuum contribution. In order to not dominate the energy density today and taking $H=10^{13}$ GeV, domain walls require nearly 100 e-foldings.  In principle, if a small amount of the energy budget today is from domain walls, a larger fraction could occur during recombination, implying a larger expansion rate in the early Universe. We leave the phenomenology of such a possibility to future work.

\section{Summary}
\label{sec:summary}



In this work, we have studied the formation, evolution, decay, and gravitational wave gastronomy signatures of hybrid topological defects. These objects, composed of two different dimensional topological defects bounded to each other, come in two flavors:  cosmic strings bounded by monopoles and domain walls bounded by cosmic strings.  As shown in Fig. \ref{fig:chains}, these hybrid defects are common in many breaking chains from $SO(10)$ to the Standard Model. As a result, if the early Universe reached sufficiently high temperatures, it is possible that hybrid defects were once part of our cosmic history. 

The relativistic motion of defects, and especially during the `devouring' of one defect by the other, leads to interesting gravitational wave signatures. We revisited the gravitational wave spectrum of strings unstable toward monopole pair creation in Sec. \ref{sec:schwingerStringsMonopole} and found a range of monopole and string symmetry breaking scales that are observable at near-future gravitational wave detectors, including within the recent NANOGrav and PPTA signal region. Similarly, we estimated the gravitational wave signal from domain walls unstable toward string holes nucleating on their surface in Sec. \ref{sec:dwconsumedbystrings}. In both nucleation cases, the gravitational wave spectrum prior to nucleation behaves as a pure string or wall network, respectively. The frequency dependence of the nucleation gastronomy scenarios are summarized in Table \ref{tab:courses}. 

Note that since nucleation is an exponentially suppressed process, the defect can be long-lived, therefore the scale size of the topological defect at decay can be large and hence emit in low frequencies observable at near future gravitational wave detectors. Nevertheless, while nucleation gastronomy scenarios typically involve easier to detect lower frequency gravitational waves, the likelihood of a nucleation gastronomy may be challenging as it requires a near degeneracy in symmetry breaking scales of the bulk and boundary defects. 

Other types of cosmological scenarios with hybrid defects, such as a monopole network becoming connected to (and eaten by) strings, and string loops becoming filled with (and eaten by) domain walls do not require any fine-tuning of the symmetry breaking energy scales. We constructed analytic models for these hybrid defects and found that they predict gravitational waves typically at high frequencies of order $10^{1-10}$ Hz, which will be explored by interferometers in some parts of parameter space, but will typically need new experimental techniques to detect the signal. Unlike nucleation, the gravitational wave signals for these gastronomy scenarios are typically high frequency because the hybrid defects decay around the time of string or domain wall formation, respectively. This can occur in the early Universe when the defects are physically small. If future high frequency detectors can observe such a signal, they may be able to see unique spectral features as shown in Table \ref{tab:courses} or even a characteristic `bump' on top of a pure string spectrum when domain walls eat strings. To confirm our analytic models describing the hybrid defects in this paper, numerical simulations will be needed.

Because all four gastronomy spectra are distinguishable by their UV and IR frequency dependence, a measurement around the peak of $\Omega_{\rm GW}$ can be used to determine the IR and UV spectral dependence. In some cases this only requires detecting the spectra over a relative small frequency domain. It may then be possible to infer which of the four types of cosmic courses generated $\Omega_{\rm GW}$.  Knowledge of the gastronomy course thus elucidates the hybrid defect which created that signal. Consequently, knowing that a certain hybrid defect existed in the early Universe can be used to narrow down the subset of GUT symmetry chains that produce that hybrid defect. Moreover, the amplitude and frequency dependence $\Omega_{\rm GW}$ can be used to infer the scales of symmetry breaking associated with \textit{both} the boundary and bulk defects. Thus, gravitational wave gastronomy has the ingredients to infer many fundamental properties of Nature.

\begin{table}
    \centering
    \begin{tabular}{m{5.5cm}|m{1cm}|m{1cm}}
      Cosmic Course & IR &  UV  \\ \hline 
      Monopoles Eating String Network (Nucleation)    & $f^2$ & $f^0$ \\ \hline  
      Strings Eating Monopole Network (Collapse)      & $f^3$ & $\ln f \rightarrow f^{-1}$ \\ \hline
      Strings Eating Domain Wall Network (Nucleation) & $f^3$ & $f^{-1}$ \\ \hline
      Domain Walls Eating String Network (Collapse)   & $f^3$ & $f^0$ \\
    \end{tabular}
    \caption{A summary of the different gastronomy signals and the characteristic fingerprints of their gravitational wave spectra at low (IR) and high (UV) frequencies. Since each gastronomy signal has a unique combination of the spectral index in the IR and UV, it is possible to map a gravitational wave spectrum to a given gastronomy scenario and hence a subset of GUT symmetry breaking chains.}
    \label{tab:courses}
\end{table}


\begin{acknowledgments}
We thank Kaustubh Agashe, Zackaria Chacko, Valerie Domcke, Lawrence Hall, Keisuke Harigaya, Takashi Hiramatsu, Anson Hook, Peera Simakachorn, Raman Sundrum, and Ofri Telem for useful discussions. HM was supported by the Director, Office of Science, Office of
High Energy Physics of the U.S. Department of Energy under the
Contract No. DE-AC02-05CH11231, by the NSF grant
PHY-1915314, by the JSPS Grant-in-Aid for
Scientific Research JP20K03942, MEXT Grant-in-Aid for Transformative Research Areas (A)
JP20H05850, JP20A203, by World Premier International Research Center Initiative, MEXT, Japan, and Hamamatsu Photonics, K.K.
YS is supported by National Natural Science Foundation of
China with Grant No.~12005309 and Young Teachers
Training Program of Sun Yat-sen University with Grant No.~20lgpy168.
\end{acknowledgments}

\appendix 
\section{Homotopy Selection Rules}
\label{ap:homotopySelectionRules}
As can be seen from Fig.~\ref{fig:chains}, all hybrid defects from the breaking of $SO(10)$ involve a symmetry breaking chain where the homotopy group of the first symmetry breaking step has a higher group number, $n$, then the succeeding one,
e.g., for monopole-bounded strings, the monopoles form prior to strings.
In order to show that the feature 
is a generic feature of any symmetry breaking group, 
we derive homotopy selection rules leading to hybrid defects.

For any groups $G \supset H \supset K$, there is a fiber bundle\footnote{For a proof, see \href{https://ncatlab.org/nlab/show/principal+bundle}{this page}.}
\begin{align}
    H/K \rightarrow G/K \rightarrow G/H.
\end{align}
It leads to the exact sequence of homotopy groups,
\begin{align}
    \cdots &\rightarrow \pi_n(H/K) \rightarrow \pi_n(G/K) \rightarrow \pi_n(G/H)
    \nonumber \\
    &\rightarrow \pi_{n-1}(H/K) \rightarrow \pi_{n-1}(G/K) \rightarrow \pi_{n-1}(G/H)
    \rightarrow \cdots
\end{align}
with each arrow indicating a homomorphism whose image is equal to the kernel of the following homomorphism.
For a topological defect of dimension $k$ in three-dimensional space, its stability is guaranteed by $\pi_{2-k}$. If a defect is stable at one step of symmetry breaking, we need a non-trivial $\pi_{2-k}(H/K)$ or $\pi_{2-k}(G/H)$, while if it is unstable in the whole theory, we need $\pi_{2-k}(G/K)=I 
$. This tells us to study a part of the exact sequence, ($n=2-k$)
\begin{align}
    I = \pi_n(G/K) 
    \rightarrow \pi_n(G/H) \rightarrow \pi_{n-1}(H/K) ,
    \label{eq:classical}
\end{align}
or 
\begin{align}
    \pi_n(G/H)
    \rightarrow \pi_{n-1}(H/K) \rightarrow \pi_{n-1}(G/K)=I. 
    \label{eq:tunneling}
\end{align}
In the first case \eqref{eq:classical}, the image of the homomorphism between $\pi_n(G/K)$ and $\pi_n(G/H)$ is $I$. Thus, the kernel of the homomorphism between $\pi_n (G/H)$ and $\pi_{n-1}(H/K)$ is $I$,
implying the homomorphism is injective and hence $\pi_n (G/H) \subseteq \pi_{n-1}(H/K)$. Therefore, any element of $\pi_n (G/H)$ at the first stage of symmetry breaking can be ``undone'' by an element of $\pi_{n-1}(G/K)$ at the second stage of symmetry breaking, and hence a $k$-dimensional defect formed at the first phase transition can be destabilized by a $(k+1)$-dimensional defect formed at the second phase transition. For example, a string can be filled with a wall, or monopoles can be connected by a string to destabilize the defect.

In the second case \eqref{eq:tunneling}, the kernel of the homomorphism between $\pi_{n-1}(H/K)$ and $\pi_{n-1}(G/K)$ is $\pi_{n-1}(H/K)$. Thus, the image of the homomorphism between $\pi_n(G/H)$ and $\pi_{n-1}(H/K)$ is $\pi_{n-1}(H/K)$ and hence $\pi_n(G/H) \supseteq \pi_{n-1}(H/K)$. Therefore, any element of $\pi_{n-1}(H/K)$ at the second stage of the symmetry breaking can be ``undone'' by an element of $\pi_n(G/H)$ at the first stage of symmetry breaking, and hence a $(k+1)$-dimensional defect formed at the second phase transition can be destroyed by the production of $k$-dimensional defect formed at the first transition. For example, a string can be cut by the nucleation of a monopole-antimonopole pair, or a domain wall can be punctured by the nucleation a string-bounded hole.

In summary,
{\it the lower dimensional topological defect (boundary defect) of a hybrid defect forms earlier than the one dimensional higher topological defect (bulk defect) that it attaches to.}
\section{Action of a String-Bounded Domain Wall}\label{ap:DWstring}
\begin{figure}[b]
    \centering
    \includegraphics[width=.35\textwidth]{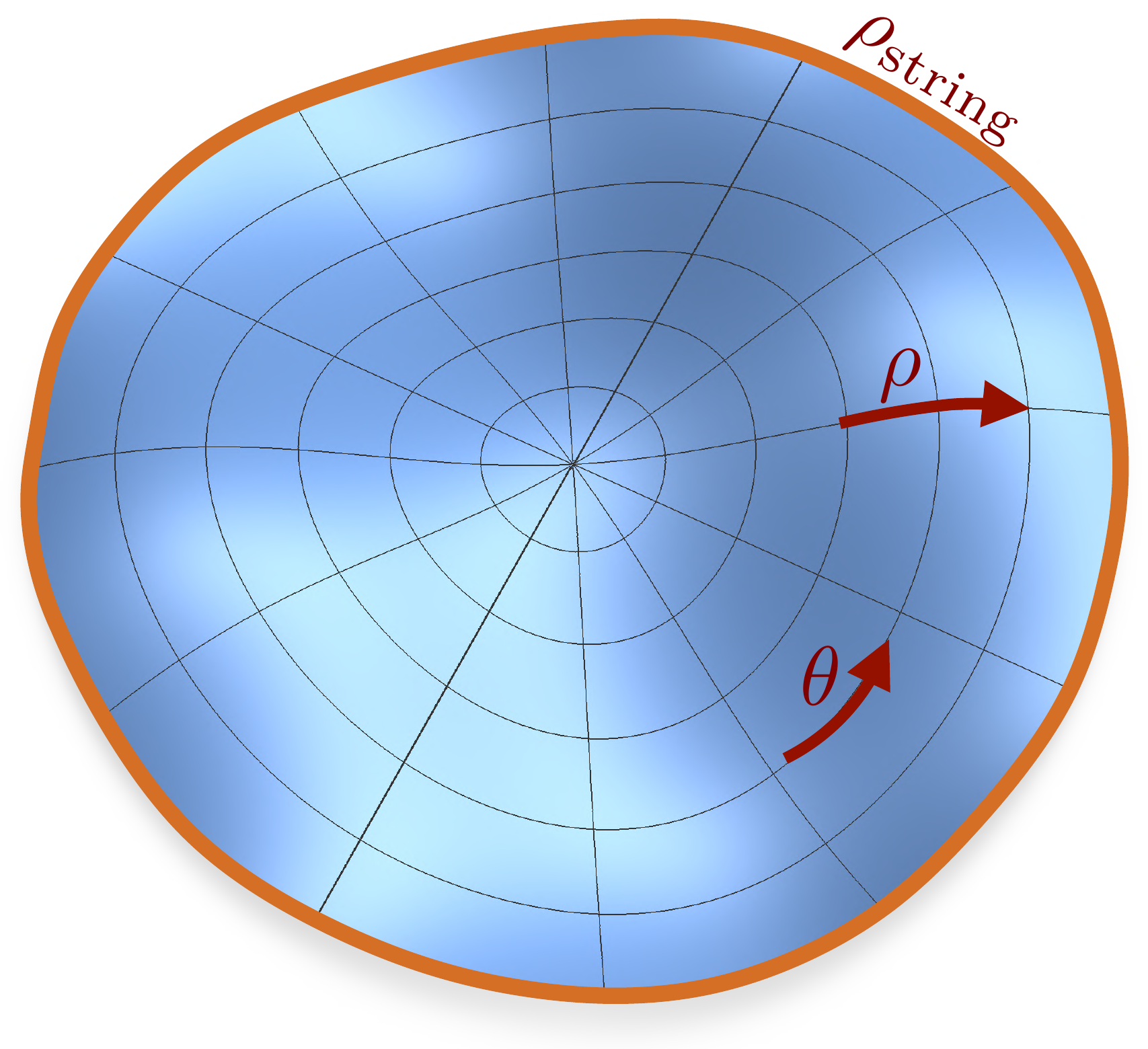}
    \caption{Coordinate parameterization of a wall-bounded string. The coordinates $\xi^1 = \rho$ and $\xi^2 = \theta$ are orthogonal and parameterize the radial and azimuthal directions in the plane of the wall, respectively. The string lies on the boundary of the wall at the coordinate $\rho_{\rm string}$.}
    \label{fig:wallParameterization}
\end{figure}

Here we derive the Lagrangian, \eqref{eq:action_subs_dw-str}, from section \ref{sec:stringboundedWalls}.
Since the worldvolumes of the string and wall are invariant under re-parameterizations of the coordinates $\zeta$, we choose a coordinate system on the wall and string such that $\zeta^0 =\eta, \zeta^1 = \theta$, $\zeta^2 = \rho$, where $0 \leq \theta < 2\pi$ parameterizes the polar direction in the plane of the wall and $0 \leq \rho \leq \rho_{\rm string}$
parameterizes the radial direction in the plane of the wall.  $\rho_{\rm string}$ is the boundary of the wall
located at the attached string, as shown in Fig. \ref{fig:wallParameterization}. In this basis, $X^\mu = (\eta, \mathbf{X})$, $Y^\mu = X^\mu \rvert_{\rho_{\rm string}}
$, $(\partial_\rho \mathbf{X})^2 = 1$, and $\partial_\rho \mathbf{X} \cdot \partial_\theta \mathbf{X} = 0$. 

The determinant of the induced metric on the wall may be written as
\begin{align}
   \label{eq:general_action_dw}
    \gamma &= a^6(\eta)
    \begin{vmatrix}
        1-(\partial_\eta \mathbf{X})^2 && - \partial_\eta \mathbf{X} \cdot \partial_\theta \mathbf{X} && - \partial_\eta \mathbf{X} \cdot \partial_\rho \mathbf{X}
        \\
        - \partial_\eta \mathbf{X} \cdot \partial_\theta \mathbf{X} && -(\partial_\theta \mathbf{X})^2 && 0
        \\
        - \partial_\eta \mathbf{X} \cdot \partial_\rho \mathbf{X} && 0 && -1
    \end{vmatrix}
    \\
    \label{eq:general_action_dw2}
    &= \frac{a^6(\eta)}{\gamma_{\perp,w}^2} (\partial_\theta \mathbf{X})^2,
\end{align}
where we define $\gamma_{\perp,w} = (1-v_{\perp, w}^2)^{-1/2}$ as the Lorentz factor for motion perpendicular to the wall. In going from Eq.~\eqref{eq:general_action_dw} to \eqref{eq:general_action_dw2}, we decompose the wall velocity into perpendicular and tangential motion, $\partial_\eta \mathbf{X} = v_{\perp,w} \hat{\theta} \times \hat{\rho} + \mathbf{v}_{\parallel}$, where 
$\mathbf{v}_{\parallel} = (\partial_\eta \mathbf{X} \cdot \partial_\rho \mathbf{X}) |\partial_\rho \mathbf{X}|^{-1}\hat{\rho} + (\partial_\eta \mathbf{X} \cdot \partial_\theta \mathbf{X}) |\partial_\theta \mathbf{X}|^{-1}\hat{\theta}$.
As indicated by Eq.~\eqref{eq:general_action_dw2}, only the motion perpendicular to the wall is physical, and any use of `$v_w$' in the text means $v_{\perp, w}$.
Similarly, the determinant of the induced metric on the string worldsheet is
\begin{align}
    \Upsilon &= a^4(\eta) 
    \begin{vmatrix}
        1-(\partial_\eta \mathbf{Y})^2 && - \partial_\eta \mathbf{Y} \cdot \partial_\theta \mathbf{Y}
        \\
        - \partial_\eta \mathbf{Y} \cdot \partial_\theta \mathbf{Y} && -(\partial_\theta \mathbf{Y})^2 
    \end{vmatrix}
    \\
    &= 
    \frac{a^4(\eta)}{\gamma_{\perp,s}^2} (\partial_\theta \mathbf{Y})^2,
    \label{eq:general_action_str}
\end{align}
where $\mathbf{Y} = \mathbf{X}(\theta,\rho_{\rm string}
)$ and $\gamma_{\perp,s}=(1-v_{\perp,s}^2)^{-1/2}$ is the Lorentz factor for motion perpendicular to the string, which will dominantly be in the tangent plane of the wall when the wall dominates the string dynamics.
As a result of Eqns.~\eqref{eq:general_action_dw2} and \eqref{eq:general_action_str}, the combined action of the domain wall and string system ~\eqref{eq:action_for_dw-str} becomes

\begin{align}
    \label{eq:stringBoundedWallAction}
    S &= -\sigma \int d\eta \int_0^{2\pi} d\theta \left|\frac{d \mathbf{X}}{d\theta}\right| \int_0^{\rho_{\rm string}} d\rho \frac{a^3(\eta)}{\gamma_{\perp,w}}
    \nonumber \\
    & -\mu \int d\eta \int_0^{2\pi} d\theta \left|\frac{d \mathbf{Y}}{d\theta} \right|\frac{a^2(\eta)}{\gamma_{\perp,s}}.
\end{align}
Because the `eating' of the wall by the string converts wall rest mass energy to string kinetic energy, we expect $\gamma_{\perp,s} > \gamma_{\perp,w} \sim 1$,
\footnote{Simulations of domain walls without strings, which do not transfer any mass energy into string kinetic energy, only have perpendicular RMS velocities mildly relativistic, $v_{\perp,w} \sim 0.3 $ \cite{Avelino:2005kn}, while pure string loops have intermediate RMS velocities $v_{\perp,s} = \sqrt{2} \simeq .707$. Strings attached to walls become even more relativistic from the conversion of wall rest mass to string kinetic energy during the `eating' process (see Fig. \ref{fig:GWBoundedStringRadius}) which makes this approximation better. Only for enclosed domain walls without strings, or `vacuum bags', which collapse relativistically under their own tension, do we expect $v_{\perp,w}$ to be significant.}
and for the string velocity, $v_{\perp,s}$ to be directed in the tangent plane of the wall.
As a result, we analyze Eq.~\eqref{eq:stringBoundedWallAction} in the limit $\gamma_{\perp,w} \rightarrow  1$ where the perpendicular wall velocity is small and subdominant compared to the string velocity. In addition, we take the string boundary to be a circular loop $\mathbf{Y}=\mathbf{r}_s$  of coordinate radius $\mathbf{r}_s$ (physical radius $\mathbf{R}_s =  \mathbf{r}_s a(\eta)$), though we do not expect more realistic loops that are not perfectly circular to behave quantitatively
different since the key relationship between wall mass and string radius, $M_{\rm DW} \sim \sigma |\mathbf{R}_s|^2$ will still hold for more complicated loop geometries, and it is this energy which is transferred to the string as kinetic energy. 
Under these assumptions, we obtain the Lagrangian for a domain wall disc with a circular string loop boundary~Eq.~\eqref{eq:action_subs_dw-str}.

For a nucleated string hole, the string forms the inner boundaries of a domain wall and the integration over the radial coordinate $\rho$ in Eq.~\eqref{eq:stringBoundedWallAction} then begins at $\rho = \rho_{\rm string}$ up to some arbitary bulk $\rho$. The effect is thus a relative minus sign in the of Lagrangian Eq.~\eqref{eq:action_subs_dw-str}  between the string and domain wall terms. In an non-expanding Universe, or for subhorizon times and distances, the solution to the Euler-Lagrange equation of motion is the relativistic rocket, of Eq.~\eqref{eq:stringBoundedWallAction}, which is in agreement with results found in \cite{vilenkin2000cosmic}.

\bibliography{DW}

\end{document}